\documentclass{tcibook}
\usepackage{fancyhea}
\usepackage{work}
\usepackage{bm}       
\usepackage{graphicx}
\usepackage{hyperref}      
\usepackage{graphicx}
\usepackage{wasysym}
\usepackage{cite}  

\newcommand{\nc}{\newcommand}  

\def\beq{\begin{equation}}
\def\eeq#1{\label{#1}\end{equation}}
\def\eeqn{\end{equation}}

\newenvironment{Eqnarray}%
   {\arraycolsep 0.14em\begin{eqnarray}}{\end{eqnarray}}
\def\beqa{\begin{Eqnarray}}
\def\eeqa#1{\label{#1}\end{Eqnarray}}
\def\eeqan{\end{Eqnarray}}

\nc{\ra}{\rightarrow}  
\nc{\slsh}{\slash\hspace*{-0.22cm}}
\def\Re{{\cal R \mskip-4mu \lower.1ex \hbox{\it e}\,}}
\def\Im{{\cal I \mskip-5mu \lower.1ex \hbox{\it m}\,}}

\nc{\vev}[1]{ \left\langle {#1} \right\rangle }
\nc{\bra}[1]{ \langle {#1} | }
\nc{\ket}[1]{ | {#1} \rangle }
\nc{\fb}{\,{\rm fb}^{-1}}
\nc{\ev}{{\rm eV}}
\nc{\kev}{{\rm keV}}
\nc{\Mev}{{\rm MeV}}
\nc{\gev}{{\rm GeV}}
\nc{\tev}{{\rm TeV}}
\nc{\mev}{{\rm MeV}}

\def\del{\partial}
\def\Dslash{\not{\hbox{\kern-4pt $D$}}}
\def\dslash{\not{\hbox{\kern-2pt $\del$}}}
\def\pslash{\not{\hbox{\kern-2pt $p$}}}
\def\ETmiss{ \not{\hbox{\kern-4pt $E$}}_T }

\def\eff{{\mbox{\scriptsize eff}}}

\def\BR{\mbox{\rm BR}}

\def\msb{{\bar{\ssstyle M \kern -1pt S}}}

\usepackage{hyperref}
\usepackage{latexsym}
\usepackage{xspace}
\usepackage{relsize}
\usepackage{subfigure}
\usepackage[section]{placeins}

\usepackage[figuresright]{rotating}

\begin{document}
\pagenumbering{none}
\def\bibname{References}

\raggedbottom

\pagenumbering{roman}

\parindent=0pt
\parskip=8pt
\setlength{\evensidemargin}{0pt}
\setlength{\oddsidemargin}{0pt}
\setlength{\marginparsep}{0.0in}
\setlength{\marginparwidth}{0.0in}
\marginparpush=0pt


\renewcommand{\chapname}{chap:intro_}
\renewcommand{\chapterdir}{.}
\renewcommand{\arraystretch}{1.25}
\addtolength{\arraycolsep}{-3pt}



\chapter{Higgs working group report}
\vspace*{10pt}
\centerline{Conveners:  Sally Dawson (BNL), Andrei Gritsan (Johns Hopkins), Heather Logan (Carleton), }
\centerline{Jianming Qian (Michigan), Chris Tully (Princeton), Rick Van Kooten (Indiana)}
\vspace*{10pt}
\noindent
Authors: 
A.~Ajaib,
A.~Anastassov,
I.~Anderson,
D.~Asner,
O.~Bake, 
V.~Barger,
T.~Barklow, 
B.~Batell, 
M.~Battaglia,
S.~Berge
% ...

\vspace*{10pt}

\vspace*{10pt}
\vspace*{10pt}
\centerline{\bf Abstract}

This report summarizes the work of the Energy Frontier Higgs Boson
working group of the 2013 Community Summer Study (Snowmass).  We
identify the key elements of a precision Higgs physics program and
document the physics potential of future experimental facilities as
elucidated during the Snowmass study.  We study Higgs couplings to
gauge boson and fermion pairs, double Higgs production for the Higgs
self-coupling, its quantum numbers and $C\!P$-mixing in Higgs couplings, 
the Higgs mass and
total width, and prospects for direct searches for additional
Higgs bosons in extensions of the Standard Model. Our report
includes projections of measurement capabilities from detailed studies
of the Compact Linear Collider (CLIC), a Gamma-Gamma Collider, the
International Linear Collider (ILC), the Large Hadron Collider
High-Luminosity Upgrade (HL-LHC), Very Large Hadron Colliders up to 100 TeV (VLHC),
a Muon Collider, and a Triple-Large
Electron Positron Collider (TLEP). 

\vspace*{10pt}

\pagenumbering{arabic}
\vspace{-0.25 truein}
\section{Introduction}\vspace*{-0.3cm}

The quest to understand the origin of mass spans at least four major energy
frontier facilities in the last 25 years -- 
from the SLC linear $e^+e^-$ collider at SLAC
and LEP circular $e^+e^-$ collider at CERN,
to the Tevatron proton-antiproton collider at Fermilab, 
and finally to the LHC at CERN.
Now, for the first time, Higgs physics is experimentally verified to be
an inextricable part of the universe and the physical laws that govern it.
While we do not know at this time whether the simplest possible incarnation
of the Higgs mechanism is what occurs in Nature, the fact that
the new boson was discovered in the search for the Standard Model Higgs boson indicates
that the basic features of the Higgs mechanism are correct.
Any significant deviation in the properties or couplings of the Higgs boson imply
fundamental changes to the understanding of elementary particles and interactions.
Furthermore, the role of the Higgs field in early universe physics and in the
unification of the forces of Nature are highly sensitive to Higgs boson properties including 
the mass, total width, spin, couplings, CP mixtures, and the existence of multiple Higgs bosons.
With this perspective, the future of the energy frontier is contemplated
with a focus on precision measurements of high statistics samples of Higgs
bosons produced in ideal conditions for laboratory studies.

The compilation of the Higgs Snowmass Report is based on input in the form of White Papers
from the particle physics community.  All major precision Higgs physics projects with
substantial representation in the US particle physics and accelerator physics communities 
have been included in this report.  These communities are listed here (in alphabetical order):
\begin{itemize}
\item[$\bullet$] Compact Linear Collider (CLIC)~\cite{Aicheler:1500095,Dannheim:2013ypa,CLICWhite:2013a}
\item[$\bullet$] Gamma-Gamma Collider~\cite{Bogacz:2012fs,Asner:2001ia}
\item[$\bullet$] International Linear Collider (ILC)~\cite{ILCWhite:2013a,Behnke:2013xla,Brau:2013mba,Baer:2013vqa,ILCprivatecomm}
\item[$\bullet$] Large Hadron Collider High Luminosity Upgrade (HL-LHC)
\item[$\bullet$] Muon Collider ($\mu$C)~\cite{MuonColliderWhitepaper,Delahaye:2013jla}
\item[$\bullet$] Triple-Large Electron-Positron Collider (TLEP)~\cite{Koratzinos:2013ncw,TLEP:2013a}
\end{itemize}
The proposed running periods and integrated luminosities at each of the center-of-mass energies
for the above facilities are listed in Table~\ref{facilities}.

\begin{table}[htb!]
\begin{center}
\caption{
Proposed running periods and integrated luminosities at each of the center-of-mass energies
for each facility.
\label{facilities}
}
\small
\begin{tabular}{lccccccc}\hline\hline
Facility                     &
HL-LHC        &   ILC          &  ILC(LumiUp)      &  CLIC           &  TLEP (4 IPs)  & HE-LHC  & VLHC  \\
$\sqrt{s}$ (GeV)             &   
14,000        &   250/500/1000 &  250/500/1000     &  350/1400/3000  &  240/350       & 33,000  & 100,000   \\
$\int{\cal L}dt$ (fb$^{-1}$) & 
3000/expt     &   250+500+1000 &  1150+1600+2500  &  500+1500+2000  &  10,000+2600   & 3000       & 3000 \\  \hline
$\int dt$ ($10^7$s)          & 
6             &   3+3+3        &  $\begin{array}{c} (\mbox{ILC 3+3+3}) \\  +\mbox{ 3+3+3} \end{array}$       &  3.1+4+3.3      &   5+5          & 6 & 6  \\
\hline\hline
\end{tabular}
\end{center}
\end{table}

The report has two primary goals.  The first is to identify the key elements of a precision
Higgs program and the fundamental importance of
this science as a human endeavor to understand the universe and the laws of physics.
The second is to document the precision and physics potential presented during Snowmass for the above listed community initiatives to develop precision Higgs programs. This report does not judge the status, maturity, feasibility, 
or readiness of any of these initiatives.  The reader should refer to the report of the Frontier Capabilities
Group for this information.
The report will also identify unique strengths of different 
collider initiatives and detector concepts and their importance to the precision Higgs program.
The detailed charge to the Higgs Snowmass committee is listed below.

\subsection{Charge to the Higgs Snowmass Committee}\vspace*{-0.3cm}

\begin{itemize}
\item[1.] Provide a compact summary of the measurements on and searches for the SM Higgs boson, including information from LEP, the Tevatron, and the LHC. Include in this summary a survey of searches for non-minimal Higgs sectors.
\item[2.] Provide a compact summary of the theoretical motivations to explore the properties of the Higgs boson to high precision.
\begin{itemize}
\item[a)] What is the full phenomenological profile of the Higgs boson? What are the predicted production modes, the final states, and the experimental observables?
\item[b)] What are the ranges of predictions for deviations from Standard Model properties that enter from new physics? Which production and decay channels and boson properties are most sensitive to these deviations?
\item[c)] What can be learned from the discovery of bosons from non-minimal Higgs sectors? What is the phenomenology of non-minimal Higgs models?
\item[d)] To what extent are properties of the Higgs boson and the Higgs sector in general important for understanding fundamental physics and the universe?
\end{itemize}
\item[3.] Organize a set of simulation studies to evaluate the level of precision that can be achieved on Higgs physics measurements for the range of choices of accelerator facilities and detector capabilities under consideration by the Facilities/Instrumentation groups. Include studies of search sensitivities for non-minimal Higgs sectors.
\begin{itemize}
\item[a)] To what degree can a particular experimental program ascertain whether the resonance at 126~GeV is the Standard Model Higgs boson? To what precision can each of the measured properties of the Higgs boson be determined and tested against SM predictions?
\item[b)] What are the search sensitivities for bosons in non-minimal Higgs sectors?
\item[c)] The studies should summarize their results in terms of these areas:
\begin{itemize}
\item[i.] Mass and width measurements;
\item[ii.] ``Couplings" in terms of production cross section by process and branching fractions by decay mode, including searches for non-SM couplings;
\item[iii.] ``Tensor structure" in terms of quantum numbers ($J^{CP}$) and effective couplings in the Lagrangian;
\item[iv.] Couplings and properties governing the Higgs potential.
\end{itemize}
\item [d)] What are intrinsic advantages of particular experimental programs? Are there unique capabilities to reconstructing particular decays or unique sensitivities to particular rare decay rates? Are there properties that can be determined in some experimental programs and not in others? To what extent can complementary programs enhance the overall Higgs physics program?
\item[e)] Provide cross-calibration for the simulation tools to provide a record of what intrinsic performances and assumptions went into these results.
\end{itemize}
\end{itemize}


\newcommand{\hp}{\hspace*{0.5cm}}

\section{Coupling Measurements}\vspace*{-0.3cm}
\label{sec:couplings}

The central question about the particle discovered at 126~GeV is whether this is ``The Higgs Boson" or
only one degree of freedom of a bigger story.  If there is more than one Higgs boson, and theories such as
supersymmetry require there to be multiple bosons at the TeV scale, then the couplings of the 126~GeV boson to matter will not directly correspond to the coupling strengths predicted from the masses of the elementary particles.  Additional parameters that describe the mixing of multiple Higgs boson states, or the ratio of vacuum expectation values, or in general the effects of additional degrees of freedom in the Higgs sector will result in deviations in the coupling measurements relative to Standard Model expectations.  This is especially true of the loop-induced decays and production modes of the 126~GeV boson where new particles can enter the loops.

The precisions that can be obtained on the coupling measurements are projected for the LHC and $e^+e^-$ machines.  A muon collider is expected to be capable of a similar program as the $e^+e^-$ machines,
but detector simulations to extract these estimates have not been completed at this time.



Extractions of the Higgs coupling constants from measured decay modes can serve to limit various new physics models, or to confirm the validity of the
Standard Model.  The conclusions derived from this exercise depend on the uncertainties in the calculation of the Standard Model cross sections 
and branching ratios.  In this subsection, we discuss the uncertainties on the theoretical predictions of the Higgs branching ratios,
which have been tabulated by the LHC Higgs cross section working group\cite{Denner:2011mq,Heinemeyer:2013tqa}.

There are two types of uncertainties that arise when computing the
uncertainties on Higgs branching ratios: parametric uncertainties and
theoretical uncertainties.  The parametric uncertainties describe the
dependence of the predictions on the input parameters.  For a $126$
GeV Standard Model Higgs boson, the parametric uncertainties arise
predominantly from the $b$-quark mass and $\alpha_s$ and we use the values
given in Table \ref{tab:pu}.  The parametric uncertainties are
combined in quadrature.  The theoretical uncertainties are estimated
from the QCD scale dependence and from higher order electroweak
interactions and are listed in Table \ref{tab:tu}.  The theory uncertainties
are combined linearly.  The errors on the
predictions for the branching ratios for a $126$ GeV Standard Model
Higgs boson are given in Table \ref{tab:brtot}.  The uncertainties on
the total widths are given in Table \ref{tab:gam_errtot}, where the
parametric errors from $\Delta \alpha_s$, $\Delta m_b$, $\Delta m_c$
and the theory uncertainties are given separately.  The dominant
source of the electroweak uncertainty is from NLO corrections which
are known but not yet implemented exactly in the partial width
calculations.  These electroweak uncertainties can be expected to be
reduced in the future.  It is also possible that the uncertainties on
the $b$-quark mass and $\alpha_s$ may be reduced by future lattice
calculations by up to a factor of 5, as shown in Table~\ref{lattice}.
  
\begin{table}[htb!]
\caption{Parametric uncertainties used by the Higgs 
Cross Section Working group to determine Higgs 
branching ratio and width uncertainties\cite{Denner:2011mq,Dittmaier:2012vm}.}
\begin{center}
\begin{tabular}{ccc}
\hline\hline
 Parameter          & Central Value & Uncertainty 
 \\ 
 \hline
$\alpha_s(M_Z)$ & $0.119$ & $\pm 0.002 \thinspace (90\%$~CL)  \\
$m_c$ &$1.42$~GeV& $\pm 0.03$~GeV\\
$m_b$ & $ 4.49$~GeV & $\pm 0.06$~GeV \\
$m_t$ &  $172.5$~GeV& $\pm 2.5$~GeV\\
\hline\hline
\end{tabular}
\end{center}
\label{tab:pu}
\end{table}

\begin{table}[htb!]
\caption{Theory uncertainties on $M_H=126$~GeV Higgs partial widths~\cite{Dittmaier:2012vm}.}
\begin{center}
\begin{tabular}{lccc}
\hline\hline
 Decay          & QCD  Uncertainty & Electroweak Uncertainty & Total 
 \\ \hline
 $H\rightarrow b {\overline b},c{\overline c}$&$\sim 0.1\%$&$\sim 1-2\%$&$\sim 2\%$\\
 $H\rightarrow \tau^+\tau^-,\mu^+\mu^-$&--& $\sim 1-2\%$&$\sim 2\%$\\
 $H\rightarrow gg$&  $\sim 3\%$&$\sim 1\%$&$\sim 3\%$\\
 $H\rightarrow \gamma\gamma$&$<1\%$&$<1\%$&$\sim 1\%$\\
 $H\rightarrow Z\gamma$&$<1\%$&$\sim 5\%$&$\sim 5\%$\\
 $H\rightarrow WW^*/ZZ^*\rightarrow 4f$&$<0.5\%$& $\sim 0.5\%$&$\sim 0.5\%$
 \\
 \hline\hline
\end{tabular}
\end{center}
\label{tab:tu}
\end{table}

\begin{table}[htb!]
\caption{Uncertainties on $M_H=126$~GeV Standard Model branching ratios arising from the parametric uncertainties on
$\alpha_s$, $m_b$, and $m_c$ and from theory uncertainties~\cite{Heinemeyer:2013tqa,Dittmaier:2012vm}.}
\begin{center}
\begin{tabular}{lcccc}
\hline\hline
 Decay         & Theory Uncertainty & Parametric Uncertainty & Total Uncertainty& Central Value
 \\
 &&&on Branching Ratios&\\
 &($\%$) &($\%$)&($\%$) &\\ 
  \hline
$H\rightarrow \gamma\gamma$ & $\pm2.7$&$\pm2.2$&$\pm4.9$ & $2.3\times 10^{-3}$\\
$H\rightarrow b {\overline b}$&$\pm1.5$&$\pm1.9$&$\pm 3.4$&$5.6\times 10^{-1}$\\
$H\rightarrow c {\overline c}$&$\pm 3.5$ &$\pm 8.7$&$\pm 12.2$&$2.8\times 10^{-2}$\\
$H\rightarrow gg$&$\pm 4.3$ &$\pm 5.8$&$\pm 10.1$&$8.5\times 10^{-2}$\\
$H\rightarrow \tau^+\tau^-$ & $\pm 3.5$&  $\pm 2.1$    & $\pm 5.6$&$6.2\times 10^{-2}$\\
$H\rightarrow WW^*$ &  $\pm 2.0$     &   $\pm 2.1$   &  $\pm 4.1$&$2.3\times 10^{-1}$\\
$H\rightarrow ZZ^*$            &  $\pm 2.1$    &    $\pm 2.1$   &$\pm 4.2$&$2.9\times 10^{-2}$\\
$H\rightarrow Z\gamma$ &$\pm 6.8$&$\pm 2.2$&$\pm 9.0$&$1.6\times 10^{-3}$\\
$H\rightarrow \mu^+\mu^-$ &$\pm 3.7$&$\pm 2.2$ &$\pm5.9$&$2.1 \times 10^{-4}$\\
\hline\hline
\end{tabular}
\end{center}
\label{tab:brtot}
\end{table}

\begin{table}[htb!]
\caption{Uncertainties on $M_H=126$~GeV Standard Model widths arising from the parametric uncertainties on
$\alpha_s$, $m_b$, and $m_c$ and from theory uncertainties~\cite{Heinemeyer:2013tqa}.  For the total uncertainty, parametric uncertainties are added in 
quadrature and the result is added linearly to the theory uncertainty.}
\begin{center}
\begin{tabular}{lccccc}
\hline\hline
 Channel      & $\Delta \alpha_s$ & $\Delta m_b$ & $\Delta m_c$ & Theory Uncertainty& Total Uncertainty
 \\ 
  \hline
$H\rightarrow \gamma\gamma$ &  $0\%$ & $0\%$ &      $ 0\%$ & $\pm 1\%$  & $\pm 1\% $\\
$H\rightarrow b {\overline b}$     & $\mp 2.3\%$ &  $^{+3.3\%}_{-3.2\%}$ & $0\%$ & $\pm 2\%$ & $\pm 6\%$\\
$H\rightarrow c {\overline c}$ &$^{-7.1\%}_{+7.0\%}$& $\mp 0.1\%$ & $^{+6.2\%}_{-6.1\%}$ & $\pm 2\%$ & $\pm 11\%$ \\
$H\rightarrow gg$ &$^{+4.2\%}_{-4.1\%}$ & $\mp 0.1\%$ & 0\% & $\pm 3\%$ & $\pm 7\%$ \\
$H\rightarrow \tau^+\tau^-$  &  $0\%$ & $0\%$&      $ 0\%$ & $\pm 2\%$  & $\pm 2\%$\\
$H\rightarrow WW^*$ &   $0\%$ & $0\%$&      $ 0\%$ & $\pm 0.5\%$  & $\pm 0.5\%$\\
$H\rightarrow ZZ^*$     &        $0\%$ & $0\%$&      $ 0\%$ & $\pm 0.5\%$  & $\pm 0.5\%$\\
\hline\hline
\end{tabular}
\end{center}
\label{tab:gam_errtot}
\end{table}

\begin{table}[htb!]
\caption{Projected future uncertainties in $\alpha_s$, $m_c$, and $m_b$, 
compared with current uncertainties estimated from various sources.  Details
of the lattice 2018 projections are given in the Snowmass QCD Working Group
report~\cite{Campbell:2013qaa}.}
\begin{center}
\begin{tabular}{cccccc}
\hline\hline
        &  Higgs X-section      & PDG\cite{Beringer:1900zz}     & non-lattice
   & Lattice  & Lattice   \\
        &   Working Group \cite{Denner:2011mq}                  &  
               &                    &  (2013) &  (2018)  \\
\hline
$\Delta \alpha_s$ &0.002        &0.0007 &0.0012 \cite{Beringer:1900zz} 
         &0.0006 \cite{McNeile:2010ji}   & 0.0004 \\
$\Delta m_c$ (GeV) &0.03        &       0.025&  0.013 \cite{Chetyrkin:2009fv}
   & 0.006 \cite{McNeile:2010ji}   & 0.004  \\
$\Delta m_b$ (GeV) &0.06        &       0.03    &     0.016 
\cite{Chetyrkin:2009fv}     & 0.023 \cite{McNeile:2010ji}    & 0.011 \\
\hline\hline
\end{tabular}
\end{center}
\label{lattice}
\end{table}

\subsection{Higgs Coupling Fits}\vspace*{-0.4cm}
\label{sec:coupling_fits}

At the LHC, the rates of Higgs boson production and decay into particular final states are parametrized 
using strength parameters $\mu$ defined as the ratios between the observed rates and the 
expected ones in the Standard Model:
$$\mu = \frac{\sigma\times {\rm BR}}{(\sigma\times {\rm BR})_{\rm SM}}.$$
The deviations from the SM are implemented as scale factors ($\kappa$'s) of Higgs 
couplings relative to their SM values~\cite{LHCHiggsCrossSectionWorkingGroup:2012nn}:
$$g_{Hff} = \kappa_f \cdot g_{Hff}^{\rm SM} = \kappa_f\cdot \frac{m_f}{v}\ \ \ {\rm and}\ \ \ 
g_{HVV} = \kappa_V \cdot g_{HVV}^{\rm SM} = \kappa_V\cdot \frac{2 m_V^2}{v}$$
such that $\kappa_f=1$ and $\kappa_V=1$ in SM.
For example, at the LHC the $gg\to H\to\gamma\gamma$ rate can be written as
$$\sigma\times {\rm BR}(gg\to H\to \gamma\gamma) = \sigma_{\rm SM}(gg\to H)\cdot {\rm BR}_{\rm SM}(
H\to\gamma\gamma)                                                               
  \cdot \frac{\kappa_g^2 \cdot\kappa_\gamma^2}{\kappa_H^2},$$
where $\kappa_g$ and $\kappa_\gamma$ are effective scale factors for the  $Hgg$ and 
$H\gamma\gamma$
couplings through loops.  Additionally, $\kappa_H^2$ is the scale factor for the Higgs width:
$$\kappa_H^2 = \sum_X \kappa_X^2  {\rm BR}_{\rm SM}(H\to X)\, , $$
where $\kappa_X$ is the scale factor for the $HXX$ coupling and ${\rm BR}_{\rm SM}(H\to X)$ 
is the SM value of the
$H\to X$ decay branching ratio. The summation runs over all decay modes in the
 SM. This parameterization assumes that there is only one Higgs resonance, that
the resonance is narrow, and that the Higgs interactions have the same
tensor structure as the Standard Model interactions. 
Non-Standard Model Higgs decay modes
will modify the total Higgs decay width and consequently rescale the branching 
ratios of all other known
decay modes. In this case,  $\kappa_H^2$ is
modified to be
$$\kappa_H^2 = \sum_X \kappa_X^2\frac{{\rm BR}_{\rm SM}(H\to X)}{1-{\rm BR}_{\rm BSM}}.$$
Here ${\rm BR}_{\rm BSM}$ is the total branching ratio of beyond-standard-model~(BSM) decays.

The loop-induced Higgs couplings can alternately be expressed in a way that
separates out potential new-physics contributions:
\begin{equation}
        \kappa_g \simeq \kappa_t + \Delta \kappa_g, \qquad \qquad
        \kappa_{\gamma} \simeq -0.28 \kappa_t + 1.28 \kappa_W + \Delta \kappa_{\gamma},
\end{equation}
where we take $m_H = 126$~GeV and keep only the dominant top and $W$
loop contributions.  In the absence of new non-SM particles contributing
to the loop, we have $\Delta \kappa_{g,\gamma} = 0$.  When the new particles
are top-partners (scalar or fermionic color triplets with charge $+2/3$), 
we have the relation $\Delta \kappa_{\gamma} \simeq -0.28 \Delta \kappa_g$.

\subsection{Non-Standard Higgs Couplings due to New Physics}\vspace*{-0.4cm}

In this section, we survey a few models that can give Higgs couplings different from those of the Standard Model. All of
these models contain new particles, so discovery of the new physics can result from direct detection of the new particles, or 
from the measurement of a deviation in the Higgs coupling from the Standard Model predictions\cite{Gupta:2012mi}.  We note
that in order to be sensitive to a deviation, $\delta$, the measurement must be made to a precision comparable to $\delta/2$ in 
order to obtain a $95 \%$ confidence level limit, or $\delta/5$ for
a $5\sigma$ discovery of new physics.  Coupling deviations in multiple production modes or final
states can provide additional sensitivity to models that predict specific patterns.

\subsubsection{One-Parameter Model}\vspace*{-0.4cm}

One of the simplest extensions of the Standard Model is to add an $SU(2)$ singlet Higgs, $S$, which
mixes with the usual Higgs doublet, $\Phi_{SM}$, through a mixing term $|\Phi_{SM}|^2 |S|^2$.
In some models that predict dark matter, the singlet, $S$, could arise from a hidden sector.
There are two mass eigenstate Higgs particles:  the observed $126$~GeV Higgs boson, $H$, and  a 
heavier Higgs particle, $H_2$. 	The Standard Model Higgs has couplings that are suppressed relative to the SM values\cite{Pruna:2013bma},
\begin{eqnarray}
\kappa_V
&=&\kappa_F=\cos\alpha
\end{eqnarray}
where $V=W,Z$ and $F$ denotes all the fermions.  
The value of $\sin\alpha$ is constrained by precision electroweak data and for $M_H\sim 1$~TeV, we
must have $\sin^2\alpha < 0.12$\cite{Gupta:2012mi},  which implies that in this model, the target for precision measurements of 
Higgs couplings is,
\begin{equation}
\kappa_V-1=\kappa_F-1 < 6~\%\, .
\end{equation}

\subsubsection{Two Higgs Doublet Models}\vspace*{-0.4cm}
\label{sec:2HDM}

One of the most straightforward extensions of the Standard Model is the two Higgs doublet model.  The 2HDMs contain five physical
Higgs bosons: two neutral scalars, $h$ and $H$, a pseudoscalar, $A$, and charged Higgs bosons, $H^\pm$. Models with a $Z_2$
symmetry can be constructed such that there are no tree level flavor changing neutral currents.
The couplings of the Higgs bosons to fermions are described by two free parameters:  
the ratio of vacuum expectation values of the two Higgs doublets, $\tan\beta \equiv \frac{v_2}{v_1}$, 
and the mixing angle that diagonalizes the neutral scalar mass matrix, $\alpha$.  
There are then four possible assignments of couplings for  the light CP-even Higgs boson, 
$h^0$, to fermions and gauge bosons
relative to the Standard Model couplings, which are given 
in Table \ref{table:coups}.  The couplings to $W$ and $Z$ are always suppressed relative to the Standard Model
couplings, while in model II the couplings to $b$'s and $\tau$'s are enhanced at large 
$\tan\beta$.

\begin{table}[htb!]
\caption{Couplings of the light Neutral Higgs, $h$, in the 2HDMs.}
\begin{center}
\begin{tabular}{ccccc}
\hline\hline
& I& II& Lepton-Specific& Flipped\\
\hline
$\kappa_V$ & $\sin(\beta-\alpha)$ & $\sin (\beta-\alpha)$ &$\sin (\beta-\alpha)$&$\sin (\beta-\alpha)$\\
$\kappa_t$&${\cos\alpha\over\sin\beta}$&${\cos\alpha\over\sin\beta}$&${\cos\alpha\over\sin\beta}$&${\cos\alpha\over\sin\beta}$\\
$\kappa_b$ &${\cos\alpha\over\sin\beta}$&$-{\sin\alpha\over\cos\beta}$&${\cos\alpha\over\sin\beta}$&$-{\sin\alpha\over \cos\beta}$\\
$\kappa_\tau$&${\cos\alpha\over \sin\beta}$&$-{\sin\alpha\over \cos\beta}$&$-{\sin\alpha\over \cos\beta}$&${\cos\alpha\over\sin\beta}$\\
\hline\hline
\end{tabular}
\end{center}
\label{table:coups}
\end{table}
Current limits on $\tan\beta$ and
$\cos(\beta-\alpha)$~\cite{Craig:2013hca}, along with projections for
the high luminosity LHC and the $\sqrt{s}=500$ GeV and $\sqrt{s}=1000$
GeV ILC (assuming no deviations from the Standard Model) are given in
Refs.~\cite{Chen:2013rba,Barger:2013ofa}.  In model II and the flipped model,
$\cos(\beta-\alpha)$ is already constrained to be near one, while
larger deviations are possible in model I and the lepton-specific
model.  Large values of $\tan\beta$ are as yet unconstrained by the
data.

\subsubsection{MSSM}\vspace*{-0.4cm}

The Higgs sector of the MSSM is a special case of the 2HDM and corresponds to model II.  In the MSSM, the mixing angle, $\alpha$,
is related to the masses of the scalars.  In the limit where the pseudoscalar $A$ is much heavier than $M_Z$, the
couplings take the simple form (called the decoupling limit)\cite{Carena:2001bg},
\begin{eqnarray}
\kappa_V&\sim& 1-{2M_Z^4\over M_A^4}\cot^2\beta\nonumber \\
\kappa_t&\sim & 1-{2M_Z^2\over M_A^2}\cot^2\beta\nonumber\\
\kappa_b=\kappa_\tau&\sim & 1+{2M_Z^2\over M_A^2}\, .
\end{eqnarray}
Studies of the MSSM suggest that with $300$~fb$^{-1}$ the LHC will be sensitive to $M_A\sim 400-500$~GeV
for all values of $\tan\beta$ not excluded by LEP\cite{Djouadi:2013vqa}, giving as a target for the coupling precisions,
\begin{eqnarray}
\kappa_V&\sim& 1-0.5\%\biggl({400~{\rm GeV}\over M_A}\biggr)^4\cot^2\beta\nonumber \\
\kappa_t&\sim & 1-{\cal O}(10\%)\biggl({400~{\rm GeV}\over M_A}\biggr)^2 \cot^2\beta\nonumber\\
\kappa_b=\kappa_\tau&\sim & 1+{\cal O}(10\%)\biggl({400~{\rm GeV}\over M_A}\biggr)^2\, .
\end{eqnarray}
For large $\tan\beta$, the Higgs coupling to $b$'s is 
enhanced and not only is the decay $h^0\rightarrow b {\overline b}$
enhanced, but the dominant production mechanism
is the production in association with $b$'s.

\subsubsection{pMSSM}\vspace*{-0.4cm}

The pMSSM is a phenomenological version of the MSSM with 19 input
parameters.  The parameters are constrained to be consistent with
current experimental limits.  A scan over input parameters looks at
regions in parameter space that can be excluded by measurements of
the Higgs couplings.  For example, a measurement of $\kappa_\tau$
(with the central value given by the Standard Model prediction) would
exclude $32.4\%$ of the parameter space at the HL-LHC, and $78\%$ of
the parameter space at the ILC500.  Combining all Higgs coupling
measurements, the HL-LHC would exclude $34\%$ of the pMSSM parameter
space, while ILC500 excludes $99.7\%$ of the parameter
space\cite{Cahill-Rowley:2013vfa}.

\subsubsection{Composite Models}\vspace*{-0.4cm}

Composite models predict deviations in Higgs couplings due to higher dimension
operators.  Typically the deviations are ${\cal O}(v^2/f^2)$, where $f$
is the scale associated with the new operators.  Typically,
\begin{eqnarray}
\kappa_V &\sim & 1-3\%\biggl({1~{\rm TeV}\over f}\biggr)^2, \nonumber \\
\kappa_F & \sim & 1-(3-9)\% \biggl({1~{\rm TeV}\over f}\biggr)^2.
\end{eqnarray} 

\subsubsection{New Couplings From Loops}\vspace*{-0.4cm}

Many models of new physics contain non-Standard Model particles that contribute via loops to 
the decays $H\rightarrow g g$, $H\rightarrow \gamma \gamma$ and/or 
$H\rightarrow Z\gamma$,\footnote{We will not
discuss $H\rightarrow Z\gamma$ here, although this decay can receive significant corrections in new physics models (see, e.g., Ref.~\cite{Maru:2013qla}).}
along with altering the $gg\rightarrow H$ production rate.
These new particles give rise to the effective interactions parameterized by $\kappa_g$ and $\kappa_\gamma$.
 Generically, one might expect these
loop corrections to be ${\cal O}\biggl({v^2\over M^2}\biggr)\sim 6\%\biggl({1~{\rm TeV}\over M}\biggr)^2$, where $M$ is the scale of the 
new physics effects.
New heavy fermions, such as top partners, and colored scalars can contribute to $H\rightarrow gg$ and $H\rightarrow\gamma\gamma$,
while electrically charged scalars and heavy leptons can contribute to $H\rightarrow \gamma\gamma$.  Below we examine 
some representative models, in order to get a feel for the size of the possible effects.

In Little Higgs models with T parity, the couplings scale with the top partner mass, $M_T$, and assuming
the Higgs couplings to Standard Model particles are not changed, the loop induced couplings are~\cite{Carmi:2012yp},
\begin{equation}
\Delta \kappa_g
\simeq -{m_t^2\over M_T^2}
\sim
{\cal O} (-8\%)\biggl({600~{\rm GeV}\over M_T}\biggr)^2\, ,
\qquad
\Delta \kappa_{\gamma} \simeq -0.28 \Delta \kappa_g
\sim {\cal O} (+2\%)\biggl({600~{\rm GeV}\over M_T}\biggr)^2\, .
\end{equation}
In this scenario the production rate from gluon fusion is suppressed, 
while the width into $\gamma \gamma$ in increased.
Adding a vector-like $SU(2)$ doublet  of heavy leptons does not change the $gg\rightarrow H$ production rate, but can give an enhancement
in  $\kappa_\gamma$  of order  $\sim 20\%$,
although large Yukawa couplings are required~\cite{ArkaniHamed:2012kq}.

Colored scalars, such as the stop particle in the MSSM, also
contribute to both $\kappa_g$ and $\kappa_\gamma$.  If we consider two
charge-${2\over 3}$ scalars as in the MSSM, then for a stop squark much
heavier than the Higgs boson~\cite{Carmi:2012yp},
\begin{equation}
\Delta \kappa_g
\simeq {1\over 4}\biggl({m_t^2\over m_{\tilde{t_1}}^2}+{m_t^2\over m_{\tilde{t_2}}^2}
-{m_t^2X_t^2\over m_{\tilde {t_1}}m_{\tilde {t_2}}}\biggr)
\sim 
{\cal O} (+17\%)\biggl(
{300~{\rm GeV}\over m_{\tilde t}}\biggr)^2\, \, (\hbox{for~}X_t=0) ,
\end{equation}
where again $\Delta \kappa_{\gamma} \simeq -0.28 \Delta \kappa_g$.
Here $X_t=\mid A_t-\mu\cot\beta\mid$ is the stop mixing parameter. 
If $X_t=0$, the Higgs couplings to gluons is always increased and the coupling
to photons decreased.  If the stops are light, and the mixing is small,
large enhancements are possible. 
In the MSSM, there are other loop contributions to the $H\gamma\gamma$ and $Hgg$ couplings which have been extensively studied.  Enhancements
in the $H\rightarrow \gamma\gamma$ coupling can be obtained with light staus and large mixing, with effects on the order of
 $\sim 25\%$~\cite{Carena:2013iba}.

In Table~\ref{tab:model_summary}, we summarize the generic size of coupling
modifications when the scale of new physics is consistently taken
to be $M\sim 1$~TeV.

\begin{table}[htb!]
\caption{Generic size of Higgs coupling modifications
from the Standard Model values when 
all new particles are $M\sim 1$~TeV and mixing angles satisfy precision
electroweak fits. The Decoupling MSSM numbers assume $\tan\beta=3.2$ 
and a stop mass of $1~TeV$ with $X_t=0$ for the $\kappa_\gamma$ prediction.}
\begin{center}
\begin{tabular}{cccc}
\hline\hline
Model  &$\kappa_V$ & $\kappa_b$& $\kappa_\gamma$\\
  \hline
Singlet Mixing   &     $\sim 6$\%  & $\sim 6$\%  &  $\sim 6\%$ \\
2HDM &   $\sim 1\%$   & $\sim 10\%$  &  $\sim 1\%$ \\
Decoupling MSSM    &     $\sim -0.0013\%$ 
& $\sim 1.6\%$ & $\sim -.4\%$\\
Composite  & $\sim -3\%$ & $\sim -(3-9)\%$ & $\sim -9\%$\\
Top Partner  &  $\sim -2\%$ & $\sim -2\%$ & $\sim +1\%$\\
 \hline\hline
\end{tabular}
\end{center}
\label{tab:model_summary}
\end{table}

\subsection{Theory Uncertainties on LHC Higgs Production}\vspace*{-0.4cm}

The uncertainty on Higgs production has been studied by the LHC Higgs cross section working group
 for the various channels
and is summarized in Table \ref{tab:lhc_errors}\cite{Dittmaier:2011ti}.  
These uncertainties must be included in extractions of the scale factors $\kappa_i$ from LHC data.
 The error includes
factorization/renormalization scale uncertainty and the correlated uncertainty
from $\alpha_s$ and the PDF choice, which are added linearly.  The scale uncertainty on the gluon fusion rate is 
$\sim \pm 10\%$, which can potentially be significantly reduced with the inclusion of recent  approximate NNNLO
results\cite{Ball:2013bra}.
In addition, there are further uncertainties from binning the Higgs data into $0,1$ and $2$-jet bins. 
The theory error on the $1$-jet bin will be significantly reduced with the inclusion of the NNLO result for Higgs plus one
 jet\cite{Boughezal:2013uia} and by resumming jet veto effects.  

\begin{table}[htb!]
\caption{Theory uncertainties for $m_H=126$~GeV Higgs Production at the LHC with $\sqrt{s}=14$~TeV\cite{Dittmaier:2011ti}. The total uncertainty is the linear sum of scale and PDF uncertainties.}
\begin{center}
\begin{tabular}{lcccc}
\hline\hline
 Process        & Cross section  & \multicolumn{3}{c}{Relative uncertainty in percent} \\ \cline{3-5}  
                &      (pb)      &   Total    &    Scale   &      PDF   
 \\ \hline
 Gluon fusion &  49.3   & $^{+19.6}_{-14.6}$  &   $^{+12.2}_{-8.4}$   &  $^{+7.4}_{-6.2}$ \\
 VBF          &   4.15   & $^{+2.8}_{-3.0}$    &   $^{+0.7}_{-0.4}$    &  $^{+2.1}_{-2.6}$ \\
 WH           &  1.474   & $^{+4.1}_{-4.4}$    &   $^{+0.3}_{-0.6}$    &  $^{+3.8}_{-3.8}$ \\
 ZH           &  0.863   & $^{+6.4}_{-5.5}$    &   $^{+2.7}_{-1.8}$    &  $^{+3.7}_{-3.7}$
 \\
 \hline\hline
\end{tabular}
\end{center}
\label{tab:lhc_errors}
\end{table}

\subsection{Theory Uncertainties on $\mathbf{e^+e^-}$ Higgs Production}\vspace*{-0.4cm}

Complete $\mathcal{O}(\alpha)$ electroweak radiative corrections to
Higgs production via $e^+e^- \to \nu \bar \nu H$ have been computed in
the SM~\cite{Denner:2003yg,Denner:2003iy}.  In the $G_{\mu}$ scheme,
the bulk of the correction to $e^+e^- \to \nu \bar \nu H$ is due to
initial state radiation and reduces the total cross section by about
10\% at high energies.  The correction is largely universal and can be
captured to within about 3\% by an improved Born
approximation~\cite{Denner:2003iy}.  The $\mathcal{O}(\alpha)$ corrections
to $e^+e^- \to e^+e^- H$ are also known~\cite{Boudjema:2004eb}.
Next-to-next-to-leading order 
electroweak and mixed electroweak-QCD corrections may need to be computed
to match the sub-percent-level experimental accuracy anticipated at 
future lepton colliders.

QCD corrections to $e^+e^- \to t \bar t H$ are large, especially when
the invariant mass of the $t \bar t$ pair is close to threshold.  The
QCD corrections have been computed up to next-to-leading
logarithm~\cite{Farrell:2006xe} with the $t \bar t$ threshold region
handled using a nonrelativistic effective
theory~\cite{Farrell:2005fk}.  Complete $\mathcal{O}(\alpha)$
electroweak corrections to $e^+e^- \to t \bar t H$ have also been
computed in the SM~\cite{Denner:2003ri}, including a resummation of
the photon initial-state radiation effects.  The electroweak
corrections reach of order 10\% and depend nontrivially on the
collision energy.



\subsection{Measurements at Hadron Colliders and Projections at LHC}\vspace*{-0.4cm}

In $pp$ or $p\bar{p}$ collisions, the Higgs boson can be produced through the following four main processes: 
gluon-gluon fusion $gg\to H$ through a heavy quark triangular loop (ggF), vector boson fusion (VBF), 
associated production with a vector boson $W$ or $Z$ ($VH$), and production in association with a pair of top 
quarks ($t\bar{t}H$). 
The cross sections of these processes in $pp$ collisions at $\sqrt{s}=7, 8$ and $14$~TeV are 
listed in Table~\ref{tab:LHCCrossSections}.

\begin{table}[htb]
\caption{Higgs boson production cross sections of different processes at 7, 8 and 14 TeV of $pp$ collisions. 
These cross sections are taken from Ref.~\cite{LHCHiggsCrossSectionWorkingGroup:2012nn}.}
\begin{center}
\begin{tabular}{cccccc}\hline\hline
 $\sqrt{s}$         & \multicolumn{5}{c}{Cross sections in pb \@ $m_H=126$~GeV} \\ \hline
 (TeV)            & \hp ggF \hp  & \hp VBF \hp  & \hp WH \hp & \hp  ZH \hp  & \hp $t\bar{t}H$\hp  \\ \hline
7          & 14.9  & 1.21 & 0.563 & 0.327 & 0.084 \\
8          & 19.0  & 1.57 & 0.686 & 0.405 & 0.126 \\
14         & 49.3  & 4.15 & 1.47  & 0.863 & 0.598 \\ \hline\hline
\end{tabular}
\end{center}
\label{tab:LHCCrossSections}
\end{table}

Since the discovery of the $\sim$126~GeV Higgs-like particle in Summer 2012, the LHC experiments
have focused on the measurements of its production rates and couplings. Both ATLAS and CMS have 
released results based on the LHC Run 1 dataset of $\sim$5~fb$^{-1}$ at 7 TeV 
and $\sim$20~fb$^{-1}$ at 8 TeV. These results strongly suggest that the new particle is a Higgs 
boson and its properties are consistent with the expectations of the SM Higgs boson.
After a two-year shutdown, the LHC is scheduled to operate again in 2015 at $\sqrt{s}=14$~TeV. It is
expected to deliver 300 fb$^{-1}$ to each experiment by 2022. With the planned high luminosity
upgrade, an integrated luminosity of 3000 fb$^{-1}$ is foreseen by 2030. The increased luminosity
will significantly increase the measurement precision of the Higgs boson properties. The current
results are briefly summarized and the projected precisions are presented below.

\subsection{Production Rates and Coupling Fits}\vspace*{-0.4cm}

Table~\ref{tab:LHCrates} summarizes the current measurements of overall rates from 
the Tevatron~\cite{Aaltonen:2013kxa}, ATLAS~\cite{Aad:2013wqa,TheATLAScollaboration:2013lia}, and CMS~\cite{CMS:yva}, 
separately for the five main decay modes. These measurements are generally in good 
agreement with the SM prediction of $\mu=1$. In addition to the measurements by decay modes, measurements
by production processes have also been done for some processes through categorizing Higgs candidate events.
From example, $VH$ candidates can be selected by the presence of additional leptons from $V$ decays while VBF 
candidates are tagged by two forward jets. Searches for rare decays of $H\to \mu\mu$~\cite{ATLAS:2013qma,CMS:2013aga} 
and $H\to Z\gamma$~\cite{ATLAS:2013rma,Chatrchyan:2013vaa} as 
well as $H\to$ invisible in $ZH$~\cite{ATLAS:2013pma,CMS:PAS-HIG-13-018} have also 
been performed. Upper limits of these searches are also shown in Table~\ref{tab:LHCrates}.

\begin{table}[htb]
\caption{Summary of the measured production rates or 95\% CL upper limits relative to their SM predictions from hadron colliders by
decay channels. The last line shows the upper limit on the branching ratio of Higgs to invisible decays from the search of $ZH$ with
$H\to$ invisible. The ATLAS combined rate includes only $H\to\gamma\gamma,\ ZZ^*$ and $WW^*$ decays.}
\begin{center}
\begin{tabular}{cccc}\hline\hline
 Decay mode          & Tevatron                  & ATLAS                        & CMS  \\
                     & ($m_H=125$~GeV)           & ($m_H=125.5$~GeV)            & ($m_H=125.7$~GeV)  \\ \hline

$H\to \gamma\gamma$  &  $5.97^{+3.39}_{-3.12}$   &  $1.55\pm 0.23 \thinspace ({\rm stat}) \pm 0.15 \thinspace ({\rm syst})$    & $0.77\pm 0.27$  \\
$H\to ZZ^*$          & $-$                       &  $1.43\pm 0.33 \thinspace ({\rm stat}) \pm 0.17 \thinspace ({\rm syst})$    & $0.92\pm 0.28$  \\
$H\to WW^*$          & $0.94^{+0.85}_{-0.83}$    &  $0.99\pm 0.21 \thinspace ({\rm stat}) \pm 0.21 \thinspace ({\rm syst})$    & $0.68\pm 0.20$  \\
$H\to \tau\tau$      & $1.68^{+2.28}_{-1.68}$    &  $-$                                                  & $1.10\pm 0.41$  \\
$H\to b\bar{b}$      & $1.59^{+0.69}_{-0.72}$    &  $0.2\pm 0.5 \thinspace ({\rm stat})\pm 0.4 \thinspace ({\rm syst})$         & $1.15\pm 0.62$  \\ \hline
Combined             & $1.44^{+0.59}_{-0.56}$    & $1.33\pm 0.14 \thinspace ({\rm stat}) \pm 0.15 \thinspace ({\rm syst})$     & $0.80\pm 0.14$  \\  \\
\multicolumn{4}{c}{95\% CL observed (expected) upper limit} \\ \hline
$H\to\mu\mu$         &  $-$                      & $<9.8\ (8.2)^\dag$                                    &                 \\
$H\to Z\gamma$       &  $-$                      & $<13.5\ (18.2)^\dag$                                  & $4-25\ (5-16)^\ddag$ \\
${\rm BR}_{\rm inv}$ &  $-$                      & $<65\ (84)\%^\dag$                                    & $<75\ (91)\%^\dag$    \\ \hline\hline
\multicolumn{4}{l}{$\dag$ at $m_H=125$~GeV; $\ddag$ for $m_H=120-160$~GeV.}
\end{tabular}
\end{center}
\label{tab:LHCrates}
\end{table}

\begin{figure}[htb]
\begin{center}
\includegraphics[width=0.40\textwidth]{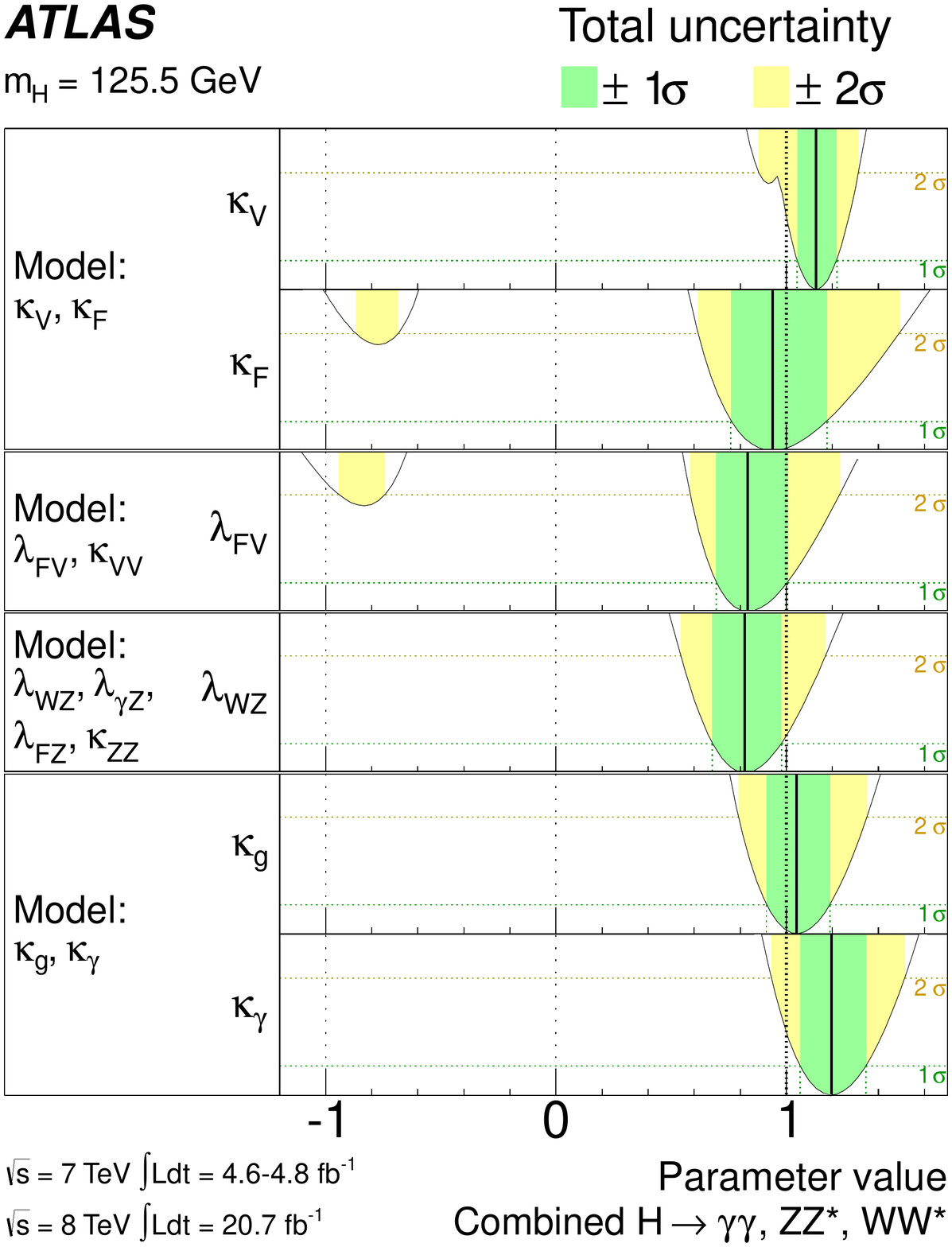}
\includegraphics[width=0.50\textwidth]{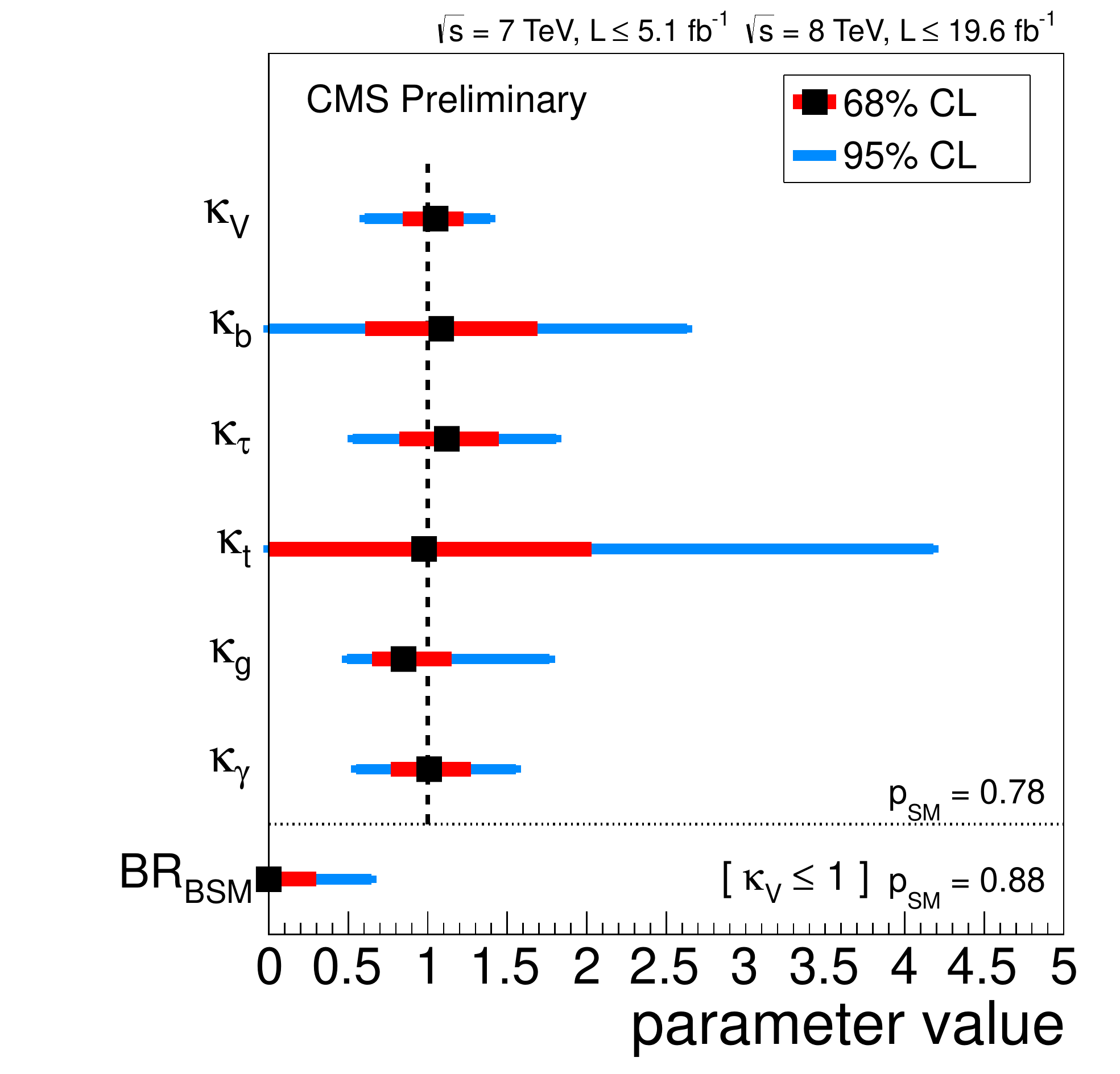}
\end{center}
\caption{Left: summary of the ATLAS coupling scale factor measurements for different models. The solid 
vertical lines are the best-fit values while the dark- and light-shaded band represent the total 
$\pm 1\sigma$ and $\pm 2\sigma$ uncertainties. The curves are distributions of the likelihood ratios. 
Right: summary of the CMS fits for deviations in the coupling for the generic six-parameter model 
including effective loop couplings. The result of the fit when extending the model to allow for
beyond-SM decays while restricting the coupling to vector bosons to not exceed unity ($\kappa_V\le 1.0$)
is also shown.}
\label{fig:LHCCouplings}
\end{figure}

Higgs couplings to fermions and vector bosons are determined following the procedure discussed in
Sec.~\ref{sec:coupling_fits}. Given the current statistics, fits to Higgs couplings to individual 
leptons, quarks and vector bosons 
are not meaningful and therefore have not been done. However fits have been performed with
reduced number of parameters under various assumptions. Results of these fits at both
the Tevatron and LHC can be found in 
Ref.~\cite{Aaltonen:2013kxa,Aad:2013wqa,CMS:yva} and
Fig.~\ref{fig:LHCCouplings} illustrates some representative results.

A note on the treatment of theoretical uncertainties is in order. The LHC Cross Section working group
recommends the linear addition of QCD scale and parametric uncertainties. However, since these two 
sources of uncertainties are represented by independent nuisance parameters in the fits at the LHC, 
the procedure effectively leads to the quadratic addition of scale and parametric uncertainties. 
This is true for the results presented in this section and for the projections discussed below.

\subsection{LHC Projections}\vspace*{-0.4cm}
\label{sec:couplings_LHC}
Precision measurements of the properties of the Higgs boson will be a central topic for the LHC physics
program in the foreseeable future. The high-luminosity LHC is not only an energy frontier machine,
it is also an intensity frontier collider. The expected large statistics will significantly improve the
precision of the current measurements of couplings to fermions and vector bosons.

The LHC is expected to deliver 300~fb$^{-1}$ at 14~TeV before the high-luminosity upgrade
and 3000~fb$^{-1}$ afterward, representing factors of 15 and 150 increases in statistics from luminosity alone
from the current 7 and 8 TeV datasets. The higher $pp$ collision energy will also increase the Higgs production 
cross sections by a factor of 2.6 or larger. The numbers of predicted Higgs events are shown in
Table~\ref{tab:LHCRates} for different production processes and decay modes. LHC experiments generally
have good sensitivities to final states with electrons, muons or photons. $H\to\tau\tau$ events
are identified through $\tau\to e\ {\rm or}\ \mu$ decays and also $\tau\to$~hadrons decays in the
case of VBF production. For $H\to b\bar{b}$ decays, most of their sensitivities are derived from the 
$VH\ (V=W,Z)$ production with the leptonic decays of $V$. The signal-background ratios~($S/B$) are strongly
dependent on analyses. For $H\to\gamma\gamma$, the $S/B$ ranges from $\sim 3\%$ for the inclusive
analysis to $\sim 20\%$ for some exclusive analyses with selection efficiencies  $\epsilon \sim 40\%$. 
The $H\to ZZ^*\to 4\ell$ analysis has a much smaller background and therefore a $S/B$ ratio of 
better than 2:1 with $\epsilon \sim 30\%$. On the other hand, the 
$H\to WW^*\to \ell\nu\ell\nu$ analysis has a $S/B\sim 15\%$ with $\epsilon\sim 5\%$. For rare decays such as 
$H\to\mu\mu$ and $H\to Z\gamma\to\ell\ell\gamma$, the $S/B$ is $\sim 0.5\%$ with $\epsilon\sim 40\%$.

\begin{table}
\caption{The numbers of predicted Higgs events produced in 3000~fb$^{-1}$ at 14~TeV in different
production processes and decay modes for $m_H=125$~GeV. Experimental sensitivity to these production modes
and decays varies widely, see text. Here $\ell=e,\mu$.}
\small
\begin{center}
\begin{tabular}{lrrrrrr}  \hline\hline
                   & ggF  & VBF & VH  & $t\bar{t}H$  && Total \\ \cline{2-7}
Cross section (pb) & 49.9 & 4.18 & 2.38 & 0.611 && 57.1 \\ \hline \\

 &\multicolumn{6}{c}{Numbers of events in 3000 fb$^{-1}$} \\ \hline
$H\to\gamma\gamma$             &    344,310 &    28,842 &    16,422 &     4,216 &&   393,790 \\
$H\to ZZ^*\to 4\ell$           &     17,847 &     1,495 &       851 &       219 &&    20,412 \\
$H\to WW^*\to \ell\nu\ell\nu$  &  1,501,647 &   125,789 &    71,622 &    18,387 && 1,717,445 \\  \\

$H\to\tau\tau$                 &  9,461,040 &   792,528 &   451,248 &   115,846 && 10,820,662 \\
$H\to b\bar{b}$                & 86,376,900 & 7,235,580 & 4,119,780 & 1,057,641 && 98,789,901 \\ \\ 
 
$H\to \mu\mu$                  &     32,934 &     2,759 &     1,570 &       403 &&     37,667 \\ 
$H\to Z\gamma\to\ell\ell\gamma$ &    15,090 &     1,264 &       720 &       185 &&     17,258 \\ \\
$H\to {\rm all}$                & 149,700,000 & 12,540,000& 7,140,000 & 1,833,000 && 171,213,000 \\ \hline\hline
\end{tabular}
\end{center}
\label{tab:LHCRates}
\end{table}

Both ATLAS and CMS experiments have projected their sensitivities to 
high luminosities~\cite{ATLAS:2013hta,ATLAS-PHYS-PUB-2013-014,CMS:2013xfa} with varying 
assumptions of detector and analysis performance. Arguably the most 
significant challenge is to deal with the high pileup that will come along with the high luminosity. 
The average number of interactions per beam crossing is expected to reach 140 compared with current 20.
However, the upgraded detectors are expected to mitigate most of the adverse impact from the higher pileup 
and maintain (in some cases exceed) the performance of the current detectors. 

ATLAS has taken the approach to estimate sensitivities using parametric simulations of the detector performance,
derived from full detector simulations that include the effect of higher pileups. 
On the other hand, CMS has taken a different approach, 
making projections based on the analyses of 7 and 8 TeV data with varying assumptions. 
Table~\ref{tab:LHCRateProjections}
summarizes the expected precisions on the signal strengths of different Higgs decay modes as well
as 95\% CL upper limit on the branching ratio of Higgs to invisible 
decay~(${\rm BR}_{\rm inv}$)~\cite{ATLAS-PHYS-PUB-2013-014,CMS:2013xfa} from both ATLAS and CMS.
An independent study of $pp\to ZH$ with $Z\to\ell\ell\ {\rm and}\ H\to{\rm invisible}$ decays
has estimated the 95\% CL upper limit on ${\rm BR}_{\rm inv}$ to be $9-22\%\ (6-10\%)$ for 
300~(3000)~fb$^{-1}$~\cite{Okawa:2013hda}, consistent with the result in the table.
These projections are based on the analysis of 7 and 8 TeV data, not all final states have been explored. 
They are expected to improve once more final states are included. 
CMS has considered two scenarios of systematic uncertainties:
\begin{itemize}
 \item {\sl Scenario 1:} all systematic uncertainties are left unchanged (note that uncertainty reductions from
                         increased statistics in data control regions are nevertheless taken into account);
 \item {\sl Scenario 2:} the theoretical uncertainties are scaled by a factor of $1/2$, while other systematic
                         uncertainties are scaled by the square root of the integrated luminosity, {\em i.e.}, 
                         $1/\sqrt{\cal L}$.
\end{itemize} 
The ranges of the projections in the table represent the cases with and without theoretical uncertainties for
ATLAS and two scenarios of systematic uncertainties for CMS.  

\begin{table}
\caption{Expected relative precisions on the signal strengths of different Higgs decay final states 
as well as the 95\% CL upper limit on the Higgs branching ratio to the invisible decay from the $ZH$ search
estimated by ATLAS and CMS. The ranges are not comparable between ATLAS and CMS. For ATLAS, they
correspond to the cases with and without theoretical uncertainties while for CMS they represent two
scenarios of systematic uncertainties.}
\small
\begin{center}
\begin{tabular}{ccccccccc}\hline\hline
$\int{\cal L}dt$ & \multicolumn{8}{c}{Higgs decay final state} \\ \cline{2-9}
 (fb$^{-1}$) & $\gamma\gamma$ & $WW^*$ & $ZZ^*$ &  $b\bar{b}$ & $\tau\tau$ & $\mu\mu$ & $Z\gamma$ & BR$_{\rm inv}$ \\ \hline
\multicolumn{9}{c}{ATLAS} \\ \hline
 300         &  $9-14\%$   &  $8-13\%$  &  $6-12\%$  &  N/A        &  $16-22\%$ & $38-39\%$ & $145-147\%$ & $<23-32\%$     \\
3000         &  $4-10\%$   &  $5-9\%$   &  $4-10\%$  &  N/A        &  $12-19\%$ & $12-15\%$ & $54-57\%$   & $<8-16\%$     \\ \hline
\multicolumn{9}{c}{CMS} \\ \hline
 300         &  $6-12\%$  &  $6-11\%$  &  $7-11\%$  &  $11-14\%$  &  $8-14\%$  & $40-42\%$ & $62-62\%$  & $<17-28\%$ \\
3000         &  $4-8\%$   &  $4-7\%$   &  $4-7\%$   &  $5-7\%$    &  $5-8\%$   & $14-20\%$ & $20-24\%$  & $<6-17\%$ \\ \hline\hline 
\end{tabular}
\end{center}
\label{tab:LHCRateProjections}
\end{table}

The estimates from ATLAS and CMS are similar for most of the final states with a few notable exceptions. ATLAS has no
estimate for $H\to b\bar{b}$ at this time, it's estimate for $H\to\tau\tau$ was based on an old study and significant
improvement is expected. The large difference between the two $H\to Z\gamma$ estimates needs to be understood. For these
reasons, CMS projections are taken as the expected LHC per-experiment precisions below.
 
Table~\ref{tab:LHCCouplingProjections} summarizes the expected precision on the Higgs couplings for the two 
aforementioned assumptions of systematic uncertainties from the fit to a generic 7-parameter model. 
These 7 parameters are $\kappa_\gamma$, $\kappa_g$, $\kappa_W$, $\kappa_Z$, $\kappa_u$, $\kappa_d$ 
and $\kappa_\ell$. In this parameter set, $\kappa_\gamma$ and $\kappa_g$ parametrize 
potential new physics in the loops of the $H\gamma\gamma$ and $H gg$ couplings. 
$\kappa_u\equiv \kappa_t=\kappa_c$, $\kappa_d\equiv \kappa_b=\kappa_s$ and $\kappa_\ell\equiv\kappa_\tau=\kappa_\mu$ 
parametrize deviations to up-and down-type quarks and 
charged leptons respectively assuming fermion family universality. Only SM production modes and decays are 
considered in the fit. The derived precisions on the Higgs total width are also included.
The expected precision ranges from $5-15\%$ for 300~fb$^{-1}$ and $2-10\%$ for
3000~fb$^{-1}$. They are limited by systematic uncertainties, particularly theoretical uncertainties
on production and decay rates. Statistical uncertainties are below one percent in most cases. Note that
the sensitivity to $\kappa_u$ is derived from the $t\bar{t}H$ production process and only 
$H\to\gamma\gamma$ and $H\to b\bar{b}$ decays have been included in the projection.

The fit is extended to allow for BSM decays while restricting the Higgs coupling to vector bosons
not to exceed their SM values ($\kappa_W,\kappa_Z\le 1$). The resulting upper limit on the branching ratio 
of BSM decay is included in the table. Note that the BR$_{\rm BSM}$ limit is derived from the
visible decays of Table~\ref{tab:LHCRateProjections} and is independent of the limit on
BR$_{\rm inv}$ from the search of $ZH$ with $H\to$ invisible.

Also listed in the Table~\ref{tab:LHCCouplingProjections} are the expected precisions on $\kappa_{Z\gamma}$
and $\kappa_\mu$, coupling scale factors for $H\to Z\gamma$ and $H\to\mu\mu$ decay vertices. 
Given the small branching ratios of the two decays in the SM, they have negligible impact on the 7-parameter fit.
With the noted differences above, ATLAS estimates are similar.  

\begin{table}[htb!]
\caption{Expected per-experiment precision of Higgs boson couplings to fermions and vector bosons with
300~fb$^{-1}$ and 3000~fb$^{-1}$ integrated luminosity at the LHC. The 7-parameter fit assumes the SM productions 
and decays as well as the generation universality of the couplings ($\kappa_u\equiv\kappa_t=\kappa_c$, 
$\kappa_d\equiv \kappa_b=\kappa_s$ and $\kappa_\ell\equiv\kappa_\tau=\kappa_\mu$). The  precision on the total 
width $\Gamma_H$ is derived from the precisions on the couplings. The range represents spread from two assumptions
of systematic uncertainties, see text.}
\begin{center}
\begin{tabular}{cccc}\hline\hline
Luminosity  & 300 fb$^{-1}$   && 3000 fb$^{-1}$ \\ \hline
Coupling parameter  & \multicolumn{3}{c}{7-parameter fit} \\ \hline
$\kappa_\gamma$     & $5-7\%$    &&  $2-5\%$ \\
$\kappa_g$          & $6-8\%$    &&  $3-5\%$ \\
$\kappa_W$          & $4-6\%$    &&  $2-5\%$ \\ 
$\kappa_Z$          & $4-6\%$    &&  $2-4\%$ \\
$\kappa_u$          & $14-15\%$  &&  $7-10\%$ \\
$\kappa_d$          & $10-13\%$  &&  $4-7\%$  \\ 
$\kappa_\ell$       & $6-8\%$    &&  $2-5\%$ \\ \hline 
$\Gamma_H$          & $12-15\%$    &&  $5-8\%$ \\ \\
   & \multicolumn{3}{c}{additional parameters (see text)} \\ \hline
$\kappa_{Z\gamma}$  & $41-41\%$  && $10-12\%$ \\
$\kappa_\mu$        & $23-23\%$  && $8-8\%$ \\ 
BR$_{\rm BSM}$          & $<14-18\%$ && $<7-11\%$ \\ \hline\hline
\end{tabular}
\end{center}
\label{tab:LHCCouplingProjections}
\end{table}

Apart from contributions from ATLAS and CMS collaborations, several independent 
studies~\cite{Onyisi:2013gta,Goncalo:2013oma,Vasquez:2013tja} have been performed.
In Ref.~\cite{Onyisi:2013gta}, authors investigate top-quark Yukawa coupling through the $t\bar{t}H$ production 
and $H\to WW^*$ decay. It is estimated that the $\kappa_t$ can be measured with a precision of $14-16\%$ and $6-9\%$
in 300 and 3000 fb$^{-1}$ from this final state alone, comparable to, but independent of, the $\kappa_u\ (\equiv\kappa_t)$ 
precision shown in Table~\ref{tab:LHCCouplingProjections}. Combining results from these independent final
states will improve the precision on $\kappa_t$. 

The rare decays $H \to V \gamma$, where $V$ denotes a vector meson such as the $J/ \psi$ or
the $\Upsilon(1S)$ which subsequently decays via $V \to \ell^+ \ell^-$, provide a handle on 
otherwise difficult-to measure properties of the Higgs boson~\cite{Bodwin:2013gca}.   Quantum interference 
between the two production mechanisms that contribute to this decay enhances its sensitivity to 
the $H\bar{Q}Q$ coupling, and potentially allows the
$H\bar{c}c$ coupling to be constrained directly by measurement of
the branching ratio for $H \to J/\psi \, \gamma$.

\begin{figure}[htb!]
\begin{center}
\includegraphics[width=0.48\textwidth]{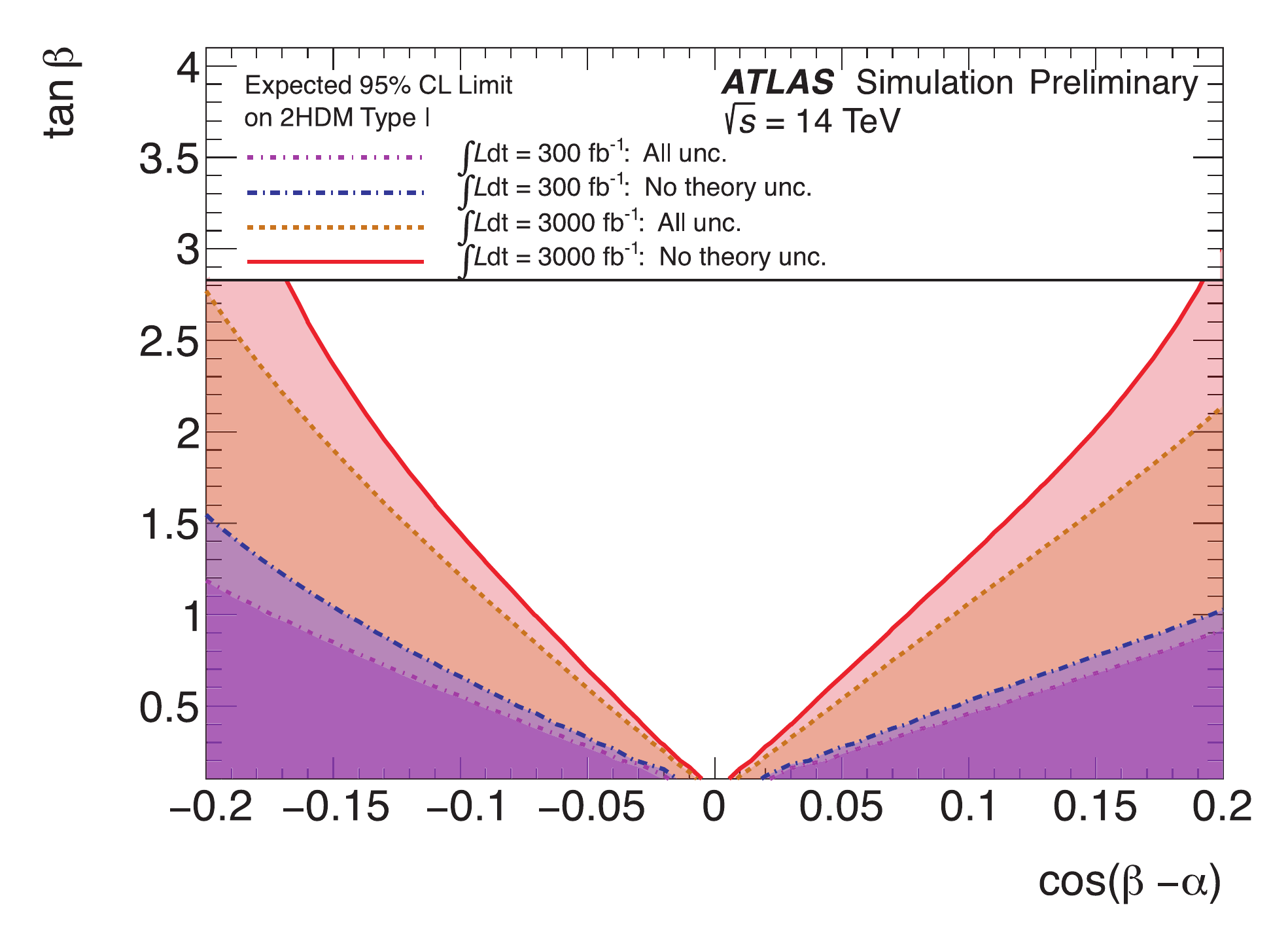}
\includegraphics[width=0.48\textwidth]{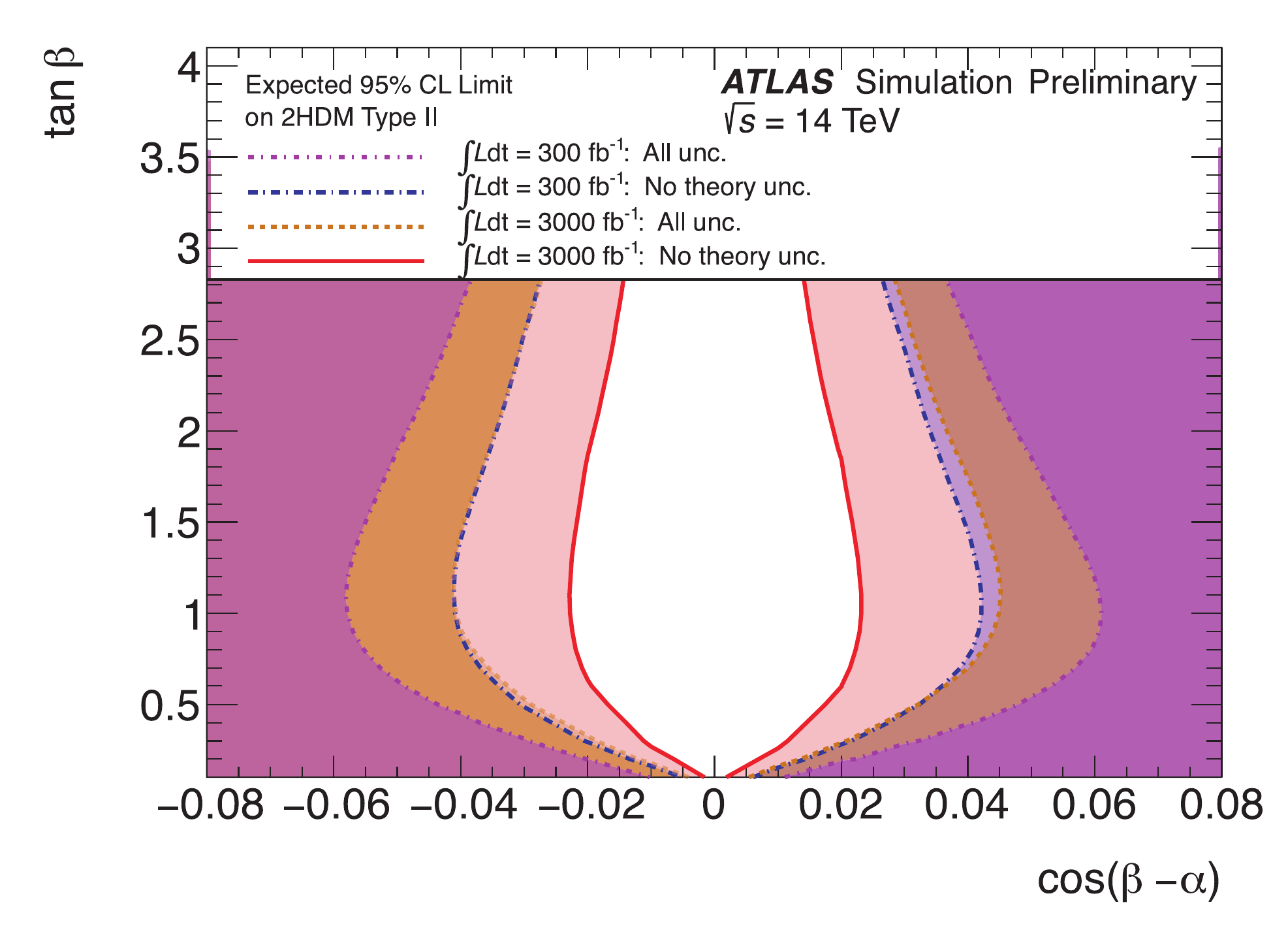}
\end{center}
\caption{Regions of the $\tan\beta$-$\cos(\beta-\alpha)$ plane of 2HDMs expected to be excluded with 300 and 3000~fb$^{-1}$ 
at $\sqrt{s}=14$~TeV for Type~I model~(left) and Type~II model~(right).}
\label{fig:ATLAS_2HDM}
\end{figure}

Higgs boson couplings to fermions and vector bosons are modified in two Higgs doublet models as discussed in Sec.~\ref{sec:2HDM}.
Therefore precise measurements of Higgs boson couplings can significantly constrain the parameter space of these models. 
Interpreting the 125~GeV particle as the light CP-even neutral Higgs boson in 2HDMs, ATLAS has estimated the  
expected limits on the $\tan\beta - \cos(\beta-\alpha)$ plane for Type~I and II models~\cite{ATLAS-PHYS-PUB-2013-015} as 
shown in Fig.~\ref{fig:ATLAS_2HDM}. The value of $\tan\beta<3$ is chosen such that $bbh$ production can be neglected.

\subsection{Projections for $\mathbf{e^+e^-}$ machines}\vspace*{-0.4cm}
\label{sec:couplings_epem}

The measurements of Higgs couplings in $e^+e^-$ collisions benefit from 
a clean experimental environment, precisely known $\sqrt{s}$ and initial state
polarization, and well predicted backgrounds many orders
of magnitude below the challenging QCD backgrounds of the hadron colliders.
The $e^+e^-$ collider is a well-studied Higgs factory.
Although most studies in the past decades focus on linear 
colliders~\cite{Murayama:1996ec,Accomando:1997wt,Dawson:2004xz,Abe:2001wn,
AguilarSaavedra:2001rg,Abe:2001gc,Djouadi:2007ik,Brau:2012hv,Baer:2013cma},
the experimentally accessible Higgs physics 
at a given center-of-mass energy
depends only weakly 
whether it is a linear or circular machine~\cite{Blondel:2012a,Azzi:2012a}, with differences
driven primarily by luminosity and possible number of detector interaction points.
In the measurement of 
Higgs couplings at a linear collider, the very small beam size at the interaction point and 
time structure of the beams allowing vertex detectors to be operated
in a pulsed mode would benefit flavor tagging in $H \rightarrow b\bar{b}$ and $c\bar{c}$ decays.
Beamstrahlung effects, resulting in collisions at less than $2E_{\mathrm{beam}}$,
tend to be less in a circular $e^+e^-$ machine than at a linear collider,
although the impact on Higgs precision measurements is small.

The measurement of couplings naturally divides according to the production process.
At relatively low $\sqrt{s}$  energies of $\simeq 250$ -- 350 GeV, the Higgs-strahlung
process $e^+e^- \rightarrow ZH$ dominates and tagging the $Z$ allows for a model-independent
separation of the recoil Higgs decays.  For $\sqrt{s} \geq 500$ GeV, the $W$-fusion mode
$e^+e^- \rightarrow \nu_e \bar{\nu}_e H$ dominates and grows with $\sqrt{s}$ allowing for
better precision of the $WW$ coupling and higher statistics for other decay modes, including
rare decays.
The cross sections of these processes in $e^+e^-$ collisions at representative collision
energies are given in Table~\ref{tab:epemxsecs}.
These higher energies also provide access to the top quark Yukawa
coupling through $e^+e^- \rightarrow t\bar{t}H$ and the Higgs trilinear self-coupling
via double-Higgs production: $e^+e^- \rightarrow ZHH$ and $\nu_e \bar{\nu}_e HH$
(discussed in section~\ref{chap:doublehiggs}).

\begin{table}[htb]
\caption{Dominant Higgs boson production cross sections at various  $e^+e^-$ collision energies.
Cross sections are calculated~\cite{Kilian:2007gr} including initial-state radiation, but not beamstrahlung effects,
for unpolarized beams and the enhancement due to polarized beams ($P(e^-,e^+) = (-0.8,0.3)$ for 250, 350, and 500 GeV, baseline for the
ILC; $(-0.8,0.2)$ for 1000 GeV, baseline for the ILC; $(-0.8,0.0)$ for 1.4 and 3.0 TeV, typical for CLIC.)} 
\begin{center}
\begin{tabular}{crccccccc}\hline\hline
                      & \multicolumn{6}{c}{Cross sections in fb \@ $m_H=125$~GeV} \\ \hline
  Mode                & & $\sqrt{s}$ (GeV) $=$ & 250  &  350  &  500  &  1000  &  1400  & 3000 \\ \hline 
  $ZH$                & unpolar. &             & 211  &  134  &  64.5 &  16.1  &  8.48  & 2.00 \\
                      &   polar. &             & 318  &  198  &  95.5 &  22.3  &  10.0  & 2.37 \\
  $\nu_e\bar{\nu}_eH$ & unpolar. &             & 20.8 &  34.1 &  71.5 &  195   &  278   & 448  \\
                      &   polar. &             & 36.6 &  72.5 &  163  &  425   &  496   & 862  \\
  $e^+e^-H$           & unpolar. &             & 7.68 &  7.36 &  8.86 &  20.1  &  27.3  & 48.9 \\ 
                      &   polar. &             & 11.2 &  10.4 &  11.7 &  24.7  &  32.9  & 56.5 \\ \hline\hline
  \end{tabular}
  \end{center}
  \label{tab:epemxsecs}
  \end{table}

\subsubsection{Collision energies 250 -- 350 GeV}\vspace*{-0.4cm}

A key production mode is $e^+e^- \rightarrow ZH$ where events can be detected inclusively,
completely independent of the Higgs decay mode by tagging the $Z$ via $Z \rightarrow \mu^+ \mu^-$
and $e^+e^-$ and requiring that the recoil mass is consistent with the Higgs boson mass.
The normalization of this rate then allows a precision measurement of 
$\sigma(ZH)$ that is in turn proportional to $\kappa_Z^2$.
With this in hand, specific $H$ decay modes can be examined and
measurements of $\sigma(ZH) \cdot {\rm BR}$ lead to {\em absolute} measurements
of {\em all} possible branching fractions, including invisible and exotic Higgs decays, as well as
decay modes undetectable 
at the LHC due to large backgrounds (e.g., $H \rightarrow c\bar{c}$ or decays to
light quark-like jets).
In many cases of the measurement of $H$ branching fractions, $Z$ decays to hadronic modes are included.
Note that the uncertainty on $\sigma(ZH)$ at $\sqrt{s} = 250$~GeV eventually limits the precisions on the 
branching fraction measurements.
Assuming a single resonance, 
\begin{equation}
\Gamma_H = \Gamma(H \rightarrow ZZ)/{\rm BR}(H \rightarrow ZZ) \propto
\sigma(ZH)/{\rm BR}(H \rightarrow ZZ),
\label{eq:Higgst_width}
\end{equation}
allowing  a model independent extraction of the width of the Higgs, free
from confusion of whether there is new physics in couplings or in new decay modes.
Note that the measurement of this recoil process is mandatory in a fully model-independent
measurement of Higgs couplings.
At increasing $\sqrt{s}$, starting at, e.g., the 350 GeV TLEP or the initial 350 GeV phase
of CLIC, there is enough rate in the $WW$-fusion process
so that $\Gamma(H \rightarrow WW^*)$ can be determined by measuring the cross section for $e^+e^-  \rightarrow \nu_e \bar{\nu}_e H$,
giving another handle on the total Higgs width, using
\begin{equation}
\Gamma_H = \Gamma(H \rightarrow WW^*)/{\rm BR}(H \rightarrow WW^*).
\label{eq:WWfusion_width}
\end{equation}
This is even more true of the higher energies at 500 GeV and beyond.
Such a rich program of Higgs physics can be carried out at any of the $e^+e^-$ machines with sufficient luminosity.

Full simulations of such events in
the ILD~\cite{ILD1} and SiD~\cite{SiD1} detectors~\cite{LCdetect} have been performed
over many years, including 
all physics backgrounds.  Overlays of $\gamma\gamma \rightarrow {\mathrm{hadrons}}$ and beam-induced backgrounds
have also been included for studies at ILC~\cite{ILCWhite:2013a,Brau:2013mba,ILCTDR:2013a} and most
CLIC~\cite{CLICWhite:2013a} studies.  
For these and higher energies, the ILC studies also include estimates of systematic uncertainties
on luminosity, polarisation, and $b$-tagging efficiencies and mistag rates.
A full simulation of the CMS detector has been used to make
projections of precisions attainable at TLEP~\cite{TLEP:2013a,Azzi:2012a,TLEP:2013b},
with extrapolations made for $H \rightarrow c\bar{c}$ and $gg$. 
Results are collated in Table~\ref{tab:epem_modelindfits} for the precision on couplings in model-independent fits,
and in Table~\ref{tab:fitinputs} for the precision on input cross sections and branching fractions.

\subsubsection{Collision energies $\mathbf{\geq}$~500~GeV}\vspace*{-0.4cm}

Collisions of $e^+e^-$ at $\sqrt{s} \geq 500$~GeV are the exclusive realm of linear colliders 
(more speculative rings such as the Very Large Lepton Collider (VLLC) with circumferences greater than 100 km are 
not considered here).
At these higher energies, large samples of events from both the $WW$ and $ZZ$ fusion processes lead to improved
precision on all the branching fractions, and allow probing of rare decays such as $H \rightarrow \mu^+\mu^-$.
Equally important, the relation of Eq.~\ref{eq:WWfusion_width} 
provides a significantly improved measurement of the total Higgs width consequently improving 
the precision on {\em all} the branching fractions and 
model-independent extraction of the associated Higgs couplings.

Higher energies also open up the production channel $e^+e^- \rightarrow t\bar{t}H$. 
Significant enhancements of this cross section near threshold
due to $t\bar{t}$ bound states~\cite{Farrell:2006xe} implies that the measurement of the top Yukawa coupling $\kappa_t$ may already
be possible at $\sqrt{s} = 500$~GeV~\cite{Yonamine:2011jg}, but has more sensitivity
at the higher energy operating points of the ILC and CLIC where the signal 
cross section is larger and $t\bar{t}$ background is smaller.

Studies using full simulations of detectors at the ILC and CLIC~\cite{CLICWhite:2013a,ILCWhite:2013a,Brau:2013mba,ILCTDR:2013a}
result in coupling precisions presented in Table~\ref{tab:epem_modelindfits} for the precision on couplings in model-independent fits,
and in Table~\ref{tab:fitinputs} for the precision on input cross sections and branching fractions.

\subsubsection{Model Independent Coupling Fits}\vspace*{-0.4cm}

To provide a true representation of the lepton-collider potential, as well as a comparison between
$e^+e^-$ options on an equal footing, Table~\ref{tab:epem_modelindfits} shows the precision on
couplings from global fits without
any assumptions on or between $\kappa_W$ and $\kappa_Z$, nor with any assumptions on the saturation of
the total width by invisible decays.
The inputs to these model-independent fits are taken from Table~\ref{tab:fitinputs}.

\begin{sidewaystable}[htb!]
 \caption{Uncertainties on coupling scaling factors as determined in a completely
   model-independent fit for different $e^+e^-$ facilities.
   Precisions reported in a given column include in the fit all
   measurements at lower energies at the same facility, and note that the model independence
   requires the measurement of the recoil $HZ$ process at lower energies.
   $^\ddag$ILC luminosity
   upgrade assumes an extended running period on top of the low luminosity program
   and cannot be directly compared to TLEP and CLIC numbers without accounting for
   the additional running period. 
   ILC numbers include a 0.5\% theory uncertainty. 
   For invisible decays of the Higgs, the number quoted is the 95\% confidence upper limit on the
   branching ratio.
}
\begin{center}
\begin{tabular}{lccccccccc}\hline\hline
Facility                     &  \multicolumn{3}{c}{ILC}    & ILC(LumiUp)    & \multicolumn{2}{c}{TLEP (4 IP)}   & \multicolumn{3}{c}{CLIC}  \\
\hline
$\sqrt{s}$ (GeV)                 & 250         &  500       &  1000    & 250/500/1000 & 240 &  350 & 350  & 1400  & 3000    \\
$\int{\cal L}dt$ (fb$^{-1}$) & 250         & +500       &  +1000   & 1150+1600+2500$^\ddag$ &  10000 & +2600 & 500 & +1500 & +2000 \\
$P(e^-,e^+)$ & $(-0.8,+0.3)$ & $(-0.8,+0.3)$ & $(-0.8,+0.2)$ & (same) & $(0,0)$ & $(0,0)$ & $(0,0)$ & $(-0.8,0)$ & $(-0.8,0)$ \\
\hline
$\Gamma_H$   & 12\%        &  5.0\%     &  4.6\%   &   2.5\%      &  1.9\%  &  1.0\% & 9.2\% & 8.5\% & 8.4\% \\ \\

$\kappa_{\gamma}$ &  18\%  & 8.4\%  & 4.0\%   &  2.4\%  & 1.7\%  & 1.5\%  & $-$ & 5.9\% & $<$5.9\%  \\
 $\kappa_{g}$           &  6.4\% & 2.3\%  & 1.6\%   &  0.9\% & 1.1\%  & 0.8\%  & 4.1\% & 2.3\%  & 2.2\% \\
 $\kappa_{W}$           &  4.9\% & 1.2\%  & 1.2\%   &  0.6\% & 0.85\% & 0.19\% & 2.6\% & 2.1\% & 2.1\% \\ 
 $\kappa_{Z}$           &  1.3\% & 1.0\%  & 1.0\%   &  0.5\% & 0.16\% & 0.15\%  & 2.1\% & 2.1\% & 2.1\% \\  \\

 $\kappa_{\mu}$     &   91\%  & 91\%    & 16\%   &  10\%  & 6.4\%  & 6.2\%  & $-$ & 11\% & 5.6\% \\
 $\kappa_{\tau}$   &  5.8\% & 2.4\%  & 1.8\%  & 1.0\% & 0.94\% & 0.54\% & 4.0\%  & 2.5\% & $<$2.5\% \\
 $\kappa_{c}$           &  6.8\% & 2.8\% & 1.8\% & 1.1\% & 1.0\% & 0.71\%  & 3.8\% & 2.4\% & 2.2\% \\ 
 $\kappa_{b}$           &  5.3\% & 1.7\% & 1.3\% &  0.8\% & 0.88\% & 0.42\% & 2.8\% & 2.2\% & 2.1\% \\ 
 $\kappa_{t}$           &  $-$ & 14\%  & 3.2\% &  2.0\%  &  $-$  & 13\%   & $-$ & 4.5\% & $<$4.5\% \\  
\hline
$BR_{\rm inv}$       & $0.9$\%   & $< 0.9$\% & $< 0.9$\% & $0.4$\% & $0.19$\%  &  $< 0.19$\%  &  &  &  \\ 
\hline
\end{tabular}
 \end{center}
\label{tab:epem_modelindfits}
 \end{sidewaystable}


\subsection{Projections for a photon collider operating on the Higgs resonance}\vspace*{-0.4cm}

A photon collider operating on the Higgs resonance could
be constructed using laser Compton backscattering off of $ee$ beams at 
$\sqrt{s_{ee}} =160-220$~GeV~\cite{Bogacz:2012fs,Asner:2001ia,Ginzburg:1981ik, Ginzburg:1981vm, Badelek:2001xb,Niezurawski:2003iu,Asner:2001vh,Chou:2013xaa},
where the higher energy is strongly preferred~\cite{Telnov:2013bpa}.
The photon collider could measure $\Gamma_{\gamma\gamma} \times {\rm BR}(H \to X)$ from event
rates in various final states. Table~\ref{tab:gamgamcollider} summarizes the anticipated sensitivities to
production times decay rates, corresponding to 50,000 raw $\gamma\gamma\to H$ events.

Model-independent Higgs coupling extraction is not possible unless
input from another collider can be provided.  Combining photon collider 
measurements with a model-independent
measurement of BR$(H \to bb)$ from an $e^+e^-$ collider yields
a 2\% measurement of $\Gamma_{\gamma\gamma}$, corresponding to 1\% precision on $\kappa_{\gamma}$.  
Combining this with the rate measurement for $\gamma\gamma \to H \to \gamma\gamma$
yields a measurement of the total Higgs width to 13\%.

\begin{table}
\caption{Photon collider precisions on Higgs production rates into
  various final states $X$, using  a sample of 50,000 $\gamma\gamma \to H$ 
  events~\cite{Bogacz:2012fs,Asner:2001ia,Badelek:2001xb,Niezurawski:2003iu,Asner:2001vh,Chou:2013xaa,Telnov:2013bpa}.
  }
\begin{center}
\begin{tabular}{c|ccccccccc}
\hline\hline
Final state & $b \bar b$ & $WW^*$ & $\tau\tau$ & $c \bar c$ & $gg$ & $\gamma\gamma$ & $ZZ^*$ & $Z\gamma$ & $\mu\mu$ \\
\hline
$\Gamma_{\gamma\gamma} \times {\rm BR}(H \to X)$ 
 & 1\% & 3\% & -- & -- & -- & 12\% & 6\% & 20\% & 38\% \\
\hline\hline
\end{tabular}
\end{center}
\label{tab:gamgamcollider}
\end{table}

\subsection{Projections for a muon collider operating on the Higgs resonance}\vspace*{-0.4cm}

A muon collider can produce the Higgs boson as an $s$-channel
resonace, $\mu^+ \mu^- \to H \to X$.  By scanning the beam energy
across the resonance, the Higgs total width can be measured directly
(see Sec.~\ref{sec:masswidthmuon}).  Combinations of production and
decay couplings can then be extracted from measurements of the event
rates in various final states.

Sensitivities have been studied for an idealized detector design
including full simulation in Ref.~\cite{MuonColliderWhitepaper}.
Important components of the detector are tungsten shielding cones at
high rapidity and precise timing to reduce beam-related backgrounds.

The studies in~\cite{MuonColliderWhitepaper} simulated Higgs events
and Drell-Yan backgrounds for a beam energy scan across the Higgs
peak.  Precisions on the $\mu\mu \to H \to X$ Higgs signal rates in
each channel and the mass and width resolution depend on the beam
energy spread, total luminosity, and scan strategy.
Table~\ref{tab:muonbrs} summarizes the precisions achievable from a
5-point energy scan centered on the Higgs resonance at $\sqrt{s} \sim
126$~GeV, with a scan point separation of 4.07~MeV.  The run scenario
assumes one Snowmass year ($10^7$~s) at $1.7 \times
10^{31}$~cm$^{-2}$s$^{-1}$ plus five Snowmass years at $8.0 \times
10^{31}$~cm$^{-2}$s$^{-1}$ and a beam energy resolution of $R = 4
\times 10^{-5}$ (the beam energy spread should be measurable to high
precision using muon precession in the accelerator field).  Perfect
$b$-tagging efficiency and purity were assumed.  An alternate strategy
of sitting on the Higgs peak increases the Higgs yield and would
slightly improve the rate measurements.

These rates are proportional to ${\rm BR}(H \to \mu\mu)\times {\rm
  BR}(H \to X) \propto \kappa_{\mu}^2 \kappa_X^2 / \Gamma_H^2$.  
Products of couplings $\kappa_{\mu} \kappa_X$ can be
extracted using the direct measurement of the Higgs width $\Gamma_H$ 
from the lineshape scan, with an estimated uncertainty $\Delta
\Gamma_H = 4.3\%$ (see also Sec.~\ref{sec:masswidthmuon}).
Model-independent Higgs coupling measurements are not possible unless
$\sigma(\mu\mu \to H \to \mu\mu) \propto \kappa_{\mu}^4/\Gamma_H^2$
can be measured.  Making the assumption of generation universality,
$\kappa_{\ell} \equiv \kappa_{\mu} = \kappa_{\tau}$, would allow
$\kappa_b$, $\kappa_W$, and $\kappa_{\ell}$ to be extracted, but the
uncertainty is dominated by the large ($\sim$60\%) uncertainty on the
$\mu\mu\to H \to \tau\tau$ rate.

\begin{table}
\caption{Muon collider statistical precisions on Higgs production
  rates into various final states $X$ from a 5-point energy scan
  centered at $m_H$ with a combined yield of 39,000 Higgs
  bosons.  The $\tau\tau$ uncertainty is an average of asymmetric
  uncertainties.  The rates are proportional to ${\rm BR}(H \to
  \mu\mu)\times {\rm BR}(H \to X) \propto \kappa_{\mu}^2 \kappa_X^2 /
  \Gamma_H^2$.}
\begin{center}
\begin{tabular}{c|ccccccccc||cc}
\hline\hline
Final state & $b \bar b$ & $WW^*$ & $\tau\tau$ & $c \bar c$ 
  & $gg$ & $\gamma\gamma$ & $ZZ^*$ & $Z\gamma$ & $\mu\mu$ & $\Gamma_H$ & $m_H$ \\
\hline
$\sigma(\mu\mu \to H \to X)$ & 9\% & 5\% & 60\% & -- & -- & -- & -- & -- & -- & 4.3\% & 0.06~MeV \\
\hline\hline
\end{tabular}
\end{center}
\label{tab:muonbrs}
\end{table}


\subsection{Comparison of Precision at Different Facilities}\vspace*{-0.4cm}


Precisions of measured cross sections and branching fractions compared across different
$e^+e^-$ Higgs factories and used as inputs to global coupling fits are presented in Table~\ref{tab:fitinputs}.

\newcommand{\ZH}{$ZH$}
\newcommand{\nnH}{ $\nu \bar{\nu} H$}
\newcommand{\nnHH}{ $\nu \bar{\nu} HH$}

\begin{table}[htb!]
\caption{Precisions of measured $\sigma \cdot {\rm BR}$ inputs for $e^+e^-$ Higgs factories 
for complete programs: ILC: 250~fb$^{-1}$ at 250 GeV, 500~fb$^{-1}$ at 500 GeV,
1000~fb$^{-1}$ at 1000 GeV; ILC LumiUp: adding 900~fb$^{-1}$ at 250 GeV, 1100~fb$^{-1}$ at 500 GeV,
1500~fb$^{-1}$ at 1000 GeV; CLIC: 500~fb$^{-1}$ at 350 GeV, 1500~fb$^{-1}$ at 1.4 TeV, 3000$^{-1}$ at 3.0 TeV;
TLEP (following luminosities the sum over 4 interaction points): 10000~fb$^{-1}$ at 240 GeV, 2600~fb$^{-1}$ at 350 GeV.
The CLIC numbers are assuming increased $WW$ cross sections above 1 TeV with $(-0.8,0)$ polarization of $(e^-,e^+)$ (effective luminosities scaled by a factor of
approximately 1.8 above the unpolarized case given in Ref.~\cite{CLICWhite:2013a}).
$^\dag$CLIC at 350~GeV. 
$^\ddag$ILC luminosity
   upgrade assumes an extended running period on top of the low luminosity program
   and cannot be directly compared to TLEP and CLIC numbers without accounting for
   the additional running period.  
}
\footnotesize
\begin{center}
\begin{tabular}{l|cc|cc|cc|c} \hline\hline
                        & \multicolumn{2}{c|}{ILC}                   & \multicolumn{2}{c|}{ILC LumiUp$^\ddag$}                   & \multicolumn{2}{c|}{CLIC}              & \multicolumn{1}{c}{TLEP} \\
		        & \multicolumn{2}{c|}{250/500/1000 GeV}      & \multicolumn{2}{c|}{250/500/1000 GeV}             & \multicolumn{2}{c|}{1.4/3.0 TeV}       & \multicolumn{1}{c}{240 \& 350~GeV} \\ \hline
		        & \ZH            & \nnH & \ZH            & \nnH                 &  \ZH$^\dag$   &        \nnH           &  \ZH $(\nu\bar{\nu}H)$     \\ \hline
Inclusive              & 2.6/3.0/$-$\%  &  $-$ & 1.2/1.7/$-$\%  &  $-$                 &  4.2\%       &  $-$                  &  0.4\%              \\
$H\to \gamma\gamma$    & 29-38\%        & $-$/20-26/7-10\% & 16/19/$-$\%     & $-$/13/5.4\%           &  $-$           &  11\%/$<11$\%        &  3.0\%   \\
$H\to gg$              & 7/11/$-$\%     & $-$/4.1/2.3\% & 3.3/6.0/$-$\%      & $-$/2.3/1.4\%            &  6\%    &  1.4/1.4\%            &  1.4\%    \\
$H\to ZZ^*$            & 19/25/$-$\%    & $-$/8.2/4.1\% & 8.8/14/$-$\%       & $-$/4.6/2.6\%            &         &  2.3/1.5\%            &  3.1\% \\
$H\to WW^*$            & 6.4/9.2/$-$\%  & $-$/2.4/1.6\% & 3.0/5.1/$-$\%      & $-$/1.3/1.0\%            &  2\%    &  0.75/0.5\% &  0.9\%        \\
$H\to\tau\tau$         & 4.2/5.4/$-$\%  & $-$/9.0/3.1\%  & 2.0/3.0/$-$\%      & $-$/5.0/2.0\%            &  5.7\%         &  2.8\%/$<2.8$\%            &  0.7\%  \\
$H\to b\bar{b}$        & 1.2/1.8/$-$\%  & 11/0.66/0.30\% & 0.56/1.0/$-$\%    & 4.9/0.37/0.30\%          &  1\%    &  0.23/0.15\%          &  0.2\%  (0.6\%)    \\
$H\to c\bar{c}$        & 8.3/13/$-$\%   & $-$/6.2/3.1\% & 3.9/7.2/$-$\%      & $-$/3.5/2.0\%            &  5\%    &  2.2/2.0\%            &  1.2\%    \\
$H\to \mu\mu$          & $-$            & $-$/$-$/31\%  & $-$            & $-$/$-$/20\%              &  $-$           &  21/12\%    &  13\%        \\
\hline
& \multicolumn{2}{c|}{$t\bar{t}H$} & \multicolumn{2}{c|}{$t\bar{t}H$}        & \multicolumn{2}{c|}{$t\bar{t}H$}         & \multicolumn{1}{c}{$t\bar{t}H$} \\
\hline
$H\to b\bar{b}$        & \multicolumn{2}{c|}{$-$/28/6.0\%} & \multicolumn{2}{c|}{$-$/16/3.8\%}        & \multicolumn{2}{c|}{8\%/$<8$\%}   &  \multicolumn{1}{c}{$-$}  \\
\hline\hline
\end{tabular}
\end{center}
\label{tab:fitinputs}
\end{table}

\begin{figure}[htb]
\begin{center}
\begin{tabular}{@{}cc@{}}
\includegraphics[width=0.48\textwidth]{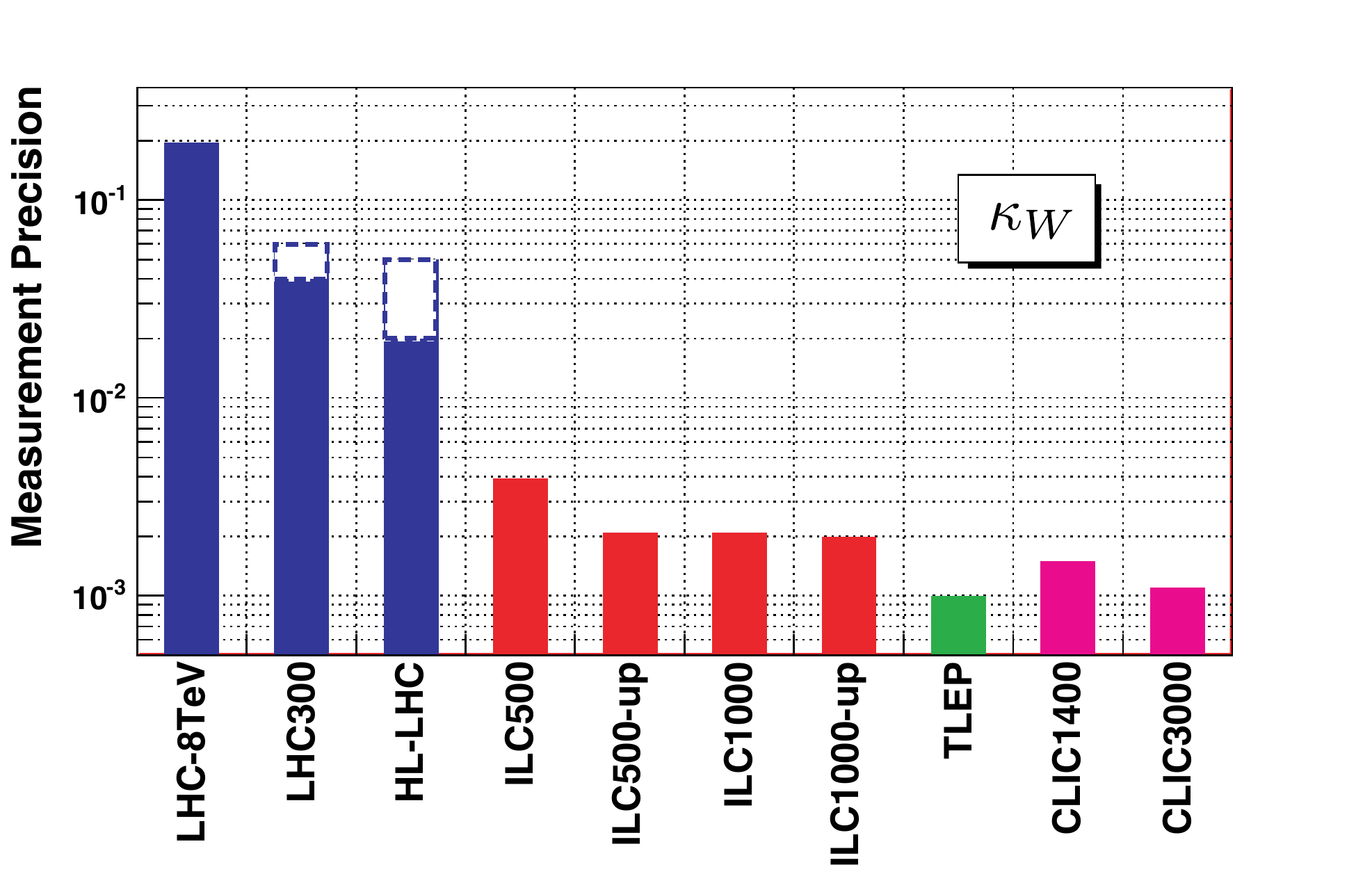} &
\includegraphics[width=0.48\textwidth]{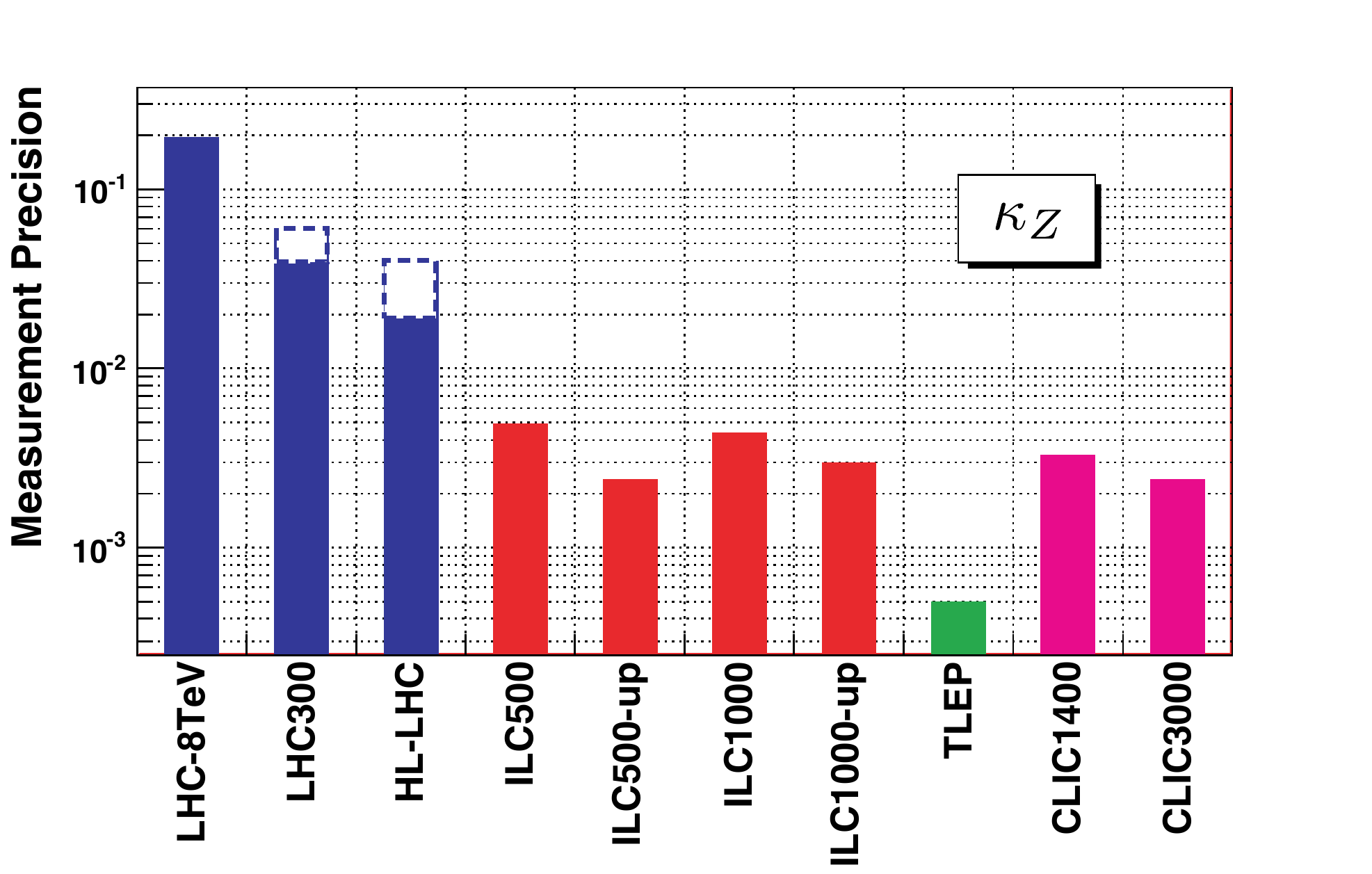} \\
\includegraphics[width=0.48\textwidth]{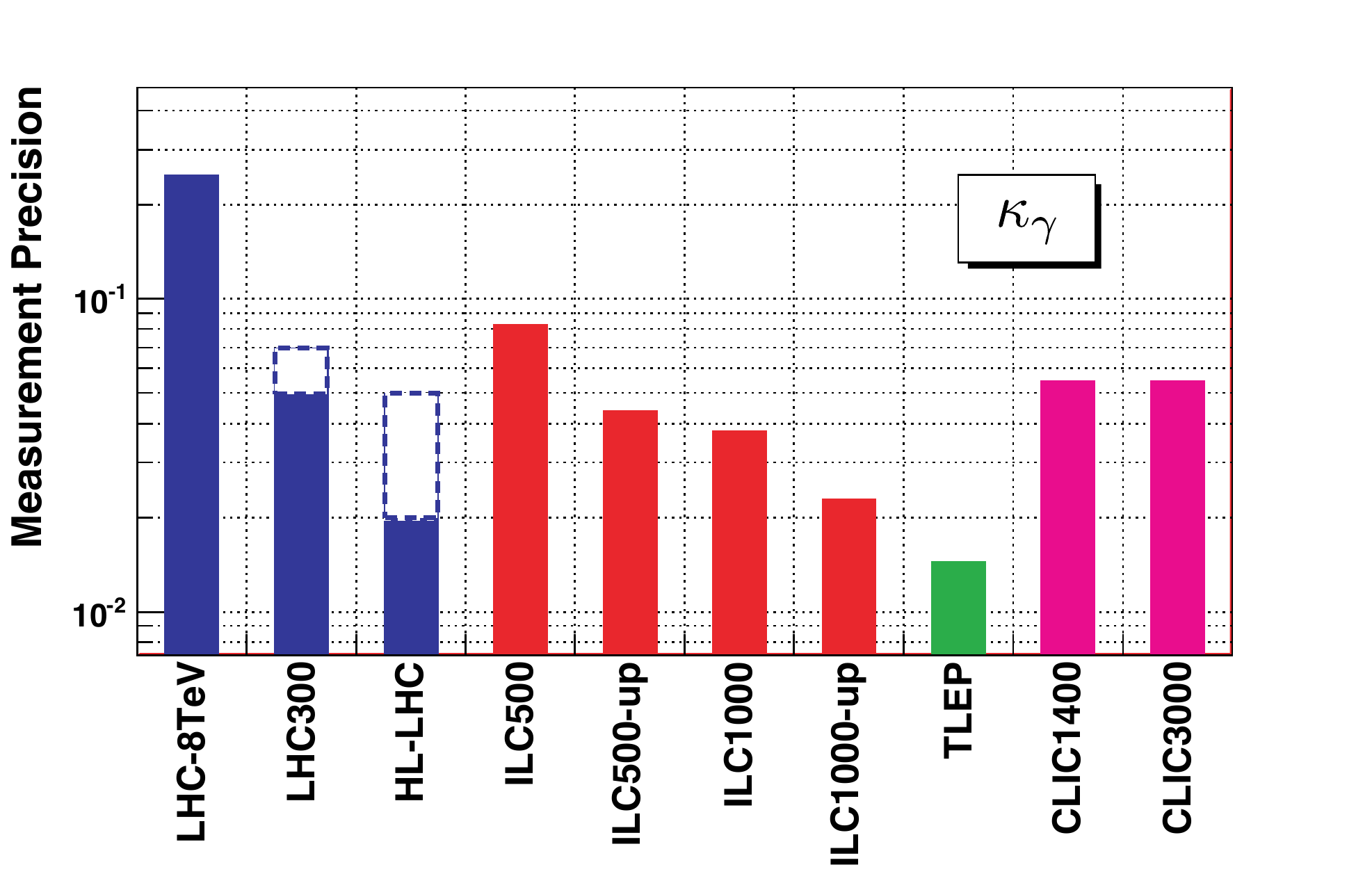} & 
\includegraphics[width=0.48\textwidth]{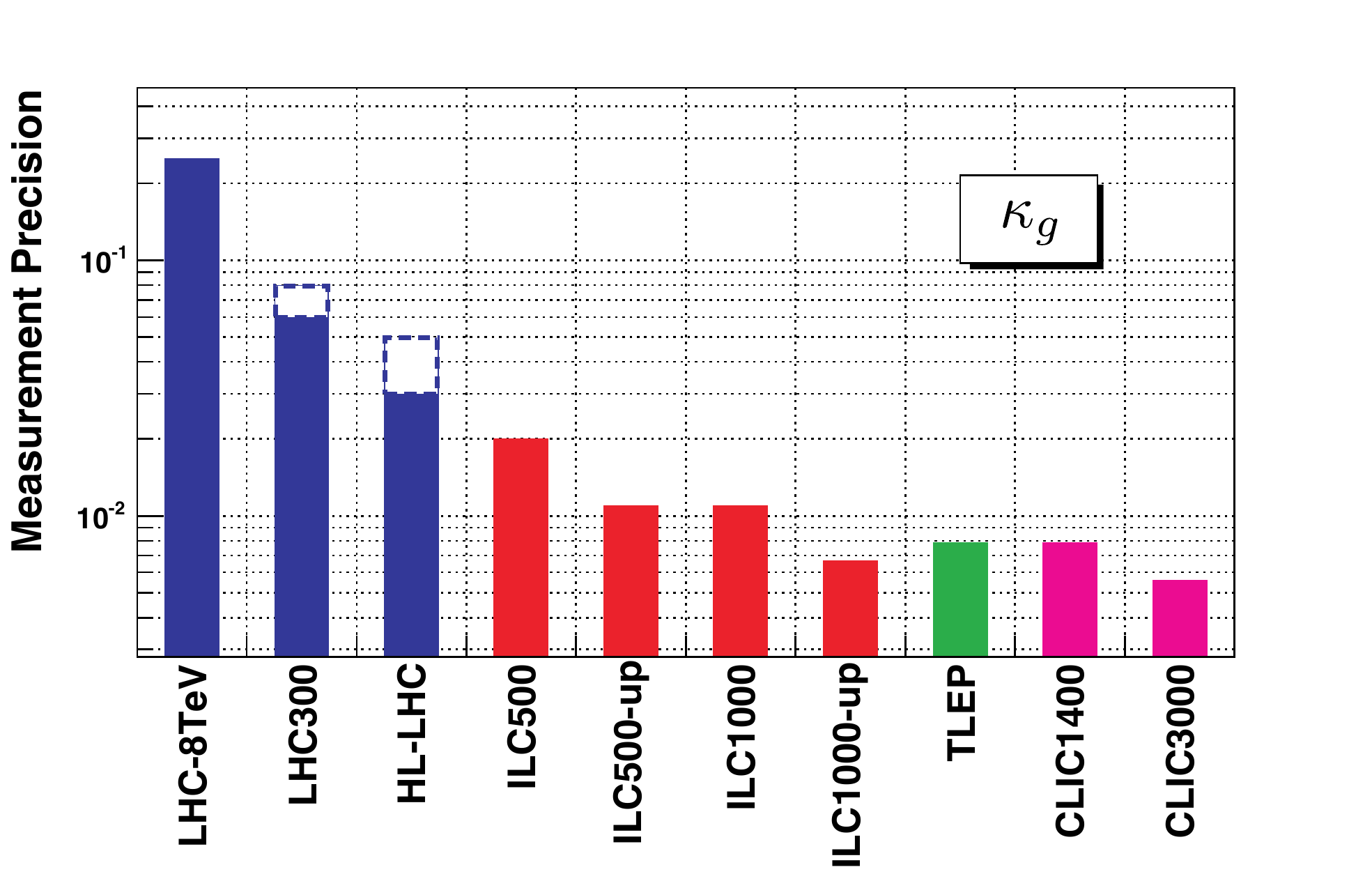} \\

\end{tabular}
\caption{Measurement precision on $\kappa_W$, $\kappa_Z$, $\kappa_{\gamma}$, and $\kappa_g$ at different facilities.\label{kappa_bosons}}
\end{center}
\end{figure}

\begin{figure}[htb]
\begin{center}
\begin{tabular}{@{}cc@{}}
\includegraphics[width=0.48\textwidth]{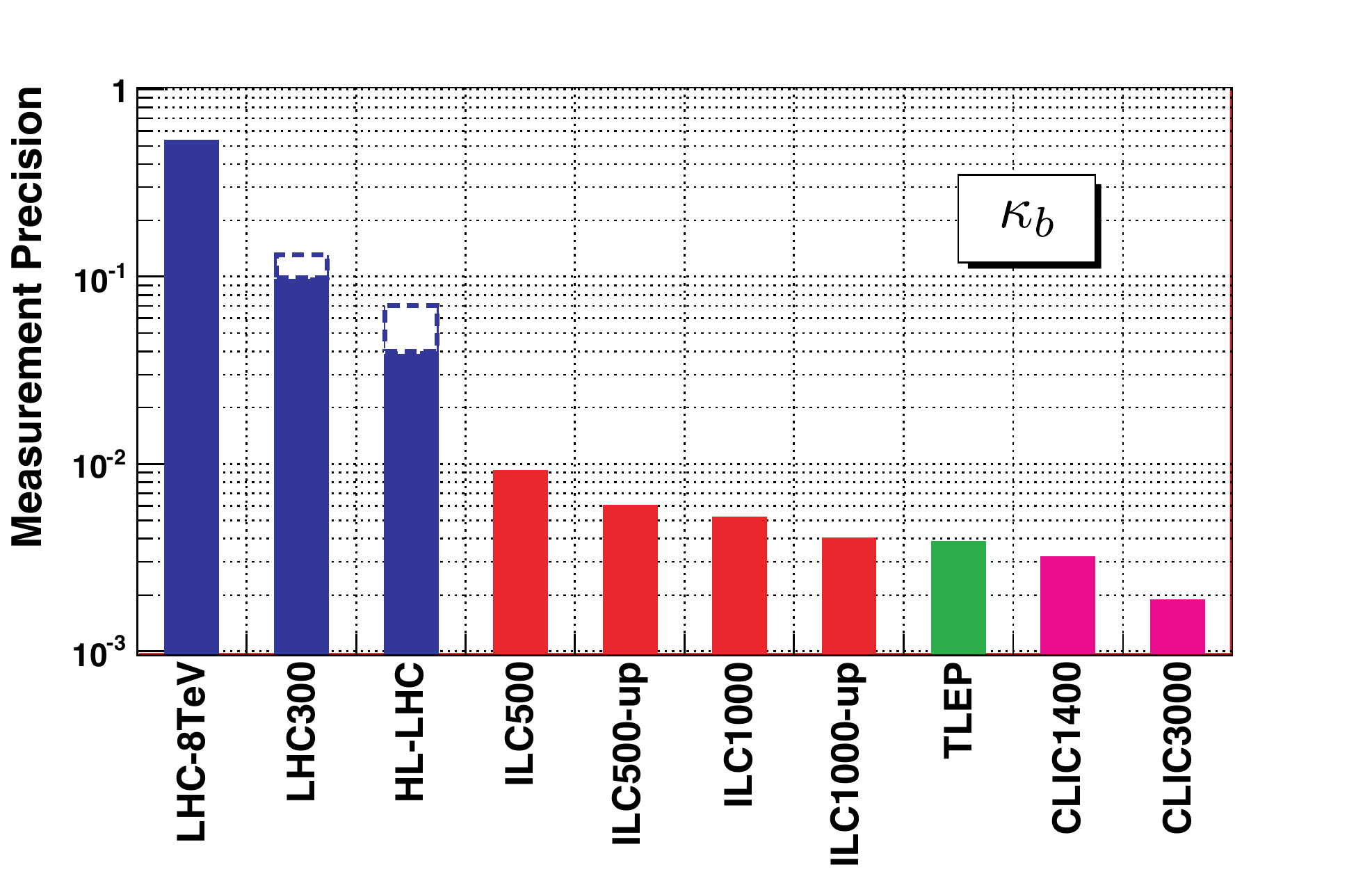} &
\includegraphics[width=0.48\textwidth]{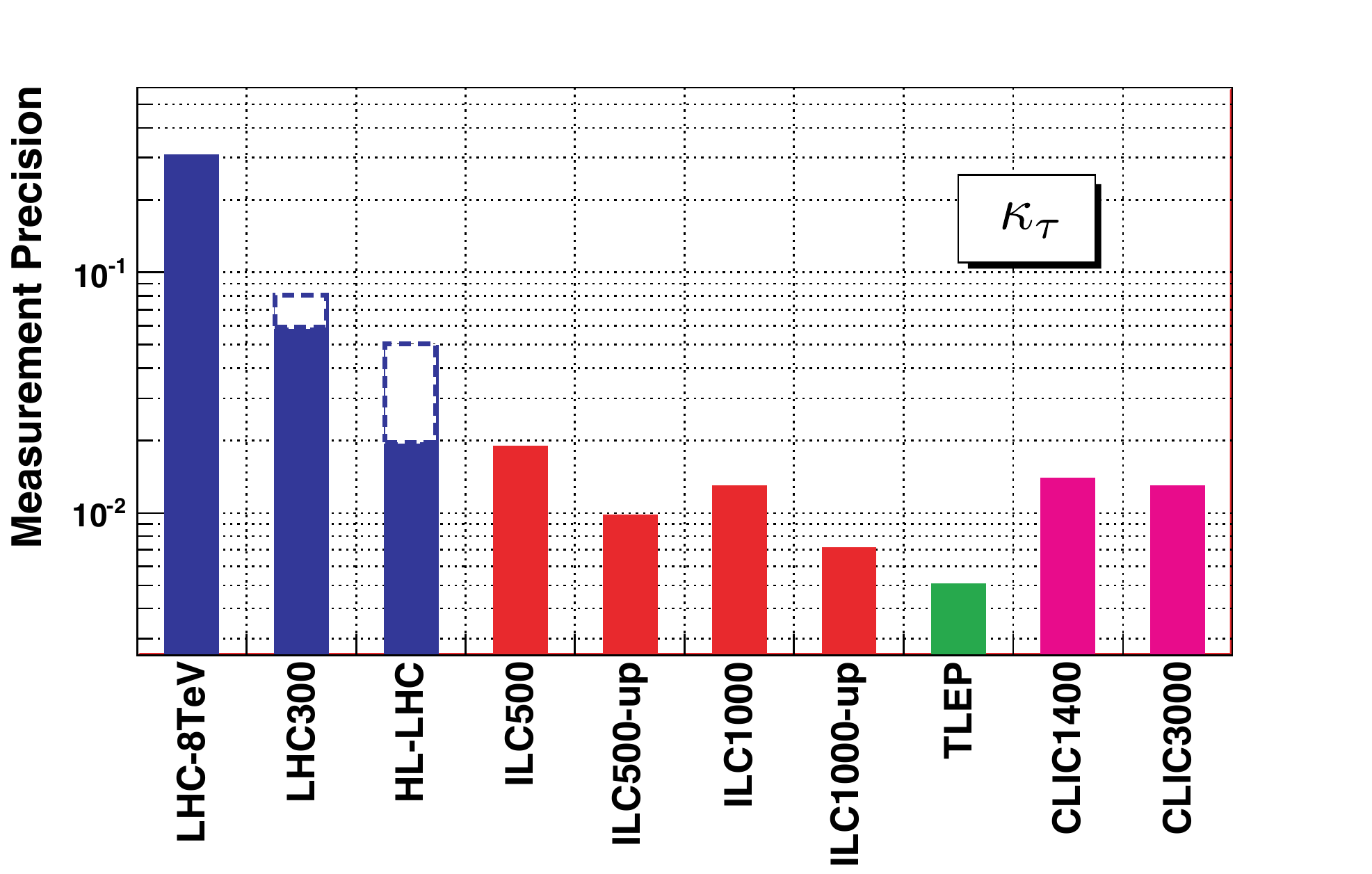} \\
\includegraphics[width=0.48\textwidth]{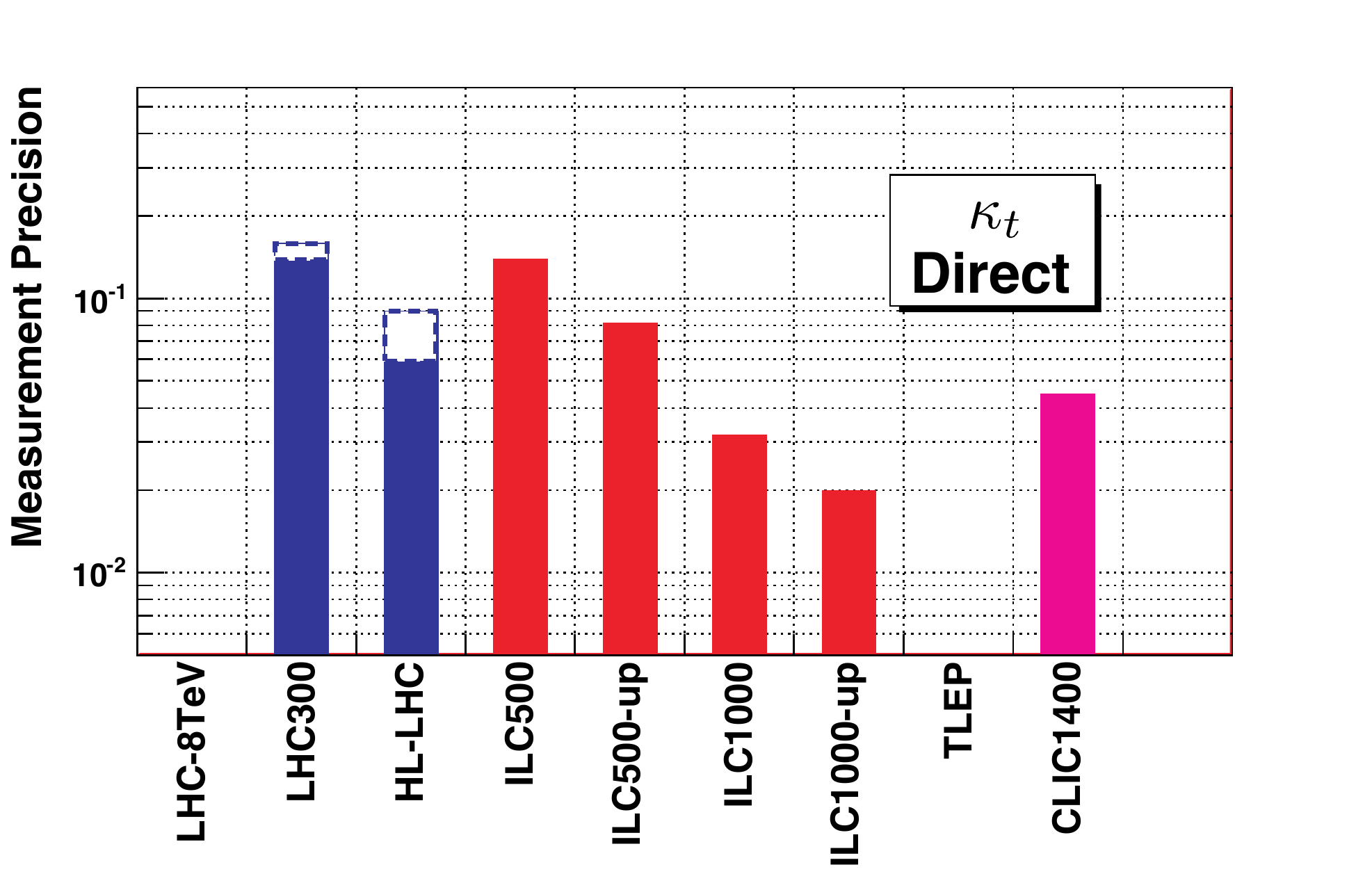} &
\includegraphics[width=0.48\textwidth]{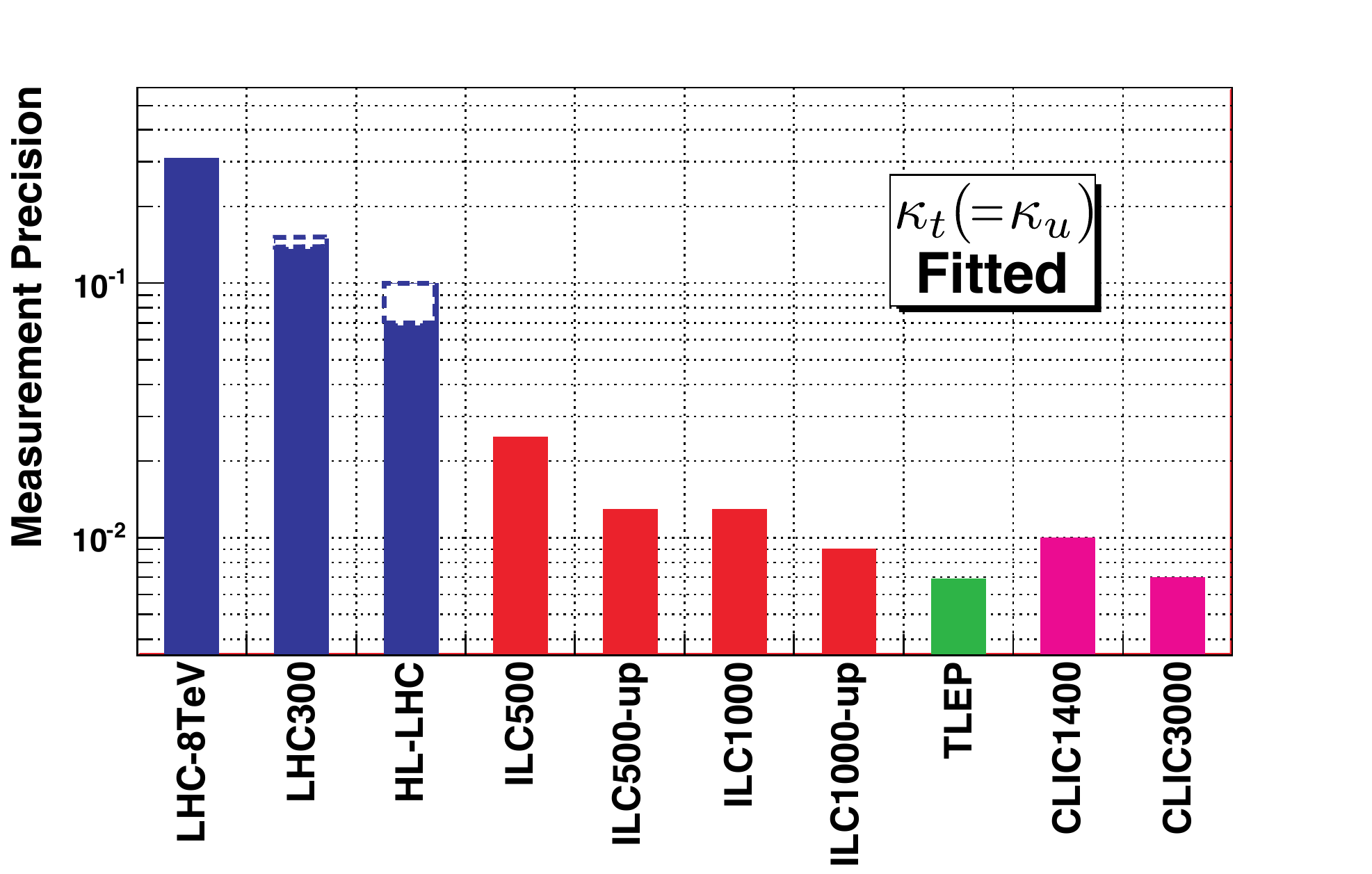} \\
\end{tabular}
\caption{Measurement precision on $\kappa_b$, $\kappa_{\tau}$, and $\kappa_t$ measured both directly
via $t\bar{t}H$ and through global fits at different facilities.\label{kappa_fermions}}
\end{center}
\end{figure}

As described earlier, the inputs of Table~\ref{tab:fitinputs} can be used to extract Higgs
couplings in a completely model-independent manner in global fits giving the results shown in 
Table~\ref{tab:epem_modelindfits} in section~\ref{sec:couplings_epem}.
Some level of model-dependence is needed to determine Higgs couplings from 
hadron collider measurements, so in order to make a comparison between facilities of the precisions that can be attained,
they are placed on equal footing using the same global 7-parameter fits as described
in section~\ref{sec:couplings_LHC} for all facilities with results summarized in Table~\ref{tab:compare_facilities}.
Comparisons of the precision on a subset of  $\kappa_x$ scale factors are also shown
in Figs.~\ref{kappa_bosons} and  \ref{kappa_fermions}. 

\begin{sidewaystable}[htb!]
 \caption{Expected precisions on the Higgs couplings and total width
   from a constrained 7-parameter fit assuming no non-SM production or
   decay modes.  The fit assumes generation universality ($\kappa_u
   \equiv \kappa_t = \kappa_c$, $\kappa_d\equiv \kappa_b = \kappa_s$,
   and $\kappa_{\ell} \equiv \kappa_{\tau} = \kappa_{\mu}$).  The
   ranges shown for LHC and HL-LHC represent the conservative and
   optimistic scenarios for systematic and theory uncertainties.  ILC
   numbers assume $(e^-,e^+)$ polarizations of $(-0.8,0.3)$ at 250 and
   500~GeV and $(-0.8,0.2)$ at 1000~GeV, plus a 0.5\% theory uncertainty.  
   CLIC numbers assume
   polarizations of $(-0.8,0)$ for energies above 1~TeV.  TLEP numbers assume
   unpolarized beams.}
\begin{center}
\begin{tabular}{lcccccccc}\hline\hline
Facility                     &   LHC            &   HL-LHC         &   ILC500  & ILC500-up & ILC1000 &   ILC1000-up &   CLIC          & TLEP (4 IPs) \\
$\sqrt{s}$ (GeV)                 &   14,000         &   14,000         &   250/500 & 250/500 & 250/500/1000 & 250/500/1000 &  350/1400/3000  &  240/350    \\
$\int{\cal L}dt$ (fb$^{-1}$) & 300/expt         & 3000/expt        &   250+500 & 1150+1600 & 250+500+1000 & 1150+1600+2500  & 500+1500+2000  & 10,000+2600  \\
\hline 
 $\kappa_{\gamma}$ &  $5-7$\%      & $2-5$\%       & 8.3\% & 4.4\% & 3.8\%  & 2.3\%            & $-$/5.5/$<$5.5\% & 1.45\% \\         
 $\kappa_g$           &  $6-8$\%       & $3-5$\%       & 2.0\% & 1.1\% & 1.1\%  & 0.67\%            & 3.6/0.79/0.56\% & 0.79\% \\
 $\kappa_W$           & $4-6$\% & $2-5$\% & 0.39\% & 0.21\% & 0.21\% & 0.2\%         & 1.5/0.15/0.11\% & 0.10\%\\
 $\kappa_Z$           & $4-6$\% & $2-4$\% & 0.49\% & 0.24\% & 0.50\% & 0.3\%        & 0.49/0.33/0.24\%& 0.05\%\\ 

 $\kappa_{\ell}$      & $6-8$\%       &  $2-5$\%      & 1.9\%  & 0.98\% & 1.3\%  & 0.72\%            & 3.5/1.4/$<$1.3\% & 0.51\%\\
 $\kappa_d =\kappa_b$ & $10-13$\%        &  $4-7$\%       & 0.93\% & 0.60\% & 0.51\% & 0.4\%           & 1.7/0.32/0.19\% & 0.39\% \\
 $\kappa_u =\kappa_t$ & $14-15$\%        &  $7-10$\%      & 2.5\%  &  1.3\% & 1.3\%   & 0.9\%            & 3.1/1.0/0.7\% & 0.69\% \\ 
\hline\hline
 \end{tabular}
 \end{center}
 \label{tab:compare_facilities}
\end{sidewaystable}

\newpage

A number of studies have presented results combining measurements from different
facilities~\cite{Klute:2013cx,Peskin:2012we}.  A general observation is that 
the precision in the measurement of many Higgs coupling at a new facility are reasonably 
or significantly improved, and these quickly dominate the combined results and overall
knowledge of the relevant coupling parameters. Exceptions are the measurements of the
branching fractions of rare decays such as
$H \rightarrow \gamma\gamma$ and $H \rightarrow \mu^+ \mu^-$ where results from
new lepton colliders would not significantly improve the coupling precisions driving these decays.
However, precision measurements of the ratio of $\kappa_Z/\kappa_{\gamma}$ at
hadron colliders combined with the high-precision and model-independent
measurements of $\kappa_Z$ at a lepton collider would substantially 
increase the precision on $\kappa_{\gamma}$.










\newpage
\section{Double Higgs production and the Higgs self-coupling}\vspace*{-0.3cm}
\


Measurement of the Higgs self-coupling allows one to probe the shape
of the Higgs potential.  In the Standard Model, the Higgs potential
can be written as (here $\langle \Phi \rangle = (0, v/\sqrt{2})^T$)
\begin{equation}
  V = - \mu^2 \Phi^{\dagger} \Phi + \lambda (\Phi^{\dagger} \Phi)^2,
\end{equation}
yielding a Higgs vev and mass of
\begin{equation}
  v = \sqrt{\mu^2/\lambda} \simeq 246~{\rm GeV}, \qquad 
  m_H = \sqrt{2 \lambda} \, v \simeq 125~{\rm GeV}.
\end{equation}
The Higgs self-interaction Lagrangian, expanded about the minimum, is
\begin{equation}
  \Delta \mathcal{L} = - \frac{1}{2} m_H^2 H^2 - \frac{g_{HHH}}{3!} H^3
  - \frac{g_{HHHH}}{4!} H^4,
\end{equation}
where the triple- and quartic-Higgs couplings are predicted in the SM in terms
of the known Higgs mass and vev,
\begin{equation}
  g_{HHH} = 6 \lambda v = \frac{3 m_H^2}{v}, \qquad
  g_{HHHH} = 6 \lambda = \frac{3 m_H^2}{v^2}.
\end{equation}
Tests of these relations probe for non-SM physics in the Higgs potential.

The triple-Higgs coupling can be probed in double Higgs production:
$gg \to HH$ at hadron colliders or $e^+e^- \to ZHH$, $\nu \bar \nu HH$
at lepton colliders.  The main challenge is the small signal cross
section.  The quartic-Higgs coupling could be probed in principle
through triple Higgs production, though the cross sections are too small
to be detectable at any foreseen future facility.

Henceforth we denote the uncertainty in the triple-Higgs coupling as
$\Delta \lambda / \lambda \equiv \Delta g_{HHH}/g_{HHH}$.

\subsection{Standard Model predictions for double-Higgs production}\vspace*{-0.4cm}

The theoretical status of double Higgs production in $pp$ collisions
has been recently summarized in Ref.~\cite{Baglio:2012np}
(Table~\ref{tab:HHxsec}).  The most interesting process, $gg \to HH$,
is currently known to next-to-leading order in QCD with a theoretical
uncertainty $\sim$30\%.  This uncertainty will need to be reduced to match
the anticipated experimental uncertainty at the HL-LHC and higher energy
$pp$ colliders.

\begin{table}
\caption{\label{tab:HHxsec} Cross sections for double Higgs production in 
$pp$ collisions, including the current estimate for the theoretical uncertainty
from Ref.~\cite{Baglio:2012np}, for $m_H = 125$~GeV.  The uncertainty on 
the $t \bar t HH$ process has not been evaluated. 
}
\begin{center}
\hspace*{-1.5cm}
\begin{tabular}{c|ccccc}
\hline\hline
 & \multicolumn{5}{c}{Cross sections (fb) and theoretical uncertainties (\%)} \\ \cline{2-6}
$\sqrt{s}$  & $gg \to HH$ & $qq \to qqHH$ & $q \bar q \to WHH$ & $q \bar q \to ZHH$  & $q \bar q/gg \to t\bar tHH$  \\
(TeV)          &   NLO        & NLO   &  NNLO  & NNLO  & LO \\
\hline
14 & $33.89^{+37.2\%}_{-29.8\%}$ & $2.01^{+7.6\%}_{-5.1\%}$ & $0.57^{+3.7\%}_{-3.3\%}$ & $0.42^{+7.0\%}_{-5.5\%}$ & $1.02$ \\
33 & $207.29^{+33.0\%}_{-26.7\%}$ & $12.05^{+6.1\%}_{-4.2\%}$ & $1.99^{+3.5\%}_{-3.1\%}$ & $1.68^{+7.9\%}_{-6.7\%}$ & $7.91$ \\
100 & $1417.83^{+29.7\%}_{-24.7\%}$ & $79.55^{+6.2\%}_{-4.1\%}$ & $8.00^{+4.2\%}_{-3.7\%}$ & $8.27^{+8.4\%}_{-8.0\%}$ & $77.82$ \\
\hline\hline
\end{tabular}
\end{center}
\end{table}

All double Higgs production processes involve not only the diagram
with the trilinear Higgs coupling $\lambda$, but also additional
diagrams that dilute the sensitivity of the cross section measurement
to $\lambda$.  This dependence has been quantified for $pp$ colliders
in Ref.~\cite{Baglio:2012np}.  Because of the different kinematic
dependences of the contributing diagrams, the two-Higgs invariant mass
$M_{HH}$ and the Higgs $p_T$ distributions depend on $\lambda$.  This
has not yet been taken into account in LHC analyses, although an
$M_{HH}$ weighting has been used in ILC studies to increase the
sensitivity to $\lambda$.

In $e^+e^-$ collisions, the full $\mathcal{O}(\alpha)$ electroweak
corrections to both the double Higgs-strahlung process $e^+e^- \to
ZHH$~\cite{Belanger:2003ya,Zhang:2003jy} and the $WW$
fusion-dominated process $e^+e^- \to \nu \bar \nu
HH$~\cite{Boudjema:2005rk} are known.  The theoretical uncertainties
in these cross sections are well below the anticipated
experimental precision.

\subsection{Models that modify the triple-Higgs coupling}\vspace*{-0.4cm}

Beyond the Standard Model, the triple-Higgs coupling is in general
modified.  The size of the modification is highly model-dependent,
potentially providing model-discriminating power.  Estimates of the 
self-coupling deviation in a variety of models were recently made
in Ref.~\cite{Gupta:2013zza}, under the constraint that no other new
physics associated with the model would be discovered by the LHC:
\begin{itemize}

\item Mixed-in singlets.  Assuming that the mixing angle and heavy
  Higgs mass are such that the heavy Higgs is not detectable at the
  LHC, $(\Delta \lambda/\lambda)^{\rm max} \simeq -18\%$.

\item Higher-dimension operators.  These can come from composite Higgs
  models or be introduced to strengthen the electroweak phase
  transition to help with baryogenesis.  Imposing precision
  electroweak constraints yields $(\Delta \lambda/\lambda)^{\rm max}
  \simeq \pm 20\%$.

\item MSSM.  The presence of the second doublet leads to mixing
  effects.  Inclusion of top quark/squark radiative corrections is
  important.  The largest deviations occur for low $\tan\beta$ and low
  $M_A$.  For $\tan\beta \sim 5$ and $M_A \sim 200$~GeV and top
  squarks in the range 1--2.5~TeV, the maximum deviation is 
  $(\Delta \lambda/\lambda)^{\rm max} \simeq -15\%$, but this number
  depends strongly on the other MSSM parameters and can be as low as
  $-2\%$.  For higher $M_A$ the coupling deviation becomes smaller in
  accordance with decoupling.

\item NMSSM.  The additional coupling parameter $\lambda_S$ from the
  singlet affects the scalar potential even when the singlet is
  decoupled.  Deviations as large as $(\Delta \lambda/\lambda)^{\rm max} 
  \simeq -25\%$ are possible for $\tan\beta
  \sim 7.5$, $M_A \sim 500$~GeV (outside the LHC reach) and top squark
  mass parameter $M_S \sim 500$~GeV, assuming that $\lambda_S$ remains
  perturbative up to at least 10~TeV.  Heavier stops lead to a smaller
  $\lambda_S$ and the deviation becomes more similar to the MSSM.

\end{itemize}

In other models, large deviations of the triple Higgs coupling from
the SM prediction can be used as characteristic signatures of the
model.  For example, a recent proposal~\cite{Galloway:2013dma} to
improve the naturalness of SUSY models by boosting the Higgs mass
using ``auxiliary'' scalar fields with tadpoles predicts a triple
Higgs coupling much smaller than in the SM, as a consequence of the
Higgs mass being generated mostly by its couplings to the auxiliary
scalars.
A separate study~\cite{Kanemura:2004ch} of electroweak baryogenesis in a
two-Higgs-doublet model or the MSSM found that successful
baryogenesis resulted in deviations of the triple Higgs coupling of at
least 10\% or 6\%, respectively.

We point out that exclusion of a coupling deviation of 20\% at 95\% CL
requires a measurement at the 10\% level; discovery of such a
deviation at 5$\sigma$ requires a measurement at the 4\% level.  This
is a seriously challenging target for both future LHC upgrades and proposed
$e^+e^-$ colliders.

\subsection{Other ways to modify the double Higgs production rate}\vspace*{-0.4cm}

The double Higgs production cross section can also be modified by new
physics separate from the triple Higgs coupling.  

At the LHC, the triple Higgs coupling is measured using the $gg \to
HH$ rate, which proceeds mainly through top quark triangle and box
diagrams.  Modification of the top quark Yukawa coupling will change
the double Higgs production rate~\cite{Goertz:2013kp}.  
The double Higgs production rate at the LHC can also be modified by 
new colored particles in the loop.  A color-octet scalar below 250~GeV 
can lead to a factor-2 enhancement of the double-Higgs production rate
even for $gg \to H$ within 25\% of its SM value~\cite{Kribs:2012kz}.
On the other hand, vectorlike singlet or mirror quarks cause only small
departures from the SM $gg \to HH$ rate once precision electroweak
and $gg \to H$ constraints are imposed~\cite{Dawson:2012mk}.  In all
cases, the kinematic distributions of the two final-state Higgs bosons
are modified with respect to the SM.
Finally, in models with a second, heavier CP-even Higgs boson (e.g., the
two-Higgs-doublet models), resonant production of the heavier Higgs
$gg \to H_2$ followed by the decay $H_2 \to HH$ can lead to large enhancements
of the double Higgs production rate at the LHC, which can be diagnosed
through the resonance peak in the $HH$ invariant mass distribution.
Note that extraction of
the double Higgs production cross section also relies on accurate
knowledge of the Higgs decay branching ratios involved for each final
state.  

At an $e^+e^-$ collider, the triple Higgs coupling is measured using
the rates for $e^+e^- \to ZHH$ (dominant at 500~GeV center-of-mass
energy) and $e^+e^- \to \nu \bar \nu HH$ (dominant at 1000~GeV or higher).
These processes are also sensitive to modifications of the $VVHH$
coupling ($V = W,Z$) caused by mixing among scalars in different
representations of SU(2).  Such effects can potentially be separated
from the triple Higgs coupling by combining measurements at different
collider energies~\cite{Killick:2013mya}.  
Double Higgs production at an $e^+e^-$ collider is also susceptible to
resonant contributions from heavier neutral Higgs states, such as $e^+
e^- \to ZH^0$ with $H^0 \to h^0h^0$ and $e^+e^- \to h^0A^0$ with $A^0 \to
Zh^0$ in two Higgs doublet models (where $h^0$ is identified with the 
discovered Higgs boson), but these processes are suppressed by
$\cos^2(\beta-\alpha) \to 0$ in the decoupling limit and will also
be constrained by the measurement of the SM-like Higgs couplings to 
$W$ and $Z$ bosons, $\kappa_V = \sin(\beta-\alpha)$.  Such contributions
can be distinguished due to their resonant kinematic structure.



\vspace*{-0.4cm}
\label{chap:doublehiggs}

\subsection{Higgs boson self-coupling at the LHC}\vspace*{-0.4cm}

The self-coupling of the Higgs boson is the consequence of the electroweak symmetry breaking of the Higgs
potential. Measurement of the triple-Higgs coupling will therefore allow for a direct probe of the potential.
This can be done through the analysis of pair production of the Higgs boson $pp\to HH+X$. At 14~TeV, the
cross section is predicted to be 34~fb in the Standard Model. 
Statistics will be limited for final states
with reasonable signal-background ratios. Combination of several final states will likely be required to
achieve meaningful results.


By far $HH\to bb\ bb$ has the highest rate. Without the signature of leptons or photons, this final state
is buried under the overwhelming QCD background. Similarly $bb\ WW$ has the second highest rate, but will
likely be shadowed by the $t\bar{t}\to bW\ bW$ production. Though only $\sim 270$ events are expected before
selection in 3000~fb$^{-1}$, the $HH\to bb\gamma\gamma$ final state is relatively clean and will likely
be the most sensitive final state for the self-coupling studies at the LHC. There have been a number of 
studies~\cite{Baglio:2012np,Goertz:2013kp,Dolan:2012rv} on $bb\gamma\gamma$, $bb\tau\tau$ and $bb WW$ 
final states. The conclusions of these studies vary widely, ranging from over a $5\sigma$ 
observation~\cite{Baglio:2012np}
of the Higgs pair production to a $\sim 30\%$ precision~\cite{Goertz:2013kp} on the Higgs self-coupling
parameter $\lambda$ with 3000 fb$^{-1}$. A recent study with a generic LHC detector simulation shows that
a $\sim 50\%$ precision on $\lambda$ can be achieved from the $HH\to bb\gamma\gamma$ channel 
alone~\cite{YaoWhitepaper}. More studies are needed to firm up these estimates. For this report, 
a per-experiment precision of $\sim 50\%$ on $\lambda$ is taken as the benchmark for HL-LHC. 
Combining the two experiments, a precision of 30\% or better can be achieved.

Note that this extraction of the Higgs self-coupling assumes that the 
effective $ggH$ coupling and the Higgs branching ratios to the final 
states used in the analysis are equal to their SM values.

\subsection{Higher-energy hadron colliders}\vspace*{-0.4cm}
The cross section for $gg \to HH$ increases with increasing hadron
collider energy due to the increase in the gluon partonic luminosity.
Even though backgrounds increase with energy at a similar rate, a
higher-energy $pp$ collider such as the HE-LHC (33~TeV) or VLHC (100~TeV)
would improve this measurement.

Results of a fast-simulation study of double Higgs production in the
$bb\gamma\gamma$ final state for $pp$ collisions at 14, 33, and
100~TeV~\cite{YaoWhitepaper} are shown in Table~\ref{tab:hhvlhc}
(14~TeV results are consistent with the European strategy study).
$bb\gamma\gamma$ is the most important channel at 14~TeV because of
large top-pair backgrounds to the $bb\tau\tau$ and $bbWW$ channels.
The simulation used Delphes with ATLAS responses~\cite{Anderson:2013kxz} 
and assumes one
detector.  The resulting uncertainty on $\Delta \lambda/\lambda$ is
extracted using the scaling of the double-Higgs cross section with
$\lambda$~\cite{Baglio:2012np}.

\begin{table}
\caption{Signal significance for $pp \to HH \to bb \gamma\gamma$ and 
percentage uncertainty on the Higgs self-coupling at
future hadron colliders, from~\cite{YaoWhitepaper}.}
\begin{center}
\begin{tabular}{cccc}
\hline\hline
& HL-LHC & HE-LHC & VLHC \\
\hline
$\sqrt{s}$ (TeV) & 14 & 33 & 100 \\
$\int \mathcal{L} dt$ (fb$^{-1}$) & 3000 & 3000 & 3000 \\
\hline
$\sigma \cdot {\rm BR} (pp \to HH \to bb\gamma\gamma)$ (fb) & 0.089 & 0.545 & 3.73 \\
$S/\sqrt{B}$ & 2.3 & 6.2 & 15.0 \\
$\lambda$ (stat) & 50\% & 20\% & 8\% \\
\hline\hline
\end{tabular}
\end{center}
\label{tab:hhvlhc}
\end{table}

\subsection{Higgs boson self-coupling at $\mathbf{e^+e^-}$ Linear Colliders}\vspace*{-0.4cm}

At an $e^+e^-$ linear collider, the Higgs trilinear self-coupling can
be measured via the $e^+e^- \rightarrow ZHH$ and $e^+e^- \rightarrow
\nu_e \bar{\nu}_e HH$ processes.  The cross section for the former
peaks at approximately 0.18~fb close to $\sqrt{s} = 500$~GeV;
however, for this channel there are many diagrams leading to the $Zhh$
final state that don't involve the Higgs boson self-coupling resulting
in a dilution of $\Delta \lambda / \lambda \simeq 1.8 \times (\Delta
\sigma_{ZHH}/\sigma_{ZHH})$.  This situation improves for the
$W$-fusion process $\nu_e \bar{\nu}_e HH$ where $\Delta \lambda /
\lambda \simeq 0.85 \times (\Delta
\sigma_{\nu\bar{\nu}HH}/\sigma_{\nu\bar{\nu}HH})$ at 1~TeV, but
requires $\sqrt{s} \geq 1.0$~TeV for useful rates.  Polarized beams
can significantly increase the signal event rate, particularly for the
$W$-fusion process.  None of the proposed $e^+e^-$ circular machines
provide high enough collision energies for sufficient rates.

The most recent full simulation study~\cite{ILCWhite:2013a,Tian:2013a}
of these two production processes including all $Z$ decay modes as
well as $HH \to bbbb$ and $HH \to bbWW^*$ final states has been
carried out using the ILD detector at the ILC where event weighting
depending on $M_{HH}$ is used to enhance the contribution of the
self-coupling diagram and improve on the dilutions above.  Results are
given in in Table~\ref{tab:expt}.

The cross section for $\nu_e \bar{\nu}_e HH$ continues to grow with
$\sqrt{s}$, and full simulation studies~\cite{CLICWhite:2013a} for
CLIC show increased sensitivity at higher collision energies of
$\sqrt{s} = 1.4$~TeV and $\sqrt{s} = 3.0$~TeV as shown in
Table~\ref{tab:expt}. 

\begin{table}[htb!]
\caption{Estimated experimental percentage uncertainties on the double
  Higgs production cross sections and Higgs self-coupling parameter
  $\lambda$ from $e^+e^-$ linear colliders.  The expected precision on
  $\lambda$ assumes that the contributions to the production cross
  section from other diagrams take their Standard Model values.  ILC
  numbers include $bbbb$ and $bbWW^*$ final states and assume
  $(e^-,e^+)$ polarizations of $(-0.8,0.3)$ at 500~GeV and
  $(-0.8,0.2)$ at 1000~GeV.  ILC500-up is the luminosity upgrade at
  500~GeV, not including any 1000~GeV running.  ILC1000-up is the
  luminosity upgrade including running at both 500 and 1000~GeV.  CLIC
  numbers include only the $bbbb$ final state.  The two numbers for
  each CLIC energy are without/with 80\% electron beam polarization.
$^\ddag$ILC luminosity
   upgrade assumes an extended running period on top of the low luminosity program
   and cannot be directly compared to CLIC numbers without accounting for
   the additional running period.
}
\small
\begin{center}
\begin{tabular}{c|cccccc}
\hline\hline
 & ILC500  & ILC500-up  & ILC1000  & ILC1000-up & CLIC1400 & CLIC3000 \\
\hline
$\sqrt{s}$ (GeV)         & 500  & 500 & 500/1000 & 500/1000  & 1400  & 3000 \\
$\int \mathcal{L} dt$ (fb$^{-1}$) & 500  & 1600$^\ddag$ & 500+1000  & 1600+2500$^\ddag$  & 1500  & +2000 \\
$P(e^-,e^+)$           & $(-0.8,0.3)$ & $(-0.8,0.3)$ & $(-0.8,0.3/0.2)$ & $(-0.8,0.3/0.2)$ & $(0,0)/(-0.8,0)$ & $(0,0)/(-0.8,0)$ \\
\hline
 $\sigma \thinspace (ZHH)$             & 42.7\% &  & 42.7\%  & 23.7\%   & --     & --  \\
 $\sigma \thinspace (\nu \bar{\nu}HH)$ & -- & --  & 26.3\% & 16.7\% &       &   \\ 
 $\lambda$                         & 83\% & 46\% & 21\% & 13\% & 28/21\% & 16/10\% \\
\hline\hline
\end{tabular}
\end{center}
\label{tab:expt}
\end{table}

\subsection{Photon collider}\vspace*{-0.4cm}
Higgs pairs can be produced at a photon collider via off-shell $s$-channel
Higgs production, $\gamma\gamma \to H^* \to HH$.  The process was studied
in Ref.~\cite{Kawada:2012uy} for an ILC-based photon collider running for
5 years, leading to 80 raw $\gamma\gamma \to HH$ events.  Jet clustering 
presents a major challenge for signal survival leading to a sensitivity
of only about $1 \sigma$.

\subsection{Muon collider}\vspace*{-0.4cm}
Double Higgs production at a muon collider can proceed via $s$-channel
off-shell Higgs production, $\mu^+ \mu^- \to H^* \to HH$.  However,
the cross section for this non-resonant process is very small, of
order 1.5~ab at the optimum energy of $\sim 275$~GeV, providing less
than one signal event in 500~fb$^{-1}$ before branching ratios and
selection efficiencies are folded in.

\subsection{Summary}
Expected precisions on the triple Higgs coupling measurement, assuming
that all other Higgs couplings are SM-like and that no other new physics 
contributes to double-Higgs production, are summarized in 
Table~\ref{tab:selfcoup}. 

\begin{table}
\caption{Expected per-experiment precision on the triple-Higgs boson 
coupling.  ILC numbers include $bbbb$ and $bbWW^*$ final states and
assume $(e^-,e^+)$ polarizations of $(-0.8,0.3)$ at 500~GeV and
$(-0.8,0.2)$ at 1000~GeV.  ILC500-up is the luminosity upgrade at
500~GeV, not including any 1000~GeV running.  ILC1000-up is the
luminosity upgrade with a total of 1600~fb$^{-1}$ at 500~GeV and
2500~fb$^{-1}$ at 1000~GeV.  CLIC numbers include only the $bbbb$
final state and assume 80\% electron beam polarization.  HE-LHC and
VLHC numbers are from fast simulation~\cite{YaoWhitepaper} and include
only the $bb\gamma\gamma$ final state.
$^\ddag$ILC luminosity
   upgrade assumes an extended running period on top of the low luminosity program
   and cannot be directly compared to CLIC numbers without accounting for
   the additional running period.
}
\scriptsize
\begin{center}
\begin{tabular}{c|ccccccccc}
\hline \hline
 & HL-LHC & ILC500 & ILC500-up & ILC1000 & ILC1000-up & CLIC1400 & CLIC3000 & HE-LHC & VLHC \\
\hline
$\sqrt{s}$ (GeV) & 14000 & 500 & 500 & 500/1000 & 500/1000 & 1400 & 3000 & 33,000 & 100,000 \\ 
$\int \mathcal{L}dt $ (fb$^{-1}$) & 3000/expt & 500 & 1600$^\ddag$ & 500+1000 & 1600+2500$^\ddag$ & 1500 & +2000 & 3000 & 3000 \\
\hline
$\lambda$ & 50\% & 83\% & 46\% & 21\% & 13\% & 21\% & 10\% & 20\% & 8\% \\
\hline \hline
\end{tabular}
\end{center}
\label{tab:selfcoup}
\end{table}

These same numbers are used to estimate
precisions possible from a  combination of facilities as shown in Table~\ref{tab:selfcomb}.
As can be seen, the precision is usually dominated by the precision achieved by one of the
collider options in the combination.

\begin{table}
\caption{Expected precision on the triple-Higgs boson
coupling for combined facilties, assuming the final states,
polarizations, and integrated luminosities assumed above in 
Table~\ref{tab:selfcoup}.
Here ``ILC-up" refers to ILC1000-up, 
and ``CLIC" refers to CLIC3000 with the two
numbers shown assuming unpolarized beams or
80\% electron beam polarization, respectively.
TLEP is in parantheses since it would not contribute
to the measurement of the self-coupling, but could be
a step along the way to the higher-energy hadron colliders.}
\begin{center}
\begin{tabular}{ccccccccc}
\hline \hline
 LHC & \multicolumn{8}{|c}{HL-LHC} \\
 \cline{2-9}
 +ILC & \multicolumn{1}{|c}{+ILC-up} & \multicolumn{3}{|c}{+(TLEP)} & \multicolumn{2}{|c}{+ILC-up} & \multicolumn{2}{|c}{+CLIC} \\
 \cline{3-9}
      & \multicolumn{1}{|c}{}        & \multicolumn{1}{|c}{+CLIC}      & +HE-LHC & +VLHC & \multicolumn{1}{|c}{+HE-LHC}      & +VLHC          & \multicolumn{1}{|c}{+HE-LHC}     & +VLHC         \\
\hline
21\%  & 12.6\%  & 15.2/9.8\% & 18.6\%  & 7.9\% & 10.9\%      & 6.8\%         & 12.5/8.9\% & 7.2/6.2\%    \\
 \hline \hline
 \end{tabular}
 \end{center}
 \label{tab:selfcomb}
 \end{table}




\section{Study of $C\!P$-mixture and spin}\vspace*{-0.3cm}
\label{sec:cp}

The discovery of the new boson with the mass around 126 GeV at the LHC~\cite{atlas:2012obs,Chatrchyan:2012obs}
opens a way for experimental studies of its properties such as spin, parity, and couplings 
to the Standard Model particles.  We split such studies into two groups
\begin{itemize} 
\item  tests of discrete spin/parity hypotheses of the new particle(s);
\item  identification and measurement of  various types of tensor couplings for a given spin assignment,
and the search for $C\!P$ violation is among the primary goals of this study.
\end{itemize}
There is a potential connection between the baryogenesis and $C\!P$ violation in the Higgs sector
and the measurements in the Higgs sector directly may be complementary to the measurements
in the EDMs~\cite{Shu:2013uua}.
The interesting level of  $C\!P$-odd state admixture angle $\theta$ is $|\sin\theta|<0.1$.

We note that several facts about the Higgs-like boson spin, parity, and its couplings have already been 
established at both the Tevatron and LHC. Indeed, we know that 
\begin{itemize} 
\item the new boson should have integer spin since it  decays to two integer-spin 
particles~\cite{atlas:2012obs,Chatrchyan:2012obs};
\item the new boson cannot have spin one because it decays to two on-shell photons \cite{Landau,Yang};
\item the spin-one assignment is also strongly disfavored by the measurement of angular distributions 
          in the decay to two $Z$ bosons~\cite{Chatrchyan:2012jja,Chatrchyan:2013mxa,Chatrchyan:2013iaa,CMS:yxa,CMS:xwa,CMS:bxa,Aad:2013xqa};
\item under the assumption of minimal coupling to vector bosons or fermions, 
the new boson is unlikely to be a spin-two particle~\cite{Chatrchyan:2012jja,Chatrchyan:2013mxa,Chatrchyan:2013iaa,CMS:yxa,CMS:xwa,CMS:bxa,Aad:2013xqa,D0spin2prel};
\item the spin-zero, negative  
  parity hypothesis is strongly disfavored~\cite{Chatrchyan:2012jja,Chatrchyan:2013mxa,Chatrchyan:2013iaa,CMS:yxa,CMS:xwa,CMS:bxa,Aad:2013xqa,D0spin0mprel};
\end{itemize}

The general amplitudes that describe the interaction of the spin-zero, spin-one, and spin-two boson can be 
found in the literature~\cite{Accomando:2006ga,Gao:2010qx}. 
In particular, the minimal coupling gravity-like coupling
of spin-2 boson to gauge boson is chosen as a benchmark spin model in the study.
For $C\!P$-mixing studies of parity and couplings of the spin-zero Higgs-like boson may 
employ either effective Lagrangians or generic parameterizations of scattering amplitudes.
For the coupling to the gauge bosons, such as $ZZ$, $WW$, $Z\gamma$, $\gamma \gamma$, or $gg$,
the scattering amplitude can be written as 
\begin{equation}
A(X_{J=0} \to VV) = v^{-1} \left ( 
  a^{}_{ 1} m_{V}^2 \epsilon_1^* \epsilon_2^* 
+ a^{}_{ 2} f_{\mu \nu}^{*(1)}f^{*(2),\mu \nu}
+ a^{}_{ 3}  f^{*(1)}_{\mu \nu} {\tilde f}^{*(2),\mu  \nu}
\right )\,.
\label{eq:fullampl-spin0} 
\end{equation}
The SM Higgs coupling at tree level (to $ZZ$ and $WW$) is described by the $a_{ 1}$ term,
while the $a_{ 2}$ term appears in the loop-induced processes, such as 
$Z\gamma$, $\gamma \gamma$, or $gg$. The $a_{3}$ term corresponds to the pseudoscalar.
Equation~(\ref{eq:fullampl-spin0}) presents the lowest orders in $q^2$-dependence 
of the three unique Lorentz structures and we assume $a_i$ to be constant and real up 
to a scale $\sim$1 TeV in the $q^2$-dependence.
The general scattering amplitude that describes the interaction of the Higgs-like boson 
with the fermions, such  as $\tau^+\tau^-$, $\mu^+\mu^-$,  $b\bar{b}$, and $t\bar{t}$,
can be written as
%
\begin{eqnarray}
&& A(X_{J=0} \to f\bar f) = \frac{m_f}{v}
\bar u_2 \left ( b_{ 1} + i b_{ 2} \gamma_5 \right ) u_1\,,
\label{eq:ampl-spin0-qq}
\end{eqnarray}
%
The two constants  $b_{ 1}$ and $b_{ 2}$ correspond to the scalar and pseudoscalar couplings.
It is important to note that each set of constants, such as $a_1$,  $a_2$,  $a_3$, is generally independent
between different coupling types ($ZZ$, $\gamma\gamma$, etc) and does not correspond directly to the mixture 
of the original state (relative strength of those could be rather different from the actual mixture). 

$C\!P$ violation in the Higgs sector could be revealed if both $C\!P$-odd and $C\!P$-even contributions
are detected. It has already been established that the $C\!P$-even contribution dominates at least in the
$HZZ$ coupling~\cite{Chatrchyan:2012jja,CMS:yxa,CMS:xwa,CMS:bxa,Aad:2013xqa}.
Therefore, measuring or setting the limit on the $C\!P$-odd contribution is a target of the study. 
We represent the couplings by fractions of the corresponding cross-sections 
(e.g., $f_{a2}$ and $f_{a3}$ for vector boson couplings).
In particular, the fraction of $C\!P$-odd contribution is defined as
($f_{a3}$ in the case of boson couplings) 
%
\begin{eqnarray}
&& f_{C\!P} =  \frac{|a_{3}|^2\sigma_3}{\sum |a^{}_{i}|^2 \sigma_i} \,.
\label{eq:fractions}
\end{eqnarray}
%
We note that $\sigma_i$ is the effective cross-section of the Higgs boson
decay process corresponding to $a^{}_{ i}=1, a_{j \ne i}=0$.
For example, for the $H\to ZZ$ decay, $\sigma_3/\sigma_1\simeq 0.160$.

\begin{figure}[htb]
\begin{center}
\begin{tabular}{@{}cc@{}}
\includegraphics[width=0.48\textwidth]{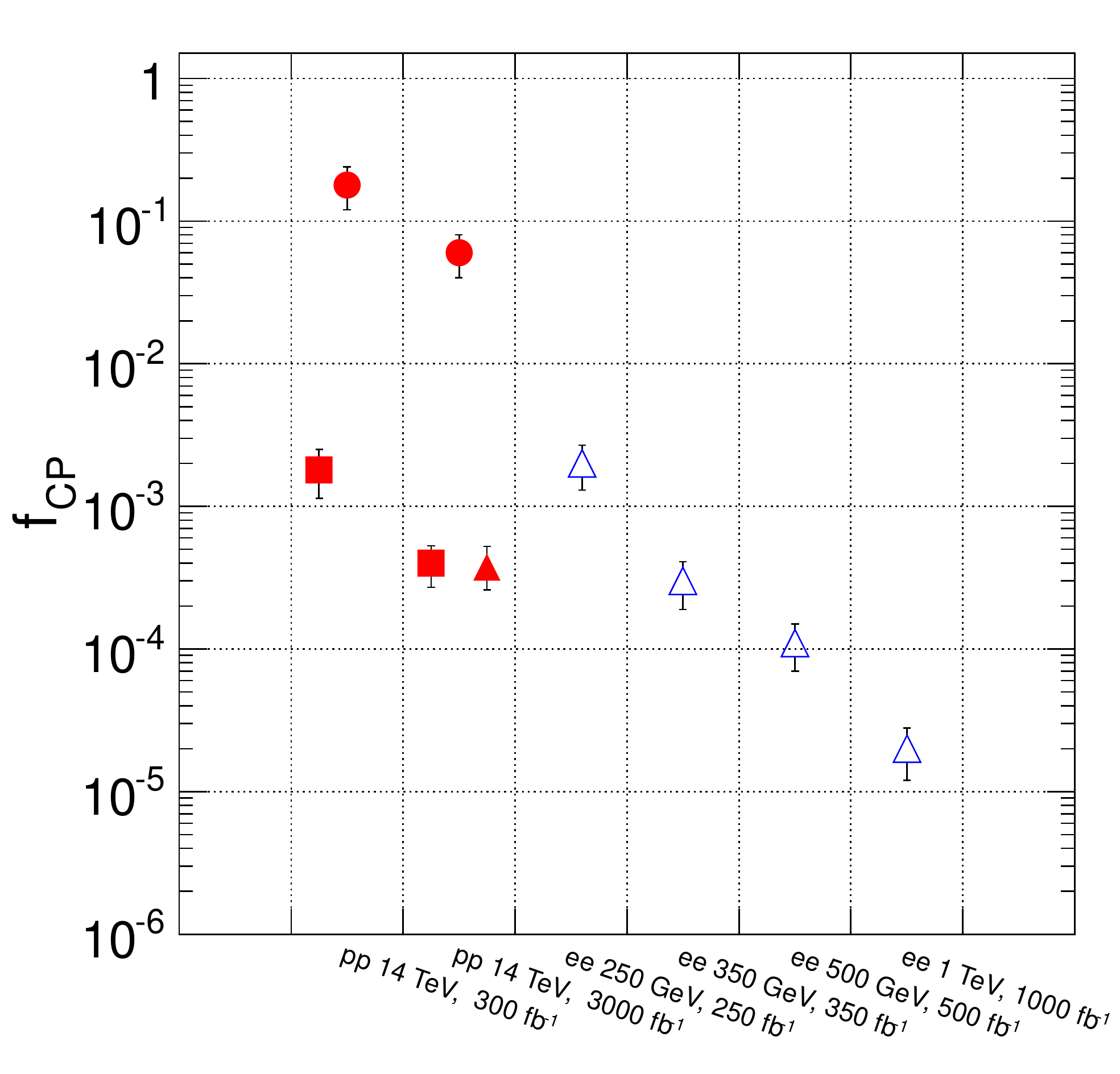} \\
\end{tabular}
\caption{
Summary of precision in $f_{C\!P}$  for $HVV$ couplings ($V=Z,W$) at the moment of $3\sigma$ measurement~\cite{Anderson:2013fba}.
Points indicate central values and error bars indicate $1\sigma$ deviations in the generated
experiments modeling different luminosity scenarios at proton (solid red) or $e^+e^-$ (open blue) colliders.
Measurements in three topologies $VH$ (triangles), VBF (squares), and decay $H\to VV$ (circles) are shown.
Different energy and luminosity scenarios are indicated on the $x$-axis.
}
\end{center}
\label{fig-cpscenarios}
\end{figure}

In Table~\ref{table-cpscenarios}
we summarize expected precision of spin and ${C\!P}$-mixture measurements
at different facilities and running conditions. 
Expectations in the $VV\!H$ couplings are illustrated in Fig.~\ref{fig-cpscenarios}.
For various effective couplings, precision is quoted on $C\!P$-odd cross-section fraction, 
such as $f_{C\!P}$ defined above. 
For the measurement precision we estimate that 10\% admixture~\cite{Shu:2013uua}
of pseudoscalar in a Higgs state is a reasonable target. 
The scalar Higgs couplings to massive vector bosons ($ZZ\!H$ and $WW\!H$) are at tree level, 
while pseudoscalar coupling is expected to be suppressed by a loop. Therefore, the 10\% admixture
of a pseudoscalar in a Higgs state would translate to a significantly suppressed 
${C\!P}$-odd contribution, with $f_{C\!P}$ smaller than $10^{-5}$ in the $H\to ZZ$ and $WW$ decays.
On the other hand, in the fermion couplings and vector boson couplings suppressed by a loop for both
scalar and pseudoscalar ($ggH$, $\gamma\gamma  H$, $Z\gamma  H$), both couplings could
be of comparable size, and the target precision on $f_{C\!P}$  is $10^{-2}$ or better. 

With the current luminosity of about 25 fb$^{-1}$ at 7 and 8 TeV, both ATLAS and
CMS experiments expect more than 2$\sigma$ separation between the minimal spin-2 model 
and SM Higgs boson~\cite{Chatrchyan:2012jja,CMS:yxa,CMS:xwa,CMS:bxa,Aad:2013xqa}.
This translates to close to 10$\sigma$ separation at high luminosity. 

The LHC expectation in ${C\!P}$ studies comes from dedicated analysis of the $H\to ZZ^*$ 
decay~\cite{CMS:2013xfa, Chatrchyan:2012jja,CMS:yxa,CMS:xwa,Aad:2013xqa,ATLAS-PHYS-PUB-2013-013} 
by CMS and ATLAS collaborations, 
as well as individual studies~\cite{Seattle-Whitbeck, Minneapolis-Gao,Anderson:2013fba}.
The CMS experiment quotes 0.40 expected error on $f_{C\!P}$ with present statistics~\cite{CMS:yxa,CMS:xwa},
which translates to $\pm 0.07$ and $\pm 0.02$ at 300 fb$^{-1}$ and 3000 fb$^{-1}$, respectively~\cite{CMS:2013xfa},
and agrees with ATLAS projections~\cite{ATLAS-PHYS-PUB-2013-013}.
These results scale well with luminosity and cross-section and match those reported in dedicated studies.

VBF production at LHC~\cite{Accomando:2006ga} offers a complementary way to measure ${C\!P}$ mixture 
in the $VVH$ coupling. Using kinematic correlations of jets in the VBF topology with the full matrix element technique,
a fraction of about 0.05 (015) of ${C\!P}$-odd cross-section contribution can be measured
at  3000 (300) fb$^{-1}$ on LHC~\cite{Minneapolis-Gao,Anderson:2013fba}.
Given different relative cross sections of the VBF production of the scalar and pseudoscalar components,
these translate to the equivalent value of $f_{C\!P}$ defined for decay in Eq.~(\ref{eq:fractions}) of 0.0005 (0.003). 
The issue of increasing pileup was not addressed in detail in this study. However, reduced precision with 
increased thresholds for jets checked in this study
would be easily compensated by considering additional final states of the Higgs
boson, since this study depends only a particular production mechanism and not the final states. 
Measurements in the $Hgg$ coupling in gluon fusion production in association with two jets
at the LHC~\cite{Anderson:2013fba} are also presented in Table~\ref{table-cpscenarios}.

The spin study at $e^+e^-$ is based on TESLA TDR studies~\cite{AguilarSaavedra:2001rg}.
A threshold scan with a luminosity of 20 fb$^{-1}$ at three centre-of-mass energies 
(215, 222, and 240 GeV for $m_H=120$ GeV)
is sufficient to distinguish the spin-1 and spin-2 hypotheses at 4$\sigma$ level.
This study has been recently updated~\cite{Seattle-Kruger} to include the Higgs boson mass 
and luminosity and energy scenarios. The typical probability for most exotic scenarios is smaller
than $10^{-6}$. This study is based on assumption of 250 fb$^{-1}$ at 250 GeV and 
 20 fb$^{-1}$ at each of three energy points below. 

\begin{table}[ht]
\caption{
List of expected precision of spin and ${C\!P}$-mixture measurements.
Spin significance is quoted for one representative model of minimal coupling KK graviton $J^P=2^+_m$.
For various effective couplings, precision is quoted on $C\!P$-odd cross-section fraction, 
such as $f_{a3}$ defined for $H\to ZZ^*$. 
Target precision is estimated to be $<10^{-5}$ for the modes with pseudoscalar coupling 
expected to be suppressed by a loop ($ZZ\!H$ and $WW\!H$), while it is estimated to be
$<10^{-2}$ for fermion couplings and vector boson couplings suppressed by a loop for both
scalar and pseudoscalar ($ggH$, $\gamma\gamma  H$, $Z\gamma  H$).
Numerical values are given where reliable estimates are provided, $\checked$ mark indicates
that some studies are done and measurement is in principle possible or
feasibility of such a measurement could be considered.
}
\begin{center}
\begin{tabular}{|l|cccccccc|c|}
\hline\hline
Collider                     &   $pp$       &   $pp$    &   $e^+e^-$    &   $e^+e^-$    &  $e^+e^-$    &  $e^+e^-$    & $\gamma\gamma$ &  $\mu^+\mu^-$  & target \\
E (GeV)                     &   14,000   &   14,000     &  250          &  350                 & 500             &   1,000       &            126               &    126                                  &   (theory) \\
${\cal L}$ (fb$^{-1}$) & 300  & 3,000 & 250          &  350                  & 500              &    1,000       &          250                  &                                     &     \\ 
%
\hline
\hline
 {spin-2$^+_m$}  & $\sim$10$\sigma$ & $\gg$10$\sigma$ & $>$10$\sigma$ & $>$10$\sigma$ & $>$10$\sigma$ &$>$10$\sigma$ &  &  & $>$5$\sigma$  \\
\hline
\hline
 $VV\!H^\dagger$ & $0.07$  & $0.02$  &  \checked &  \checked  &  \checked &  \checked &  \checked & \checked & $<10^{-5}$  \\
 $VV\!H^\ddagger$ &  $4\!\cdot\!10^{-4}$ &  $1.2\!\cdot\!10^{-4}$& $7\!\cdot\!10^{-4}$ & $1.1\! \cdot\!10^{-4}$ & $4\!\cdot \!10^{-5}$ & $8\!\cdot\!10^{-6}$  & --  &  -- & $<10^{-5}$  \\
$VV\!H^\Diamond$ & $7\!\cdot\!10^{-4}$  & $1.3\!\cdot\!10^{-4}$ &  \checked &  \checked  &  \checked &  \checked & --  & -- & $<10^{-5}$  \\
\hline
\hline
  $ggH$ & 0.50 & 0.16 & -- & -- & -- & -- & -- & -- & $<10^{-2}$   \\
\hline
  $\gamma\gamma  H$ & --  & -- & -- & --  &-- & -- & 0.06 & -- & $<10^{-2}$  \\
\hline
  $Z\gamma  H$ & --  & \checked & -- & --  & -- & --  & -- & -- & $<10^{-2}$   \\
\hline
\hline
 $\tau\tau H$ &  \checked & \checked& $0.01$ & $0.01$  &  ${0.02}$  & ${0.06}$  & \checked & \checked & $<10^{-2}$  \\
\hline
 $ttH$ & \checked  & \checked & -- & -- & 0.29 & 0.08  & -- & -- & $<10^{-2}$  \\
\hline
 $\mu\mu H$ & --  & -- & -- & -- & -- & --  & -- & \checked & $<10^{-2}$   \\
\hline
\hline
\multicolumn{10}{l}{$^\dagger$ estimated in $H\to ZZ^*$ decay mode}\\
\multicolumn{10}{l}{$^\ddagger$ estimated in $V^*\to HV$ production mode}\\
\multicolumn{10}{l}{$^\Diamond$ estimated in $V^*V^*\to H$ (VBF) production mode}\\
\end{tabular}
\end{center}
\label{table-cpscenarios}
\end{table}

The $C\!P$ mixture study at an $e^+e^-$ collider was shown based on 500 fb$^{-1}$ at a centre-of-mass energy
of 350 GeV and $m_H=120$ GeV~\cite{AguilarSaavedra:2001rg}. 
Recent studies~\cite{Seattle-Whitbeck, Minneapolis-Gao,Anderson:2013fba} compare expected performance of an $e^+e^-$ collider 
and LHC. Precision on $C\!P$-odd cross-section fraction of 0.036 (0.044) is obtained at 250 GeV (500 GeV)
scenarios. However, these fractions correspond to different  $f_{C\!P}$ values in the $H\to ZZ$ decay,
due to different relative strength of  $C\!P$-odd and  $C\!P$-even couplings. The corresponding 
precision on $f_{C\!P}$ is $0.0007$ ($0.00004$)~\cite{Seattle-Whitbeck, Minneapolis-Gao,Anderson:2013fba},
 assuming that no strong momentum dependence of couplings occurs at these energies.  

A promising channel to study ${C\!P}$ violation is the decay $H\to \tau^+\tau^-$. 
Spin correlations are possible to use in the $\tau$ decay. For example, the 
pion is preferably emitted in the direction of the $\tau$ spin in the $\tau$ rest frame.
These studies are performed in the clean $e^+e^-$ environment, while it is extremely
difficult in proton collisions. Several studies have been performed, 
in the decays $\tau\to\pi\pi\nu$~\cite{Desch:2003rw,Harnik:2013aja}, 
and all final states~\cite{Seattle-Berge, Berge:2012wm,Berge:2013jra}.
All studies agree on a similar precision of about $5^\circ$ for the typical scenarios
in Table~\ref{table-cpscenarios}.
The above estimate translates to approximately 0.01 precision on $f_{C\!P}$.
The precision becomes somewhat worse with increased collider energy due to
reduced $ZH$ production cross-section, and this technique relies on the knowledge
of the $Z$ vertex. 
A recent study~\cite{Harnik:2013aja} indicates that with 3000 fb$^{-1}$ at LHC,
the $C\!P$ phase could be measurable to an accuracy of about 11$^\circ$. 
 
A study of ${C\!P}$-odd contribution in the $ttH$ coupling has been studied in the context 
of ILC~\cite{Minneapolis-Fujii}. Cross-section dependence on the coupling has been employed
and an uncertainty of 0.08 (0.29) at 1000 (500) GeV center-of-mass energy has been estimated. 
A beam polarization of $(+0.2,-0.8)$ and $(+0.3,-0.8)$ is assumed at 1000 and 500 GeV, respectively.
These estimates further improve to 0.05 (0.16) for the luminosity upgrade of the ILC. 
Interpretation of a cross-section deviation as an indication of ${C\!P}$-odd coupling contribution
is strongly model-dependent, but allows access to anomalous $ttH$ couplings.

Beam polarization in the photon and muon colliders would be essential for $C\!P$ measurements in the
$\gamma\gamma H$ and $\mu\mu H$ couplings.  
Three parameters ${\cal A}_1, {\cal A}_2, {\cal A}_3$ sensitive to $C\!P$ violation
have been defined in the context of the photon collider~\cite{Grzadkowski:1992sa, Kramer:1993jn, Gunion:1994wy}.
The ${\cal A}_1$ parameter can be measured as an asymmetry in the Higgs boson production cross-section
between the $A_{++}$ and $A_{--}$ circular polarizations of the beams. This asymmetry 
is the easiest to measure, but it is proportional to $\Im(a_2a_3^*)$ and is zero
when in Eq.~(\ref{eq:fullampl-spin0}) $a_2$ and $a_3$ are real,
as expected for the two loop-induced couplings with heavier particles in the loops. 
A more interesting parameter:
%
\begin{eqnarray}
{\cal A}_3 
= \frac{|A_{\parallel}|^2-|A_{\perp}|^2}{|A_{\parallel}|^2+|A_{\perp}|^2}
= \frac{2\Re(A_{--}^*A_{++})}{|A_{++}|^2+|A_{--}|^2}
= \frac{|a_2|^2-|a_3|^2}{|a_2|^2+|a_3|^2}
= (1-2f_{C\!P})
\label{eq:photoncp}
\end{eqnarray}
%
can be measured as an asymmetry between two configurations with the linear
polarization of the photon beams, one with parallel and the other with orthogonal 
polarizations. 
In Ref.~\cite{Asner:2001ia} careful simulation of the process has been performed. 
The degree of linear polarization at the maximum energies is 60\% for an electron
beam of energy $E_0 \approx 110$ GeV and a laser wavelength $\lambda \approx 1\, \mu{m}$.
The expected uncertainty on ${\cal A}_3$ is 0.11 for $2.5\cdot 10^{34} \times
10^7$ = 250 fb$^{-1}$ integrated luminosity. This translates to a $f_{C\!P}$
uncertainty of 0.06. 
The $C\!P$ mixture study at a photon collider was also shown based on a sample of 50,000 raw
$\gamma\gamma\to h$ events assuming 80\% circular polarization of both electron beams~\cite{Chou:2013xaa}.  
This study corresponds to a ${\cal A}_1$ asymmetry measurement, 
with expected precision on ${\cal A}_1$ of about 1\%. 
However, this asymmetry is expected to be zero with real coupling constants $a_2$ and $a_3$
and is therefore of limited interest compared to $f_{C\!P}$.

At the muon collider,
the $C\!P$  quantum numbers of the states can be determined if the muon
beams can be transversely polarized.  
The cross section for production of a resonance~\cite{Grzadkowski:2000hm}
depends on $P_T$ ($P_L$), the degree of transverse (longitudinal)
polarization of each of the beams and $\zeta$ is the angle of the
$\mu^+$ transverse polarization relative to that of the $\mu^-$
measured using the direction of the $\mu^-$ momentum as the $z$ axis.
In particular, muon beams polarized in the same transverse direction
selects out the $C\!P$-even state, while muon beams polarized in opposite
transverse directions (i.e., with spins $+1/2$ and $-1/2$ along one
transverse direction) selects out the $C\!P$-odd state. 

Several other measurements are possible on $pp$, $e^+e^-$, photon, and muon colliders,
which are left to future feasibility studies. 
In Table~\ref{table-cpscenarios} we summarize various couplings where
$C\!P$ measurements are possible.


\section{Mass and Total Width Measurements}\vspace*{-0.3cm}

A broadening of the total width of the Higgs boson relative to
Standard Model predictions is the clearest, model-independent
discovery mode for new physics.  The experimental challenge is to make
a model-independent measurement of the total width that reaches the
level of the theoretical uncertainty on this quantity in the Standard
Model.  The total width of the Standard Model Higgs boson is predicted
to be approximately 4~MeV for a boson with a mass of 125~GeV.  The
Standard Model decay modes to $b\bar{b}$, $W^+W^-$, $\tau^+\tau^-$,
$gg$, $c\bar{c}$, and $ZZ$ account in total for over 99\% of the total
width.  The Higgs to $b\bar{b}$ branching fraction at roughly
$(58 \pm 3)\%$ is the single largest contribution to the theoretical
uncertainty on the total width at this time.  With the anticipated
improvements in the precision of input parameters, especially
$\alpha_s$ and $m_b$ from lattice QCD, as well as full implementation
of the NLO electroweak radiative corrections to $h \to f \bar f$, the
Standard Model prediction on the total width should achieve an
accuracy around 1\%.  The experimental measurement of the Higgs to
$b\bar{b}$ branching fraction in $ZH$ production to sub-percent
accuracy will independently reduce the uncertainty on the Higgs total
width prediction to 1\%.

There are three proposed techniques to access the total width of the
Higgs boson: interferometry in $gg (\to H) \to \gamma\gamma$ or $gg
(\to H) \to ZZ$, measuring a partial width via a cross section and the
corresponding branching fraction, and a direct lineshape scan.

The mass of the Higgs boson provides an important self-consistency
test of the electroweak Standard Model at the quantum level: radiative
corrections involving the Higgs boson contribute to the SM prediction
for the $W$ mass.  At current precisions, the electroweak fit
indirectly predicts $m_H = 94^{+29}_{-24}$~GeV (1$\sigma$
range)~\cite{Freitas:2013xga}.  Foreseeable improvements of $m_t$ to
0.1~GeV, $M_W$ to 5--6~MeV, and $\sin^2 \theta^{lept}_{\eff}$ to $1.3
\times 10^{-5}$ achievable with the ILC/GigaZ option would tighten this
constraint to $\delta m_H \approx \pm 10$~GeV~\cite{Freitas:2013xga}.
A discrepancy between the SM prediction for $m_H$ extracted from the
precision electroweak fit and the directly measured mass would
constitute clear evidence for new physics.  The current sub-GeV
uncertainty in the Higgs mass from the LHC
experiments~\cite{Aad:2013wqa,CMS:yva} is already much better than the
precision needed to make this test.

An important issue is that the Higgs mass uncertainty also feeds into the uncertainty on the 
Higgs couplings to SM fermions and gauge bosons through the kinematic 
dependence of the Higgs branching ratios.  A 100~MeV uncertainty in 
$m_H$ corresponds to a 0.5\% uncertainty in the ratio of couplings 
$\kappa_b/\kappa_W$.

The mass of the Higgs boson is the most sensitive parameter in
determining whether the electroweak vacuum is stable.  A
vacuum-to-vacuum decay of the Higgs vacuum expectation value from the
electroweak scale to the Planck scale would cause a massive increase
in the relative strength of the gravitational forces on elementary
particles and cause catastrophic changes to the large-scale structures
in the universe.
Assuming only the SM up to the Planck scale, the Higgs mass needed for
vacuum stability is given by~\cite{Gupta:2013zza,Degrassi:2012ry}
\begin{equation}
  m_H > 129.4~{\rm GeV} 
  + 1.4~{\rm GeV} \left( \frac{m_t - 173.2~{\rm GeV}}{0.7~{\rm GeV}} \right)
  - 0.5~{\rm GeV} \left( \frac{\alpha_s(M_Z) - 0.1184}{0.0007} \right)
  \pm 1~{\rm GeV},
\end{equation}
where the last $\pm 1$~GeV is the theoretical uncertainty coming
mainly from the low-energy matching scale for the quartic coupling in
the Higgs potential.  The top quark mass uncertainty plays an
important role.  To match a Higgs mass uncertainty of $\delta m_H \sim
150$~MeV, the top mass uncertainty must be below 100~MeV, comparable
to the expected uncertainty from an $e^+e^-$ threshold scan~\cite{Asner:2013hla}.

Another use of the Higgs mass is to test parameter relations in
theories beyond the SM, such as the MSSM, in which the Higgs mass is
predicted in terms of the $Z$ boson mass, $\tan\beta$, $M_A$, and
radiative corrections coming mainly from top-quark and top-squark
loops.  The latter depend strongly on $m_t$, leading to an uncertainty
in the predicted Higgs mass of order 100~MeV for $\delta m_t \sim
100$~MeV~\cite{Gupta:2013zza}.  The usefulness of a Higgs mass
measurement at the 100~MeV level or below thus depends on a precision
measurement of the top quark mass.

Results from different facilities below are summarized in Table~\ref{tab:masswidthsummary}.

\subsection{Hadron colliders}\vspace*{-0.4cm}
\label{sec:masswidthLHC}

The Higgs boson mass can be measured directly from the $H\to\gamma\gamma$ and $H\to ZZ^*\to 4\ell$ decays
at the LHC. Based on the dataset taken during the  2011-2012 running of a combined luminosity of 
$\sim 25$~fb$^{-1}$ at 7 and 8 TeV, ATLAS and CMS have measured the 
mass to be $125.5\pm 0.2 \thinspace ({\rm stat}) ^{+0.5}_{-0.6} \thinspace ({\rm syst})$~GeV~\cite{Aad:2013wqa} and 
$125.7\pm 0.3 \thinspace ({\rm stat})\pm 0.3 \thinspace ({\rm syst})$~GeV~\cite{CMS:yva} 
respectively, with a poor-man's average of 0.25~GeV uncertainty from the statistics and 0.45~GeV for the
systematics.

The statistical uncertainty on the mass of the current measurement is already smaller (or comparable) than
the systematic uncertainty. With the expected $\times 2.5$ increase in Higgs cross section from 8~TeV 
to 14~TeV, the statistical uncertainty is expected to reduce to $\sim 50$~MeV and $\sim 15$~MeV with 
300 and 3000~fb$^{-1}$ at 14~TeV, repectively. Thus the precision of the future measurement will likely be dominated by 
systematics. The largest contribution is the knowledge of the energy/momentum scale of photons, electrons 
and muons, which should improve with increasing statistics. If one makes the optimistic assumption that
that the systematics also scales with statistics, the expected systematic uncertainty is $\sim 70$~MeV 
and $\sim 25$~MeV at 300 and 3000~fb$^{-1}$. At this precision, theoretical issues such as the 
interference between the $H\to\gamma\gamma$ resonance and the continuum will need to be taken into account. 
However,
the $H\to ZZ^*\to 4\ell$ decay is not expected to suffer from such interference, in particular the
$H\to ZZ^*\to 4\mu$ decay should be free of the limitations from many of experimental and systematic 
uncertainties. These arguments suggest that a precision of $\sim 100$~MeV and $\sim 50$~MeV per experiment 
on the Higgs mass should be achievable at the LHC with 300 and 3000 fb$^{-1}$.

Both ATLAS and CMS have studied Higgs mass precisions in their technical design 
reports~\cite{ATLAS:1999vwa,Ball:2007zza}~(Fig. 19-45 of the ATLAS report and Fig. 10.37 of CMS report). 
ATLAS estimates that a relative precision of 0.07\% is 
achievable with 300~fb$^{-1}$ while CMS projects a statistical uncertainty of 0.1\% with 30~fb$^{-1}$, 
both for $m_H=125$~GeV. These results are consistent with the estimates above.

The most direct method of measuring of the Higgs width is to fit the mass distributions of the 
observed $H\to \gamma\gamma$ and $H\to ZZ^*\to 4\ell$ resonances. However, since the experimental mass 
resolutions are much larger than the expected SM Higgs width, this method is not expected to be sensitive 
to $\Gamma_H^{SM}\ (\approx 4.2~{\rm MeV})$. Indeed the expected upper bound on $\Gamma_H$ from the CMS 
$H\to\gamma\gamma$ analysis of the 7 and 8 TeV datasets is 5.9~GeV~\cite{CMS:PAS-HIG-13-016}, or 
$1400\times\Gamma_H^{SM}$.  With the assumption that
the bound scales with the number of Higgs candidates and combining with the $H\to ZZ^*\to 4\ell$ analysis, 
an upper limit of $\sim 200$~MeV should be achievable with 3000~fb$^{-1}$, corresponding to 
$\sim 50\times \Gamma_H^{SM}$.

New theoretical 
developments~\cite{Martin:2012xc,Martin:2013ula,Dixon:2013haa}
open the possibility to significantly improve the sensitivity 
using the Higgs mass measurement in the $\gamma\gamma$ channel.
Interference between $gg \to H \to \gamma\gamma$ and the continuum $gg
\to \gamma\gamma$ background cause a shift in the reconstructed Higgs
mass in the $\gamma\gamma$ final state of about
$-70$~MeV~\cite{Dixon:2013haa}, which grows with increasing Higgs
total width. ATLAS has studied the mass shifts for two different
Higgs $p_T$ ranges of the $H\to\gamma\gamma$ decays and estimates that an 
upper bound of $40\times \Gamma_H^{SM}$ on the total Higgs decay width
can be obtained with 3000~fb$^{-1}$~\cite{ATLAS-PHYS-PUB-2013-014}. 
A direct comparison of the Higgs mass determination in $ZZ \to
4\mu$ and $\gamma\gamma$ should have better sensitivity. It is
estimated that an $\mathcal{O}(100\%)$ measurement of the Higgs total width may be 
possible, although no
study of the possible future precision has been done.  Because the
sign of the mass shift depends on the sign of $\kappa_g
\kappa_{\gamma}$, this measurement also has the potential to determine
the relative sign of these two loop-induced couplings.

Another approach was very recently proposed in
Ref.~\cite{Caola:2013yja} using off-shell $gg \to H \to ZZ$ production. 
This process is strongly enhanced above the $ZZ$ (and $WW$)
pair-production threshold~\cite{Kauer:2012hd}.  
Based on the current LHC data for the $ZZ$ production
cross section, Refs.~\cite{Caola:2013yja,Campbell:2013una} estimate a bound $\Gamma_H
\leq (20-45)\times\Gamma_H^{\rm SM}$ at 95\% CL, depending on the $4\ell$
invariant mass range studied and assuming that $\kappa_g \kappa_Z$ has
the same sign as in the SM.  The ultimate precision of this method
will be limited by the uncertainty on the $ZZ$ cross section; taking
an optimistic 3\% ultimate precision on the cross section would yield a
Higgs width bound of $\Gamma_H \leq (5-10)\times \Gamma_H^{\rm SM}$ at 
95\% CL~\cite{Caola:2013yja}.  Off-shell $gg \to H \to WW$ can provide 
a complementary measurement~\cite{Campbell:2013wga}.

A more stringent limit on $\Gamma_H$ can be set from the coupling fit with 
the assumption of $\kappa_W, \kappa_Z \le 1$. From Table~\ref{tab:LHCCouplingProjections}, 
an upper bound of $\Gamma_H^{SM}/(1-{\rm BR}_{\rm BSM}) \approx 1.12\times\Gamma_H^{SM}$ 
can be obtained with 3000~fb$^{-1}$. Assuming only SM production modes and decays, the total Higgs 
width can be measured with a $5-8\%$ precision at the HL-LHC. 

\subsection{{$\mathbf{e^+e^-}$} colliders}\vspace*{-0.4cm}

Using the process $e^+e^- \rightarrow ZH$ with $Z \rightarrow \mu^+\mu^-$ and $Z \rightarrow e^+e^-$,
plus a measurement of the beamstrahlung energy distribution, a
precision measurement of the Higgs mass can be made from
the shape and distribution of the invariant mass recoiling against the
reconstructed $Z$~\cite{CLICWhite:2013a,Azzi:2012a,ILD1,Li:2012taa}.
Indeed, the specification of required momentum resolution in the linear collider detector designs,
particularly for muons, is optimized on the precision of the mass measurement.

To address the total Higgs width, as described in Sec.~\ref{sec:couplings}, at lower energies of
$\sqrt{s} \sim 250$~GeV
involves a measurement
of the total production cross section
 in $e^+e^- \to ZH$, independent of branching ratios, which can be
 done using the recoil mass technique.
 The partial width $\Gamma(H \rightarrow ZZ)$ is directly proportional to
 the cross section, and from a measurement of the complementary branching fraction ${\rm BR}(H \rightarrow ZZ)$,
 a totally  model-independent total Higgs width can be extracted: 
 $\Gamma_H = \Gamma(H \rightarrow ZZ)/{\rm BR}(H \rightarrow ZZ)$.
 At higher energies, a measurement of the cross section for
 the $WW$-fusion process $e^+e^- \rightarrow \nu_e \bar{\nu}_e H$ along with
 ${\rm BR}(H \rightarrow WW^*)$ provides a further improvement on the extracted width,
 as summarized in Table~\ref{tab:masswidthsummary}.

\subsection{Muon collider}\vspace*{-0.4cm}
\label{sec:masswidthmuon}

A direct lineshape scan of the Higgs boson in $s$-channel production
will achieve sub-MeV precision on the mass. This precision is
unmatched using any other known technique.  The beam spread and beam
energy resolution at a muon collider is good enough to resolve the SM
Higgs width of $\sim 4$~MeV directly through a line scan with a
precision of 4.3\%.

The Higgs total width is predicted to be $3\times 10^{-5}$ of the
resonance center-of-mass energy.  Therefore, a beam energy scan will
be needed to locate the central value of the Higgs resonance.  Input
on the mass value from previous measurements will be important to
reduce the scan range.  The muon collider
proposal~\cite{MuonColliderWhitepaper} envisions measuring the Higgs
mass, total width, and production rates in the $bb$, $WW^*$ and
$\tau\tau$ final states with a 5-point energy scan centered on the
Higgs resonance at $\sqrt{s} \sim 126$~GeV, with a scan point separation
of 4.07~MeV.  The run scenario assumes one Snowmass year ($10^7$~s) at
$1.7 \times 10^{31}$~cm$^{-2}$s$^{-1}$ plus five Snowmass years at
$8.0 \times 10^{31}$~cm$^{-2}$s$^{-1}$ and a beam energy resolution of
$R = 4 \times 10^{-5}$.  Achievable precisions on the Higgs mass and
width are $\Delta m_H = 0.06$~MeV and $\Delta \Gamma_H = 0.18 \ {\rm
MeV} = 4.3\%$.


\subsection{Summary}\vspace*{-0.4cm}

A summary of the Higgs mass and width measurement capabilities for the 
facilities is given in Table~\ref{tab:masswidthsummary}.

\begin{table}[htb!]
\caption{Summary of the Higgs mass and total width measurement precisions
of various facilities.  
$^\ddag$ILC luminosity
   upgrade assumes an extended running period on top of the low luminosity program
   and cannot be directly compared to TLEP and CLIC numbers without accounting for
   the additional running period.  The ILC assumes 0.1\% theory uncertainties.
}
\scriptsize
\begin{center}
\begin{tabular}{lcccccccc}
\hline\hline
Facility & LHC & HL-LHC & ILC500 & ILC1000 & ILC1000-up & CLIC & TLEP (4 IP) & $\mu$C \\
$\sqrt{s}$ (GeV) & 14,000 & 14,000 & 250/500 & 250/500/1000 & 250/500/1000 & 350/1400/3000 & 240/350 & 126 \\
$\int \mathcal{L} dt$ (fb$^{-1}$) & 300 & 3000 & 250+500 & 250+500+1000 & 1150+1600+2500$^\ddag$ & 500+1500+2000 & 10,000+2600 & 4.2 \\
\hline
$m_H$ (MeV) & 100 & 50 & 32 & 32 & 15 & 33 & 7 & 0.06 \\
$\Gamma_H$ & -- & -- & 5.0\% & 4.6\% & 2.5\% & 8.4\% & 1.0\% & 4.3\% \\
\hline\hline
\end{tabular}
\end{center}
\label{tab:masswidthsummary}
\end{table}


\section{Direct searches for BSM Higgs bosons $\mathbf{H^0}$, $\mathbf{A^0}$, $\mathbf{H^{\pm}}$}\vspace*{-0.3cm}

\subsection{Theory}\vspace*{-0.4cm}
Many well-motivated extensions of the SM contain a second Higgs
doublet, including the MSSM.  Including a second doublet introduces an
additional four scalar degrees of freedom beyond the SM-like Higgs
boson $h$.  These are commonly denoted $H^0$ ($C\!P$-even neutral), $A^0$
($C\!P$-odd neutral), and $H^{\pm}$ (charged Higgs pair).  If $C\!P$ is
violated in the Higgs sector, the three neutral states (including $h$)
can be $C\!P$ mixtures.

Because electroweak symmetry breaking is shared between the two
doublets, the couplings of the additional scalars depend on the
couplings of the discovered SM-like Higgs boson $h$.  Here we assume
that $h$ is nearly SM-like and the new scalars are heavier than $h$.
The two Higgs doublets can be written as $\Phi_1$ and $\Phi_2$ with
\begin{equation}
  \Phi_j = \left( \begin{array}{c} \phi_j^+ \\ \phi_j^0 \end{array} \right)
  = \left( \begin{array}{c} \phi_j^+ \\ 
    (v_j + \phi_j^{0,r} + i \phi_j^{0,i})/\sqrt{2} \end{array} \right),
\end{equation}
where $v_1$ and $v_2$ are the vacuum expectation values of the two
doublets normalized according to $v_1^2 + v_2^2 = v_{\rm SM}^2 \simeq
246$~GeV.  Their ratio is a free parameter $\tan\beta \equiv v_2/v_1$.
The second free parameter $\alpha$ is the mixing angle in the
$h$--$H^0$ sector.  The ``mismatch'' between the two mixing angles,
$\cos(\beta - \alpha)$, controls how SM-like the Higgs $h$ is.

In the \emph{decoupling limit}, $M_{H^0} \simeq M_{A^0} \simeq
M_{H^{\pm}} \gg M_Z$, the properties of $h$ approach those of the SM
Higgs boson.  This limit occurs in the MSSM when $M_A \gg M_Z$.  It
also occurs in more general two Higgs doublet models (2HDMs) when the
scalar quartic couplings are not allowed to become too large.  In this
limit, discovery of the heavy Higgs particles becomes difficult
because of kinematic suppression of cross sections.

\subsection{LHC constraints and projections}\vspace*{-0.4cm}
The heavy Higgs bosons $H^0$, $A^0$, and $H^{\pm}$ have been searched for
at the LHC in the context of the MSSM, as well as a more general search
for $H^0$ in $WW$ and $ZZ$ final states in a two-Higgs-doublet model.

$H^0$ and $A^0$ are produced through gluon fusion as well as $b \bar
b$ fusion at large $\tan\beta$.  Production cross sections are
rescaled from the SM gluon-fusion calculations of the LHC Higgs Cross
Sections Working Group~\cite{Heinemeyer:2013tqa} and the $b \bar b$
fusion code {\tt bbh@nnlo}~\cite{Harlander:2003ai}.  $b \bar b$ fusion
dominates the production cross section for moderate to large $\tan\beta$,
leading to highest sensitivity at large $\tan\beta$.
Decay branching ratios are calculated assuming the MSSM $m_h^{\rm
max}$ scenario with $M_{\rm SUSY} = 1$~TeV.  This implies that the
$H^0$ and $A^0$ signals are both present and their mass splitting is
fixed point-by-point in the parameter space.

From a search for the $\tau\tau$ final state using 17~fb$^{-1}$ at 7
and 8~TeV, CMS excludes $M_A$ values below 800~GeV for $\tan\beta =
50$, falling to 250~GeV for $\tan\beta = 5$ (no exclusion is made for
$\tan\beta < 5$)~\cite{CMS:gya}.
Searches in the $\mu\mu$ final state have much better mass resolution
but are currently less sensitive due to the smaller branching ratio.

LHC searches for heavy Higgs states in a generic 2HDM are also
underway.  ATLAS searched for $H^0 \to WW$ in the context of Type-I
and -II 2HDMs for various $\tan\beta$ values as a function of the
mixing angle $\alpha$, and excludes ranges of $\alpha$ for $M_{H^0}$
as high as 250~GeV (13~fb$^{-1}$ at 8~TeV)~\cite{ATLAS:2013zla}.

A recent study of the decays $H\rightarrow ZZ$
 and $A\rightarrow Zh$ demonstrates strong complementarity
 between direct search and precision measurements 
of the observed Higgs-like boson couplings, 
in terms of ability to constrain 2HDM parameter space\cite{Brownson:2013lka}.
  Precision measurements are unable to constrain 2HDMs 
near the alignment limit\cite{Brownson:2013lka,chen:2013qda}.  
However, if nature can be described by a type II 2HDM, 
with $3000$~fb$^{-1}$ of data at $\sqrt{s} = 14$ TeV,
 a $300$ GeV scalar could be discovered 
via direct search for $\tan\beta = 1$ and 
$|\cos(\beta-\alpha)|$ as small as $0.005$, 
and in the pseudo-scalar case with $|\cos(\beta-\alpha)|$
 as small as $0.012$. 


The direct search for heavy Higgs bosons  at the LHC in the MSSM excludes
regions in the $\tan\beta$-$M_A$ plane.  At relatively low $M_A$, the region
at high $\tan\beta$ is excluded by searches for $A\rightarrow
\tau^+\tau^-$, while at smaller $\tan\beta$ and small $M_A$ the exclusion
results from the heavier $H$ decaying to $ZZ$ and $W^+W^-$.  With
25~fb$^{-1}$, Ref.~\cite{Djouadi:2013vqa} estimates that for all
values of $\tan\beta$ the region below about $M_A\sim 200$~GeV can be
excluded.  The location of this wedge is increased to $\sim 380$~GeV
with 300~fb$^{-1}$\cite{Lewis:2013fua}.

If $A^0$ and $H^{\pm}$ are not too heavy, they will be pair produced
at LHC via the electroweak processes $pp\to H^{\pm} A^{0},\
H^{+}H^{-}$, with the decays $A^0 \to bb$ and $H^{\pm} \to \tau\nu$
dominating in large parts of the MSSM and 2HDM parameter space.  The
cross section depends only on the particle masses.  A recent
parton-level study~\cite{Christensen:2012si} estimated that both
processes should be discoverable at the 14~TeV LHC with less than
20~fb$^{-1}$ if $M_A \sim 95$--130~GeV and, in fact, could already be 
discoverable in the current 8~TeV LHC data-set.

At low $\tan\beta$, signals of interest are decays of $A^0/H^0$ to
charginos or neutralinos (this depends strongly on SUSY model spectrum
assumptions).  Decays $A^0 \to Z h$~\cite{Coleppa:2013xfa}, $H^0 \to hh$,
$H^0/A^0 \to \gamma\gamma$, and $H^0 \to WW$ are also of interest.

\subsection{Projections for $\mathbf{e^+e^-}$ machines}
\begin{figure}[h]
\begin{center}
\includegraphics[width=0.60\textwidth]{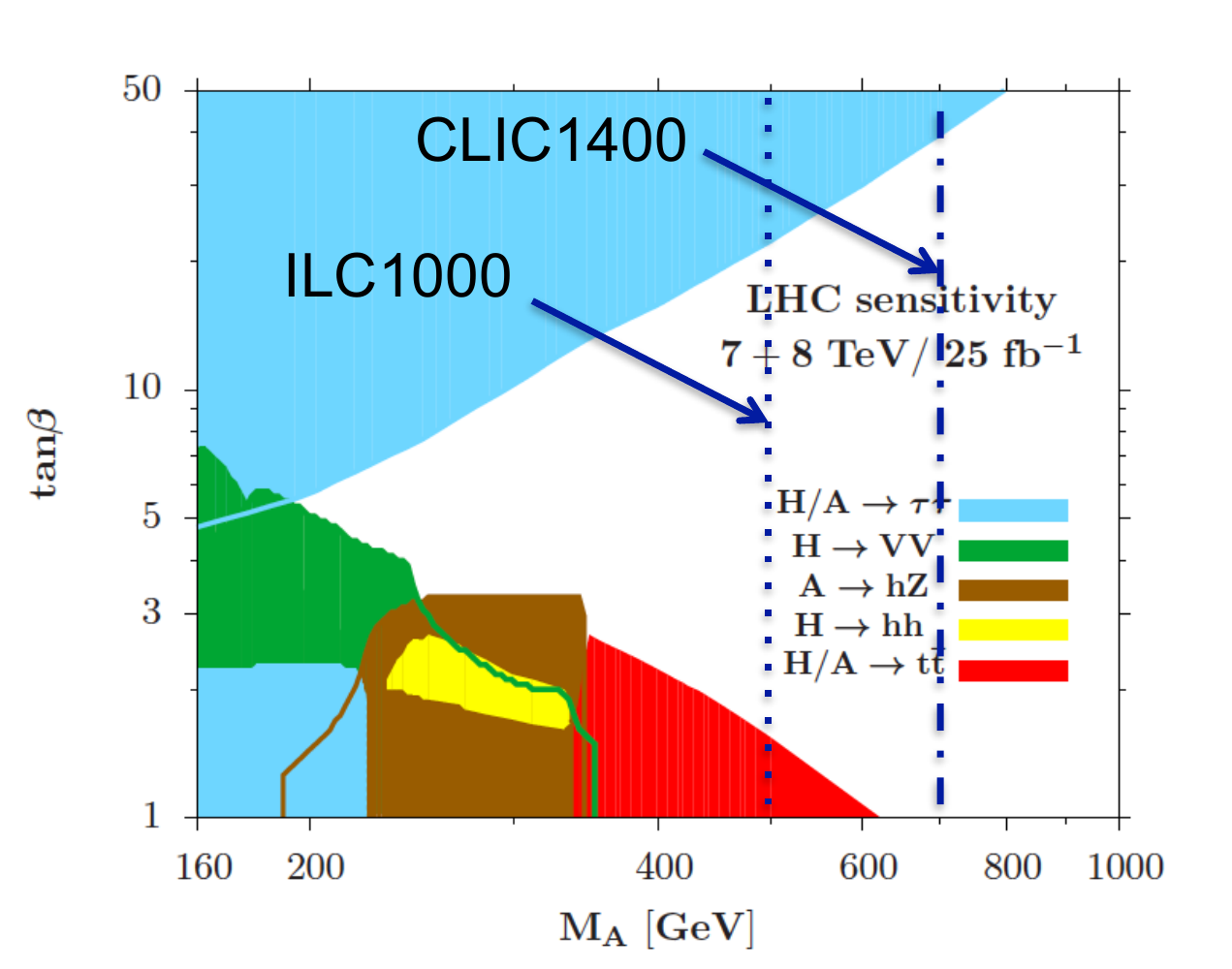}
\caption{MSSM Higgs sector search reach in the $m_A$--$\tan \beta$ plane for $e^+e^-$ colliders compared to the expected LHC 7+8~TeV upper limits (95\% C.L.).\label{MSSMwedge}}
\end{center}
\end{figure}

\vspace*{-0.4cm}
At an $e^+e^-$ collider, the cross section for $e^+e^- \to Z^* \to H^0
Z$ is suppressed by $\cos^2(\beta - \alpha)$ compared to the cross
section for a SM Higgs with the same mass as $H^0$.  In the decoupling
limit, the associated production cross section for $e^+e^- \to Z^* \to
H^0 A^0$, which is proportional to $\sin^2(\beta - \alpha)$, is
maximal, but requires associated production of two heavy particles,
limiting the kinematic reach to half the collider center-of-mass
energy (similarly for $e^+e^- \to H^+ H^-$).
Charged Higgs pair production, $e^+e^- \to H^+H^-$, is a pure gauge
process and hence also unsuppressed.  With sufficient luminosity, the
discovery reach for these states at an $e^+e^-$ collider is thus close
to the kinematic limit:
\begin{equation}
  M_{H^+} < \sqrt{s}/2, \qquad 
  M_{H^0} + M_{A^0} < \sqrt{s}.
\end{equation}
Since the mass splitting between $H^0$ and $A^0$ is typically small
in the decoupling region, the reach for either of them is roughly
$\sqrt{s}/2$~\cite{CLICCDR_vol2}, as shown in Fig.~\ref{MSSMwedge}.

\subsection{Resonant production at a muon collider}\vspace*{-0.4cm}
The neutral heavy Higgs bosons $H^0$ and $A^0$ can be produced as
$s$-channel resonances in $\mu^+ \mu^-$ or $\gamma\gamma$ collisions.
They can also be pair produced via electroweak processes as at $e^+e^-$
machines.

If the heavy Higgs bosons $H^0$ and $A^0$ are not very light, resonant
production at a muon collider may be the best opportunity to study
their properties in detail.  This was studied in
Ref.~\cite{Eichten:2013ckl} for the ``Natural Supersymmetry''
benchmark point of Ref.~\cite{Baer:2012yj}, which has $M_{A^0} \simeq
M_{H^0} \simeq 1.55$~TeV and $\tan\beta = 23$.  The mass difference
between $A^0$ and $H^0$ is about 10~GeV and their decay widths are
around 20~GeV.  

The parton-level analysis~\cite{Eichten:2013ckl} was based on a
center-of-mass energy scan over a 200~GeV range centered at 1550~GeV
in 100~steps, collecting a total of 500~fb$^{-1}$.  Signal and
background cross sections in the $b \bar b$ and $\tau\tau$ final
states were computed using PYTHIA6 modified to include a Gaussian beam
energy spread of 0.1\%.  The overlapping lineshapes were then fitted
with two Breit-Wigners in the $b \bar b$ final state (a single
Breit-Wigner is ruled out at high confidence) allowing extraction of
the masses to $\pm 0.5$~GeV, the widths to $\sim 3.5$\%, and 
the peak $\sigma \times {\rm BR}(b \bar b)$ to 9\% for the two
states.  The $\tau\tau$ channel can then be used to measure ${\rm
  BR}(\tau\tau)/{\rm BR}(b \bar b)$ with an uncertainty of about 10\%.
As a bonus, decays of $H^0$ and $A^0$ to charginos or neutralinos
may provide the largest sample of the heavier ones of these particles,
whose direct production cross sections can be quite small at lepton
colliders~\cite{Eichten:2013ckl}.

The $C\!P$  quantum numbers of the states can be determined if the muon
beams can be transversely polarized, see Sec.~\ref{sec:cp} for details. 
This would allow identification of the two resonances as $A^0$ and $H^0$ as well
as probing for $C\!P$-violating mixing between the states.
Similar techniques are possible at a photon collider.

\subsection{Resonant production at a photon collider}\vspace*{-0.4cm}
The neutral heavy Higgs bosons $H^0$ and $A^0$ can be produced as $s$-channel
resonances in high-energy $\gamma\gamma$ collisions.  Such a high-energy
photon collider could be realized as an option at a high-energy $e^+e^-$
collider such as ILC or CLIC.  The photon collider offers the following
features:
\begin{itemize}

\item High mass reach, $M_{H^0,A^0} \simeq \sqrt{s_{\gamma\gamma}}$.

\item Selective production of CP-even or CP-odd states through photon
  beam polarization.

\item Capability to measure ratios of couplings by taking ratios of rates
into different final states.

\item Sensitivity to the loop-induced $H^0\gamma\gamma$ and $A^0\gamma\gamma$
couplings through the production cross section, measured in the form 
$\sigma \times {\rm BR}$.


\end{itemize}




\section{Conclusions}\vspace*{-0.3cm}

A precision Higgs physics program is compelling because the Standard Model
precisely predicts all Higgs boson couplings and properties with no free
parameters, now that the Higgs mass is known.  Any deviation from these
predictions therefore represents clear evidence for new physics.
The current outlook on high-$p_{\rm T}$ physics from 8~TeV LHC
operation and the first measurements of the boson at 126~GeV indicate that the standard for 
discovering new physics in the Higgs sector beyond the 300~fb$^{-1}$, 14~TeV LHC program
requires an order of magnitude improvement in the measurement of Higgs couplings and properties.
This requirement translates to percent-level precisions in the leading couplings of the Higgs to matter.
The primary and perhaps most fundamental conclusion of this report is that a precision Higgs program
requires a combined program of high statistics production of the 126~GeV boson in an lepton collider
environment and a comprehensive evaluation of the fundamental parameters that enter the 
theory-experimental comparisons.

The following list are bulleted conclusions, highlighting the main outcomes of this report.

\begin{itemize}

\item 
The Higgs boson, a new state of matter, has been discovered at the LHC.
Understanding the properties of this new state is of fundamental importance
and deserves further investigation in the form of a precision experimental
program.
Any deviation in the predicted properties of the Higgs boson is a strong,
unabiguious signature for new physics.
Comparisons for different Higgs physics programs are provided in terms of the measurement 
precision on the mass, total width, spin, 
couplings, $C\!P$ mixtures, and the searches for multiple Higgs bosons.

\item Full exploitation of LHC and HL-LHC Higgs measurements will require improvements
in theoretical calculations of the gluon fusion Higgs production cross section, both
inclusive and with jet vetoes. To match sub-percent experimental uncertainties on
Higgs partial widths from Higgs factories will require consistent inclusion of higher
order electroweak corrections to Higgs decays, as well as an improvement of the
bottom quark mass determination to below $\pm$0.01 GeV.

\item LHC is the place to study Higgs boson in the next decade. The expected
precision of Higgs couplings to fermions and vector bosons, assuming only SM decay modes, are 
estimated to be $4-15$\% for 300~fb$^{-1}$ and $2-10$\% for 3000~fb$^{-1}$ at 14~TeV. Better precisions 
can be achieved for some coupling ratios.

\item  Given sufficient integrated luminosity, all
Higgs boson decays, including invisible or exotic final states, as well as those undetectable
at the LHC, are accessible in the $ZH$ production mode
at an $e^+e^-$ collider through the model-independent recoil mass technique.

\item  Precision tests of Higgs boson couplings to one-percent will require complementary
precision programs. Proposed Higgs factories such as linear or circular $e^+e^-$ colliders
and potentially a muon collider will be able to achieve these precisions for many of
the absolute couplings, and in a model-independent way.

\item HL-LHC can measure the Higgs boson mass with a precision of $\sim$50 MeV per experiment, however has limited 
sensitivity to the Higgs decay total width, even with SM assumptions.  Higgs factories such as ILC, CLIC, 
or TLEP will achieve a mass precision of about 35 MeV and measure the Higgs decay width up 
to $\sim$1.3\% in precision. Through a line-shape scan, a muon collider can measure 
the total width directly to 4.3\% and the mass to sub-MeV precision. 

\item Direct $ttH$ coupling measurements can be done at LHC, ILC, CLIC and muon colliders. 
The expected precisions are 7--10\% at HL-LHC per experiment, $\sim$2--3\% at ILC with luminosity upgrade
and $\sim$3\% at CLIC.  A high-energy 
muon collider is expected to have the comparable precision as CLIC per experiment.

\item  Higgs self-coupling is difficult to measure at any of these facilities. 
A 50\% measurement per experiment is expected from HL-LHC and 13\% from linear
$e^+e^-$ colliders at 1 TeV.  Improvement would need higher collision energies, with CLIC achieving 10\% at 3 TeV
and VLHC achieving 8\% at 100~TeV.

\item   The spin of the 126 GeV boson will be constrained by the LHC. A~limited parameter space 
of spin-two couplings may be left to be constrained by the data from the future facilities. 

\item Potential $C\!P$-odd fraction in $H\to ZZ^*$ cross-section ($f_{C\!P}$) will be measured 
by LHC to a few percent precision, with further improvement in VBF production. 
The $e^+e^-$ machines can measure this to a greater precision in the $ee\to ZH$. 
The $C\!P$ admixture in fermion couplings is not expected to suffer from loop 
suppression and can be studied in $H\to\tau\tau$ decay and $ttH$ production, 
leading to about 1\% precision on $f_{C\!P}$ in $\tau\tau H$ coupling at an $e^+e^-$ machine.
The photon and muon colliders are unique in their capability to probe $C\!P$ violation 
directly with polarized beams. 

\item  There are strong theoretical arguments for physics beyond the Standard Model. 
The LHC and CLIC have the highest discovery potential for heavy Higgs bosons as predicted by many 
Standard Model extensions.  At the LHC, the mass reach can be 1 TeV or higher with 3000 fb$^{-1}$ 
at 14 TeV, but is strongly model dependent. The mass reach is generally limited 
to less than half the collision energy for $e^+e^-$ colliders and potentially up to 
the collision energy for a muon collider through s-channel processes.

\end{itemize}

We have also arrived to the following facility comparison:

\begin{itemize}

\item LHC at 14 TeV with 300~fb$^{-1}$ of data is essential to firmly establish the five major production mechanisms
of a Higgs boson ($ggH$, VBF, $WH$, $ZH$, $t\bar{t}H$) and the main bosonic and fermionic decay modes
 ($b\bar{b}$, $WW^*$, $\tau^+\tau^-$, $ZZ^*$, $\gamma\gamma$). 
This will lead to about a factor of 3--5 improvement in the most precise measurements
compared to the 8 TeV run of LHC. This will also lead to about 100 MeV precision on the Higgs boson mass
and the measurement of the boson spin. 

\item  
HL-LHC provides unique capabilities to measure rare statisically limited SM decay modes such as $\mu^+\mu^-$,
$\gamma\gamma$, and $Z \gamma$ and make the first measurements of the Higgs self-coupling.
The high luminosity program increases the precision on the couplings compared to the LHC with 300~fb$^{-1}$
by roughly a factor of 2--3 and has a high discovery potential for heavy Higgs bosons.

\item  TeV-scale $e^+e^-$ linear colliders (ILC and CLIC) offer the full menu of measurements
of the 126~GeV Higgs boson with better precision than the LHC, though their mass reach for heavy Higgs bosons 
are generally weaker than high-energy $pp$ colliders, except for CLIC running at 3 TeV.
The two linear colliders have different
capabilities -- the ILC can run on the $Z$ peak while CLIC has a higher mass
reach and better precision in Higgs self-coupling measurement when operating at 3~TeV.

\item  TLEP offers the best precisions for most of the Higgs coupling measurements because of its
projected integrated luminosity and multiple detectors.  This program also includes high luminosity
operation at the $Z$ peak and top threshold.
There is no sensitivity to $ttH$ and $HHH$ couplings at these center-of-mass energies. 

\item A higher energy $pp$ collider such as a 33~TeV(HE-LHC) or 100~TeV(VLHC) hadron collider
provides high sensitivity
to the Higgs self-coupling as well as the highest discovery potential for heavy Higgs bosons.

\item  A TeV-scale muon collider should have similar physics capabilities as the ILC and
CLIC with potentially higher energy reach, but this needs to be demonstrated with more complete
studies.
The muon collider also has the potential for resonant production of heavy Higgs bosons.
$C\!P$ measurements are possible if a beam polarization option is included.

\item  A $\gamma\gamma$ collider is able to study $C\!P$ mixture and violation in the 
Higgs sector with polarized photon beams. It can improve the precision of the effective 
$\gamma\gamma H$ coupling measurement through s-channel production.

\end{itemize}

\section*{Acknowledgments}\vspace*{-0.4cm}

We gratefully acknowledge the ATLAS, CMS, ILC, and CLIC
collaborations, as well as the proponents of TLEP, the Muon
Collider, and photon colliders, without whose simulation work this report could not have
been written.  We are also grateful to the many theorists and
experimentalists who contributed to the community understanding of
Higgs physics reflected in this report.

Contributors acknowledge support from the U.S. Department of Energy, 
the U.S. National Science Foundation,
and the Natural Sciences and Engineering Research Council of Canada.


Finally, we thank the American Physical Society's Division of Particles and Fields 
for setting the charge for these studies and helping in their organization.

\bibliographystyle{h-physrev}

\bibliography{HiggsSnowmassReport} 


\end{document}